\documentclass[submission,Phys]{SciPost}
\usepackage{color,ifthen,amsthm,amsmath,amsxtra,amsfonts,dsfont,graphicx,bm,tikz,scalerel,wasysym,bbm,graphicx,amsthm,mathtools, tensor, mathrsfs,physics, comment}

\newcommand{\be}{\begin{equation}}
\newcommand{\ee}{\end{equation}}
\newcommand{\bea}{\begin{eqnarray}}
\newcommand{\eea}{\end{eqnarray}}
\newcommand{\bse}{\begin{subequations}}
\newcommand{\ese}{\end{subequations}}
\DeclareMathOperator{\Sp}{Sp}

\theoremstyle{plain}
\newcommand{\1}{\mathbbm{1}}

\renewcommand{\vec}[1]{{\bm{\mathrm{#1}}}}

\usetikzlibrary{arrows,calc,matrix,backgrounds,fit,positioning,decorations.pathmorphing,decorations.markings,patterns,decorations.pathreplacing,shapes.misc,external}
\definecolor{FcolU}{rgb}{0.71,0.78,0.91}
\definecolor{colVMPSgen}{rgb}{0.71,0.78,0.91}
\definecolor{IcolVMPSgen}{rgb}{0.71,0.8,0.76}
\definecolor{IcolVMPSgenc}{rgb}{0.71,0.41,0.42}
\definecolor{colMPS}{rgb}{0.27,0.45,0.77}
\definecolor{colSMPS}{rgb}{0.5,0.5,0.5}
\definecolor{colVMPS}{rgb}{0.96,0.74,0.59}
\definecolor{colLines}{rgb}{0.31,0.31,0.31}
\definecolor{colMPSLines}{rgb}{0,0.01,0.18}
\definecolor{colVMPSLines}{rgb}{0.11,0.11,0.11}
\definecolor{IcolUc}{rgb}{0.71,0.41,0.42}
\definecolor{colITensor}{rgb}{0.81,0.77,0.78}
\definecolor{colTensor}{rgb}{0.71,0.8,0.76}
\definecolor{IcolU}{rgb}{0.71,0.8,0.76}
\definecolor{IcolVMPSc}{rgb}{0.73,0.69,0.7}
\definecolor{IcolVMPS}{rgb}{0.81,0.77,0.78}
\definecolor{colObs}{rgb}{1.,1.,1.}
\definecolor{myred}{RGB}{232,102,102}
\definecolor{myblue}{RGB}{187,187,255}

\definecolor{full}{rgb}{0,0,0}
\definecolor{border}{rgb}{0.3,0.3,0.3}
\def\dx{0.3}
\def\r{0.08}
\def\sqrtThree{1.7320508075688772}

\newcommand\emptyrectangle[2]{
  \draw[border] ({(#1)},{(#2-1)})  -- ({(#1+1)},{(#2)})  -- ({(#1)},{(#2+1)})  
  -- ({(#1-1)},{(#2)})  -- cycle;
}

\newcommand\fullrectangle[2]{
  \draw[border,fill=full] ({(#1)},{(#2-1)})  -- ({(#1+1)},{(#2)})  -- ({(#1)},{(#2+1)})  
  -- ({(#1-1)},{(#2)})  -- cycle;
}

\newcommand\rectangle[3]{
  \ifthenelse{\equal{#3}{1}}{\fullrectangle{#1}{#2}}{\emptyrectangle{#1}{#2}};
}

\newcommand\mpsWire[4]{
	\draw [very thick,colMPSLines] ({(#2)*\dx},{-(#1)*\dx}) -- ({(#4)*\dx},{-(#3)*\dx});
}

\newcommand\IbottomHook[3]{
    \draw [semithick,colLines] ({(#2)*\dx-2*(#3)*\dx},{-(#1)*\dx}) arc (0:180:{-(#3)*\dx});
}

\newcommand\ItopHook[3]{
  \draw [semithick,colLines] ({(#2)*\dx},{-(#1)*\dx}) arc (0:180:{(#3)*\dx});
}

\newcommand\IrightHook[3]{
  \draw [semithick,colLines] ({(#2)*\dx},{-(#1)*\dx}) arc (90:-90:{(#3)*\dx});
}

\newcommand\IleftHook[3]{
  \draw [semithick,colLines] ({(#2)*\dx},{-(#1)*\dx}) arc (90:270:{(#3)*\dx});
}

\newcommand\gridLine[4]{
  \draw [semithick,colLines] ({(#2)*\dx},{-(#1)*\dx}) -- ({(#4)*\dx},{-(#3)*\dx});
}

\newcommand\griddashedLine[4]{
  \draw [semithick,dotted,colLines] ({(#2)*\dx},{-(#1)*\dx}) -- ({(#4)*\dx},{-(#3)*\dx});
}

\newcommand\vmpsWire[4]{
	\draw [double,colVMPSLines]({(#2)*\dx},{-(#1)*\dx}) -- ({(#4)*\dx},{-(#3)*\dx});
}

\newcommand\vmpsV[3]{
	\draw [thick,rounded corners=0.5,colVMPSLines,fill=#3] ({\dx*#2},{-\dx*#1})
    +({\r/2.},{-\sqrtThree*\r/2.}) -- +({\r/2.},{\sqrtThree*\r/2.}) 
	-- +({-\r},0) -- cycle;
}

\newcommand\vmpsW[3]{
	\draw [thick,rounded corners=0.5,colVMPSLines,fill=#3] ({\dx*#2},{-\dx*#1})
    +({-\r/2.},{-\sqrtThree*\r/2.}) -- +({-\r/2.},{\sqrtThree*\r/2.}) 
    -- +({\r},0) -- cycle;
}

\newcommand\mpsBvecV[3]{
	\draw [thick, rounded corners=0.5,colMPSLines,fill=#3] ({\dx*#2},{-\dx*#1})
    +({\r/2.},{-\sqrtThree*\r/2.}) -- +({\r/2.},{\sqrtThree*\r/2.}) 
	-- +({-\r},0) -- cycle;
	\node at ({\dx*#2},{-\dx*#1}) {$\cdot$};
}

\newcommand\mpsBvecW[3]{
	\draw [thick, rounded corners=0.5,colMPSLines,fill=#3] ({\dx*#2},{-\dx*#1})
    +({-\r/2.},{-\sqrtThree*\r/2.}) -- +({-\r/2.},{\sqrtThree*\r/2.}) 
	-- +({\r},0) -- cycle;
	\node at ({\dx*#2},{-\dx*#1}) {$\cdot$};
}

\newcommand\mpsBvec[2]{
  \draw [very thick,colMPSLines] ({\dx*#2},{-\dx*#1}) + ({-0.75*\r},0) -- + ({0.75*\r},0);
}

\newcommand\TleftHook[2]{
  \draw [double,colLines] ({#2*\dx},{-\dx*#1}) arc (90:240:{0.125*\dx});
}

\newcommand\TrightHook[2]{
  \draw[double,colLines] ({#2*\dx},{-\dx*#1}) arc (90:-50:{0.125*\dx});
}

\newcommand\TbottomHook[2]{
  \draw [double,colLines] ({#2*\dx},{-\dx*#1}) arc (-180:0:{0.125*\dx});
}

\newcommand\TtopHook[2]{
  \draw[double,colLines] ({#2*\dx},{-\dx*#1}) arc (180:0:{0.125*\dx});
}

\newcommand\dtensorLine[4]{
  \draw [double,colLines] ({(#2)*\dx},{-(#1)*\dx}) -- ({(#4)*\dx},{-(#3)*\dx});
}

\newcommand\dtensordottedLine[4]{
  \draw [double, dotted,colLines] ({(#2)*\dx},{-(#1)*\dx}) -- ({(#4)*\dx},{-(#3)*\dx});
}

\newcommand\tensorM[3]{
	\draw [thick,rounded corners=0.5,colMPSLines,fill=#3] 
	({\dx*#2},{-\dx*#1}) +({1.25*\r},{1.25*\r}) rectangle +({-1.25*\r},{-1.25*\r});
}

\newcommand\mpsA[3]{
	\draw [thick,rounded corners=0.5,colMPSLines,fill=#3,rotate around={45:({\dx*#2},{-\dx*#1})}] 
	({\dx*#2},{-\dx*#1}) +({0.75*\r},{0.75*\r}) rectangle +({-0.75*\r},{-0.75*\r});
}

\newcommand\mpsB[3]{
	\draw [thick,rounded corners=0.5,colMPSLines,fill=#3] ({\dx*#2},{-\dx*#1}) +({0.75*\r},{0.75*\r}) rectangle +({-0.75*\r},{-0.75*\r});
}

\newcommand\bCircle[3]{
	\draw [thick,colLines,fill=#3] ({\dx*#2},{-\dx*#1}) circle ({1.25*\r});
}

\newcommand\sCircle[3]{
	\draw [thick,colLines,fill=#3] ({\dx*#2},{-\dx*#1}) circle ({0.5*\r});
}

\newcommand\proj[4]{
    \draw[thick,colLines] ({\dx*#3},{-\dx*#2}) -- ({\dx*#3},{-\dx*#1});
    \sCircle{#1}{#3}{#4}
    \sCircle{#2}{#3}{#4}
}

\newcommand\ngridLine[4]{
	\draw [thick,colLines] ({(#2)*\dx},{-(#1)*\dx}) -- ({(#4)*\dx},{-(#3)*\dx});
}

\newcommand\leftHook[2]{
  \draw[semithick,colLines] ({#2*\dx},{-\dx*#1}) arc (90:240:{0.15*\dx});
}

\newcommand\rightHook[2]{
  \draw[semithick,colLines] ({#2*\dx},{-\dx*#1}) arc (90:-50:{0.15*\dx});
}

\newcommand\TEsheet[2]{
  \draw [thick,colLines,fill=IcolU,rounded corners=0.5] ({(#2)*\dx},{-(#1)*\dx}) rectangle ({((#2)+7)*\dx},{(-(#1)+4)*\dx});
  \draw [thick,colVMPSLines,fill=IcolU,rounded corners=0.5] ({((#2)-0.01)*\dx},{(-(#1)-0.01)*\dx}) rectangle ({((#2)+7+0.01)*\dx},{(-(#1)+0.31)*\dx});
}

\newcommand\TECsheet[2]{
  \draw [thick,colLines,fill=IcolUc,rounded corners=0.5] ({(#2)*\dx},{-(#1)*\dx}) rectangle ({((#2)+7)*\dx},{(-(#1)+4)*\dx});
  \draw [thick,colVMPSLines,fill=IcolUc,rounded corners=0.5] ({((#2)-0.01)*\dx},{(-(#1)-0.01)*\dx}) rectangle ({((#2)+7+0.01)*\dx},{(-(#1)+0.31)*\dx});
}

\newcommand\sconnection[4]{
  \draw [thick,colLines,fill=white,rounded corners=0.5] ({((#4)-0.01+1.75)*\dx},{-((#3)+0.01-4)*\dx}) rectangle ({((#4)+0.01)*\dx},{-((#3)-0.25-4)*\dx});
  \draw [thick,colLines,fill=white,rounded corners=0.5] ({((#4)-0.01+1.75)*\dx},{-((#3)-0.25-4)*\dx}) -- ({((#2)-0.01+1.75)*\dx},{-((#1)-0.25-4)*\dx}) -- ({((#2)+0.01)*\dx},{-((#1)-0.25-4)*\dx}) -- ({((#4)+0.01)*\dx},{-((#3)-0.25-4)*\dx}) -- cycle;
  \draw [thick,colLines,fill=white,rounded corners=0.5] 
  ({((#2)-0.01+1.75)*\dx},{-((#1)-0.25-4)*\dx}) --
  ({((#2)+0.01)*\dx},{-((#1)-0.25-4)*\dx}) -- 
  ({((#2)+0.01)*\dx},{-((#1)-0.01-4)*\dx}) -- 
  ({((#2)-0.01+1.75)*\dx},{-((#1)-0.01-4)*\dx}) -- cycle;
}

\newcommand\connection[4]{
  \draw [thick,colLines,fill=white,rounded corners=0.5] ({((#4)-0.01+3.5)*\dx},{-((#3)+0.01-4)*\dx}) rectangle ({((#4)+0.01)*\dx},{-((#3)-0.25-4)*\dx});
  \draw [thick,colLines,fill=white,rounded corners=0.5] ({((#4)-0.01+3.5)*\dx},{-((#3)-0.25-4)*\dx}) -- ({((#2)-0.01+3.5)*\dx},{-((#1)-0.25-4)*\dx}) -- ({((#2)+0.01)*\dx},{-((#1)-0.25-4)*\dx}) -- ({((#4)+0.01)*\dx},{-((#3)-0.25-4)*\dx}) -- cycle;
  \draw [thick,colLines,fill=white,rounded corners=0.5] 
  ({((#2)-0.01+3.5)*\dx},{-((#1)-0.25-4)*\dx}) --
  ({((#2)+0.01)*\dx},{-((#1)-0.25-4)*\dx}) -- 
  ({((#2)+0.01)*\dx},{-((#1)-0.01-4)*\dx}) -- 
  ({((#2)-0.01+3.5)*\dx},{-((#1)-0.01-4)*\dx}) -- cycle;
}

\newcommand\bConnection[4]{
  \draw [thick,colLines,fill=white,rounded corners=0.5] ({((#4)-0.01+3.5)*\dx},{-((#3)-0.25-4)*\dx}) -- ({((#2)-0.01+3.5)*\dx},{-((#1)-0.25-4)*\dx}) -- ({((#2)+0.01)*\dx},{-((#1)-0.25-4)*\dx}) -- ({((#4)+0.01)*\dx},{-((#3)-0.25-4)*\dx}) -- cycle;
  \draw [thick,colLines,fill=white,rounded corners=0.5] 
  ({((#2)-0.01+3.5)*\dx},{-((#1)-0.25-4)*\dx}) --
  ({((#2)+0.01)*\dx},{-((#1)-0.25-4)*\dx}) -- 
  ({((#2)+0.01)*\dx},{-((#1)-0.01-4)*\dx}) -- 
  ({((#2)-0.01+3.5)*\dx},{-((#1)-0.01-4)*\dx}) -- cycle;
}

\newcommand\fConnection[4]{
  \draw [thick,colLines,fill=white,rounded corners=0.5] ({((#4)-0.01+3.5)*\dx},{-((#3)+0.01-4)*\dx}) rectangle ({((#4)+0.01)*\dx},{-((#3)-0.25-4)*\dx});
  \draw [thick,colLines,fill=white,rounded corners=0.5] ({((#4)-0.01+3.5)*\dx},{-((#3)-0.25-4)*\dx}) -- ({((#2)-0.01+3.5)*\dx},{-((#1)-0.25-4)*\dx}) -- ({((#2)+0.01)*\dx},{-((#1)-0.25-4)*\dx}) -- ({((#4)+0.01)*\dx},{-((#3)-0.25-4)*\dx}) -- cycle;
}

\newtheorem{lemma}{Lemma}
\theoremstyle{plain}
\newtheorem{property}{Property}
\theoremstyle{plain}

\begin{document}

\begin{center}{\Large \textbf{
Entanglement dynamics in Rule $54$: Exact results and quasiparticle picture
}}
\end{center}

\begin{center}
    Katja Klobas\textsuperscript{1*} and 
    Bruno Bertini\textsuperscript{1}
\end{center}

\begin{center}
{\bf 1} Rudolf Peierls Centre for Theoretical Physics, Oxford University, Parks Road, Oxford OX1 3PU, United Kingdom
\\
* katja.klobas@physics.ox.ac.uk
\end{center}

\begin{center}
\today
\end{center}

\section*{Abstract}
{
We study the entanglement dynamics generated by quantum quenches in the quantum cellular automaton Rule $54$. We consider the evolution from a recently introduced class of \emph{solvable} initial states. States in this class relax (locally) to a one-parameter family of Gibbs states and the thermalisation dynamics of local observables can be characterised exactly by means of an evolution in space. Here we show that the latter approach also gives access to the entanglement dynamics and derive exact formulas describing the asymptotic linear growth of all R\'enyi entropies in the thermodynamic limit and their eventual saturation for finite subsystems. While in the case of von Neumann entropy we recover exactly the predictions of the quasiparticle picture, we find no physically meaningful quasiparticle description for other R\'enyi entropies. Our results apply to both homogeneous and inhomogeneous quenches.
}

\vspace{10pt}
\noindent\rule{\textwidth}{1pt}
\tableofcontents\thispagestyle{fancy}
\noindent\rule{\textwidth}{1pt}
\vspace{10pt}

\section{Introduction}

The growth of entanglement is arguably the most universal phenomenon observed so far in studies of quantum many-body dynamics. Whenever a quantum many body system with short-range interactions is prepared in a non-equilibrium state with low entanglement and then let to evolve unitarily, the entanglement among neighbouring spatial regions is observed to grow linearly in time. For instance, this behaviour has been reported in conformal field theories, both rational~\cite{calabrese2005evolution} and holographic~\cite{liu2014entanglement}, in free systems of fermions~\cite{fagotti2008evolution} and bosons~\cite{alba2018entanglement}, as well as in interacting integrable~\cite{alba2017entanglement, alba2018entanglement} and non-integrable systems~\cite{laeuchli2008spreading, kim2013ballistic, pal2018entangling, bertini2019entanglement, piroli2020exact, gopalakrishnan2019unitary}. Remarkably, the growth of entanglement has even been measured in cold-atom experiments~\cite{Greiner2015,Greiner2016,Greiner2019}. In essence, the only exceptions to this empirical rule are systems exhibiting localisation~\cite{dechiara2006entanglement, znidaric2008many, nandkishore2015many}, confinement~\cite{kormos2017real}, or when the dynamics is not purely unitary, for example if the evolution is monitored with measurements~\cite{li2019measurement, skinner2019measurement, vasseur2019entanglement}.

Given such a universal phenomenology a natural direction for the theoretical research has been to find an equally universal description and identify the seemingly very general emergent laws describing it. Recent years have witnessed important progress in this direction with the proposal of two alternative effective descriptions of the spreading of entanglement which are believed to work in  integrable and chaotic systems respectively. The first, known as the \emph{quasiparticle picture}~\cite{calabrese2005evolution}, explains the growth of entanglement by imagining that correlations are transported by quasiparticle excitations. These excitations, stable because of integrability, are created when the system is driven out of equilibrium and are correlated with those created nearby. During the evolution correlated quasiparticles move far apart, effectively spreading correlations and entanglement throughout the system. The second effective description is known as the \emph{membrane picture}~\cite{nahum2017quantum} and interprets the entanglement geometrically. In essence it claims that the entanglement between two complementary regions is given by the tension of the minimal spacetime surface that separates the two. 

A quantitative verification of these pictures and their predictive power in genuinely interacting systems, however, has proven to be a daunting task. This is ultimately due to the fact that the out-of-equilibrium dynamics of interacting many-body quantum systems are generically too complicated to be characterised analytically and, moreover, the growth of entanglement provides a great limitation to the most efficient numerical methods at our disposal to treat quantum many-body systems~\cite{schollwock2011density}. For this reason, the benchmark provided by exact solutions in minimal solvable cases is of rare value. 

Surprisingly, such a benchmark has recently become available in the case of quantum chaotic systems with the discovery of dual-unitary circuits~\cite{bertini2019exact}. In these systems one can exploit a duality between space and time to compute exactly the time-evolution of many relevant quantities~\cite{bertini2019entanglement, piroli2020exact, gopalakrishnan2019unitary, claeys2021ergodic, suzuki2021computational, bertini2020operatorI, bertini2020operatorII, claeys2020maximum, reid2021entanglement}, including that of entanglement, by performing an evolution in space (or in the ``time-channel") rather than in time. Up to very recently, however, no such solvable benchmark was known for the case of interacting integrable models. The situation changed recently, when Ref.~\cite{klobas2021exact} presented an exact characterisation of the growth of entanglement in the quantum cellular automaton Rule 54, which can be considered one of the simplest examples of interacting integrable models (see also~\cite{pozsgay2014quantum, pozsgay2016real,zadnik2021foldedI,zadnik2021foldedII,pozsgay2021integrable,pozsgay2021yang,gombor2021integrable}). The result was again based on a time-channel approach and lead to an exact characterisation of the growth of entanglement when the system is initialised in a particular class of initial states.

The objective of this paper is to extend the exact results presented in Ref.~\cite{klobas2021exact} to a larger class of initial states. This is the second of two papers dedicated to this task. While in the first part of our work~\cite{klobas2021exactI}, which in the following we will refer to as ``Paper I'', we focussed on the dynamics of local observables, here we consider the evolution of the entanglement. The extension that we present bares a remarkable physical significance. Indeed, while the states considered in Ref.~\cite{klobas2021exact} all relax (locally) to the Gibbs state with infinite temperature, here we show that exact results can be obtained also for states relaxing to richer Gibbs ensembles (characterised by an arbitrary  chemical potential). This allows us, for instance, to study exactly inhomogeneous quenches giving rise to a non-trivial (generalised) hydrodynamic regime at late times~\cite{bertini2016transport, castroalvaredo2016emergent}. We use our exact results to test the predictions of the quasiparticle picture for the von Neumann entanglement entropy, both for homogeneous~\cite{alba2017entanglement} and inhomogeneous~\cite{bertini2018entanglement, alba2019entanglement} quenches, providing what is, to the best of our knowledge, the first exact confirmation of this picture in the presence of either inhomogeneity or interactions. Using our exact results we also argue that no consistent quasiparticle picture can be designed in the case of R\'enyi entropies. 

The rest of the paper is organised as follows. In Sec.~\ref{sec:timechannelent} we introduce the time-channel approach to the entanglement dynamics in generic systems. In Sec.~\ref{eq:solvablestates} we specialise the treatment to the case of Rule 54 and recall some of the results of Paper I that are necessary for our discussion. Sec.~\ref{sec:mainresult} contains the derivation of our main results, i.e.\ exact formulae for the stationary values eventually approached by the entropies of finite regions and for the rate of entanglement entropies after a quench from a solvable state. In Sec.~\ref{sec:qp} we derive the predictions of the quasiparticle picture for the cases of interest and compare them with our findings. Finally Sec.~\ref{sec:conclusions} contains our conclusions. Some more technical points and proofs are reported in the two appendices.

\section{Entanglement dynamics in the time-channel}
\label{sec:timechannelent}
In this section we show that the time-channel description of the dynamics introduced in~\cite{banuls2009matrix, muller2012tensor} can also be applied to study of entanglement. As discussed in the aforementioned references (see also Paper I), this approach is based on the simple idea of evolving a many-body system in space, rather than in time, and can be applied whenever the time-evolution operator is represented as a matrix product operator (MPO). This approach has a very general scope, since essentially any evolution generated by a short-range Hamiltonian can be efficiently represented by a unitary MPO~\cite{osborne2006efficient, cirac2020matrix}, but it does not generically give a computational advantage. On the contrary in certain special cases it leads to exact results. In particular, concerning the entanglement dynamics, it provides exact results in dual-unitary quantum circuits~\cite{bertini2019exact, bertini2019entanglement,piroli2020exact,gopalakrishnan2019unitary} and in Rule 54~\cite{klobas2021exact}. 

To describe the main ideas let us consider the setting described in Paper I: $2L$ qudits (with $d$ internal states) are arranged along a one-dimensional chain and driven out of equilibrium through a standard quantum quench protocol~\cite{calabrese2006time, calabrese2007quantum}. Specifically, we prepare the system in a two-site shift invariant {product state} denoted by $\ket{\Psi_0}$ (note that here, differently from Paper I, we do not consider more general matrix product states) and evolve it with a unitary MPO (with bond dimension $\chi^2$), which we indicate by $\mathbb{U}$. The regime of interest is ${L\gg t}$ and we will eventually take the thermodynamic limit $L\to\infty$.

Making use of the graphical representation introduced in Paper I we depict initial state and time-evolution operator as follows   
\begin{align}
\ket{\Psi_0} & = 
\mkern14mu
\begin{tikzpicture}
    [baseline={([yshift=1.2ex]current bounding box.center)},scale=1.75]
        \foreach \x in {0,...,11}{\gridLine{0}{\x}{-1}{\x}}
        \foreach \x in {0,2,...,10}{
            \vmpsV{0}{\x}{IcolVMPSgen}
            \vmpsW{0}{(\x+1)}{IcolVMPSgen}
        }
        \draw [decorate,decoration={brace,amplitude=5pt},xshift=0pt,yshift=0pt] 
    ({11.25*\dx},{-0.25*\dx})--({-0.25*\dx},{-0.25*\dx}) node [midway,yshift=-10pt]
    {$2L$};
\end{tikzpicture},\label{eq:initialstate}\\ 
\mathbb{U}  &=
    \begin{tikzpicture}[baseline={([yshift=1.2ex]current bounding box.center)},scale=1.75]
    \gridLine{-2}{0.25}{-2}{12.75};
        \leftHook{-2}{0.25}
        \rightHook{-2}{12.75}
        \griddashedLine{-1.75}{0.25}{-1.75}{12.75};
        \foreach \x in {1,...,12}{
          \gridLine{-1.25}{\x}{-2.75}{\x};
        }
            \foreach \x in {1,...,12}{
          \gridLine{-.25}{\x}{-1.75}{\x};
        }
        \foreach \x in {1,3,...,12}{
          \bCircle{-2}{\x}{IcolU}
            }
        \foreach \x in {2,4,...,12}{
          \sCircle{-2}{\x}{IcolU}        
          }
        \gridLine{-1}{0.25}{-1}{12.75};
        \leftHook{-1}{0.25}
        \rightHook{-1}{12.75}
        \griddashedLine{-0.75}{0.25}{-0.75}{12.75};
        \foreach \x in {1,3,...,12}{
          \sCircle{-1}{\x}{IcolU}
            }
        \foreach \x in {2,4,...,12}{
          \bCircle{-1}{\x}{IcolU}        
          }
        \draw [decorate,decoration={brace,amplitude=5pt},xshift=0pt,yshift=0pt] 
    ({12.25*\dx},{0.25*\dx})--({0.75*\dx},{0.25*\dx}) node [midway,yshift=-10pt]
    {$2L$};
      \end{tikzpicture}\,, \label{eq:UeUoMPO}
\end{align}
where we assumed periodic boundary conditions and, for the time being, the tensors 
\be
\begin{tikzpicture}
    [baseline={([yshift=-0.6ex]current bounding box.center)},scale=1.75]
    \gridLine{1}{0}{-1}{0};
    \gridLine{0}{1}{0}{-1};
    \bCircle{0}{0}{IcolU}
    \node at ({-1.3*\dx},{0}) {$\alpha$};
    \node at (0,{-1.3*\dx}) {$s$};
    \node at ({1.3*\dx},{0}) {$\beta$};
    \node at (0,{1.3*\dx}) {$r$};
\end{tikzpicture}
,\qquad
\begin{tikzpicture}
    [baseline={([yshift=-0.6ex]current bounding box.center)},scale=1.75]
    \gridLine{0.75}{0}{-0.75}{0};
    \gridLine{0}{0.75}{0}{-0.75};
    \sCircle{0}{0}{IcolU}
    \node at ({-1.25*\dx},{0}) {{$\alpha$}};
    \node at (0,{-\dx}) {{$s$}};
    \node at ({1.25*\dx},{0}) {{$\beta$}};
    \node at (0,{\dx}) {{$r$}};
\end{tikzpicture},
\qquad
\begin{tikzpicture}
    [baseline={([yshift=-1.6ex]current bounding box.center)},scale=1.5]
    \gridLine{0}{0}{-1}{0};
    \vmpsV{0}{0}{IcolVMPSgen}
    \node at (0,{1.5*\dx}) {{$s$}};
\end{tikzpicture},\quad
\begin{tikzpicture}
    [baseline={([yshift=-1.6ex]current bounding box.center)},scale=1.5]
    \gridLine{0}{0}{-1}{0};
    \vmpsW{0}{0}{IcolVMPSgen}
    \node at (0,{1.5*\dx}) {{$s$}};
\end{tikzpicture}\,,
\qquad r,s=1,\ldots,d,\quad\alpha,\beta=1,\ldots,\chi\,,
\label{eq:tensors}
\ee
can be considered generic (the only constraints on them are that $\mathbb U$ must be unitary and $\ket{\Psi_0}$ normalised). As discussed in Paper I, two remarks are in order at this point: (i) here we are interested in MPOs describing local interactions and hence we should impose additional constraints on \eqref{eq:tensors}. However, since the upcoming discussion does not rely upon these constrains, we ignore them for the sake of simplicity; (ii) the space-time staggering in \eqref{eq:initialstate} and \eqref{eq:UeUoMPO} is inessential and can be easily removed by appropriately merging tensors and local sites. Nevertheless, here we keep it because it arises naturally in Rule 54 which is the case of interest in this paper.

As a result of the unitary evolution, the state
\be
\ket{\Psi_t} = \mathbb{U}^t \ket{\Psi_0}, 
\ee
becomes increasingly more entangled as time advances. The growth of entanglement between a finite region $A$ and the rest of the system is quantitatively characterised by the \emph{R\'enyi entropies}
\be
S^{(\alpha)}_A(t) = \frac{1}{1-\alpha}
\log\left[\tr\big(\rho^\alpha_A(t)\big)\right], \qquad \alpha\in\mathbb R\,,
\label{eq:Srenyi}
\ee  
where $\rho_A(t)$ is the density matrix of the system reduced to the subsystem
$A$. In particular, the limit $\alpha\to1$ of~\eqref{eq:Srenyi} gives the \emph{von
Neumann entropy} or \emph{entanglement entropy}
\be
S_{A}(t) = -\tr\left[\rho_A(t) \log \rho_A(t)\right] ,
\ee
which is the standard measure of bipartite entanglement for pure
states~\cite{amico2008entanglement}. The latter, however, is not the only
interesting member of the family. Although R\'enyi entropies for ${\alpha\neq1}$
are not entanglement measures in the strict sense, they are attracting
increasing attention. This is because they characterise
the spectrum of $\rho_A(t)$ --- the \emph{entanglement spectrum} --- which
contains non-trivial information about the system~\cite{laflorencie2016quantum}
(e.g.\ on its topological properties~\cite{li2008entanglement}). Moreover, and perhaps more importantly, they have recently become experimentally accessible~\cite{Greiner2015,Greiner2016,Linke2018,Greiner2019,Elben2020mixed, zhou2020single}. Even though physically very relevant, these quantities are notoriously hard to compute. This is especially true when considering interacting integrable models, where, up to very recently~\cite{klobas2021exact}, they could be accessed only in the limit $t\gg A$. Indeed, as pointed out in Ref.~\cite{alba2017quench}, in this limit one can assume that the state of the subsystem $A$ is described by a generalised Gibbs ensemble and compute the R\'enyi entropies using thermodynamic Bethe ansatz (TBA)~\cite{yang1969thermodynamics,
takahashi1999thermodynamics}. The goal of this section is to derive an alternative representation for these quantities, which, as we will see,  for Rule 54 allows us to access the regime $t< A$.

Let us start by looking more closely at the expression \eqref{eq:Srenyi}. Employing the diagrammatic representation in \eqref{eq:initialstate} and \eqref{eq:UeUoMPO}, we can depict the reduced density matrix at time~$t$ as 
\be\label{eq:rhoAtn}
\rho_A(t)= \tr_{\bar A}[\mathbb{U}^t \ketbra{\Psi_0} \mathbb{U}^{-t}]=
\begin{tikzpicture}
  [baseline={([yshift=-3ex]current bounding box.center)},scale=1.75]
  \newcommand\bCross[2]{
    \draw [thick,colLines] ({\dx*#2-0.5*\r},{-\dx*#1-0.5*\r}) --
    ({\dx*#2+0.5*\r},{-\dx*#1+0.5*\r});
    \draw [thick,colLines] ({\dx*#2-0.5*\r},{-\dx*#1+0.5*\r}) --
    ({\dx*#2+0.5*\r},{-\dx*#1-0.5*\r});
  }
  \foreach \t in {-1,...,-4}{
    \gridLine{\t}{0.25}{\t}{12.75};
    \IleftHook{\t}{0.25}{0.1}
    \IrightHook{\t}{12.75}{0.1}
    \griddashedLine{(\t+0.2)}{0.25}{(\t+0.2)}{12.75};
  }
  \foreach \t in {1.75,...,4.75}{
    \gridLine{\t}{0.25}{\t}{12.75};
    \IleftHook{\t}{0.25}{0.1}
    \IrightHook{\t}{12.75}{0.1}
    \griddashedLine{(\t+0.2)}{0.25}{(\t+0.2)}{12.75};
  }
  \foreach \x in {1,...,12}{
    \gridLine{0}{\x}{-4.75}{\x};
    \gridLine{0.75}{\x}{5.5}{\x};
  }
  \foreach \x in {1,3,...,12}{
    \vmpsV{0}{\x}{IcolVMPSgen}
    \vmpsV{0.75}{\x}{IcolVMPSgenc}
    \sCircle{1.75}{\x}{IcolUc}
    \bCircle{2.75}{\x}{IcolUc}
    \sCircle{3.75}{\x}{IcolUc}
    \bCircle{4.75}{\x}{IcolUc}
  }
  \foreach \x in {2,4,...,12}{
    \vmpsW{0}{\x}{IcolVMPSgen}
    \vmpsW{0.75}{\x}{IcolVMPSgenc}
    \bCircle{1.75}{\x}{IcolUc}
    \sCircle{2.75}{\x}{IcolUc}
    \bCircle{3.75}{\x}{IcolUc}
    \sCircle{4.75}{\x}{IcolUc}
  }
    \foreach \x in {2,4,...,12}{
    \bCircle{-1}{\x}{IcolU}
    \sCircle{-2}{\x}{IcolU}
    \bCircle{-3}{\x}{IcolU}
    \sCircle{-4}{\x}{IcolU}
  }
  \foreach \x in {1,3,...,12}{
    \sCircle{-1}{\x}{IcolU}
    \bCircle{-2}{\x}{IcolU}
    \sCircle{-3}{\x}{IcolU}
    \bCircle{-4}{\x}{IcolU}
  }
  \foreach \x in {1,...,4}{
    \gridLine{-4.75}{(-0.1-0.2*\x)}{5.5}{(-0.1-0.2*\x)};
    \draw [semithick,colLines,rounded corners=2] ({(-0.1-0.2*\x)*\dx},{4.75*\dx}) --
    ({(-0.1-0.2*\x)*\dx},{(4.9+0.1*\x)*\dx}) --
    ({\x*\dx},{(4.9+0.1*\x)*\dx}) -- ({\x*\dx},{4.75*\dx});
    \draw [semithick,colLines,rounded corners=2] ({(-0.1-0.2*\x)*\dx},{-5.5*\dx}) --
    ({(-0.1-0.2*\x)*\dx},{-(5.6+0.1*\x)*\dx}) --
    ({\x*\dx},{-(5.6+0.1*\x)*\dx}) -- ({\x*\dx},{-5.5*\dx});
  }
   \foreach \x in {9,...,12}{
    \gridLine{-4.75}{(13.4-0.2*(\x-12))}{5.5}{(13.4-0.2*(\x-12))};
    \draw [semithick,colLines,rounded corners=2] ({(13.4-0.2*(\x-12))*\dx},{4.75*\dx}) --
    ({(13.4-0.2*(\x-12))*\dx},{(4.9-0.1*(\x-12))*\dx}) --
    ({\x*\dx},{(4.9-0.1*(\x-12))*\dx}) -- ({\x*\dx},{4.75*\dx});
   \draw [semithick,colLines,rounded corners=2] ({(13.4-0.2*(\x-12))*\dx},{-5.5*\dx}) --
    ({(13.4-0.2*(\x-12))*\dx},{-(5.6-0.1*(\x-12))*\dx}) --
    ({\x*\dx},{-(5.6-0.1*(\x-12))*\dx}) -- ({\x*\dx},{-5.5*\dx});
  }
    \draw [decorate,decoration={brace,amplitude=5pt},xshift=0pt,yshift=0pt] 
    ({14.125*\dx},{4*\dx})--({14.125*\dx},{1*\dx}) node [midway,xshift=15pt]
    {$2t$};
    \draw [decorate,decoration={brace,amplitude=5pt},xshift=0pt,yshift=0pt] 
    ({0.75*\dx},{5.375*\dx})--({12.125*\dx},{5.375*\dx}) node [midway,yshift=15pt]
    {$2L$};
\end{tikzpicture}\mkern-32mu,\mkern12mu
\ee
where we introduced the symbols
\be 
\begin{tikzpicture}
  [baseline={([yshift=-0.6ex]current bounding box.center)},scale=1.75]
  \gridLine{1}{0}{-1}{0};
  \gridLine{0}{1}{0}{-1};
  \bCircle{0}{0}{IcolUc}
\end{tikzpicture}=
  \left.\begin{tikzpicture}
  [baseline={([yshift=-0.6ex]current bounding box.center)},scale=1.75]
  \gridLine{1}{0}{-1}{0};
  \gridLine{0}{1}{0}{-1};
  \bCircle{0}{0}{IcolU}
  \end{tikzpicture}\right.^{\ast}
  ,\qquad
 \begin{tikzpicture}
  [baseline={([yshift=-0.6ex]current bounding box.center)},scale=1.75]
  \gridLine{0.875}{0}{-0.875}{0};
  \gridLine{0}{0.875}{0}{-0.875};
  \sCircle{0}{0}{IcolUc}
\end{tikzpicture}=\left.\begin{tikzpicture}
  [baseline={([yshift=-0.6ex]current bounding box.center)},scale=1.75]
  \gridLine{0.875}{0}{-0.875}{0};
  \gridLine{0}{0.875}{0}{-0.875};
  \sCircle{0}{0}{IcolU}
 \end{tikzpicture}\right.^{\ast},
\label{eq:redballs}
\ee
for the complex conjugate of the tensors \eqref{eq:balls}.   

We now interpret the tensor network \eqref{eq:rhoAtn} as the result of an evolution in space rather than in time. Specifically, by defining the  space transfer matrices
\be
\label{eq:Wmatrix}
\tilde{\mathbb{W}}_{s_1\,s_2}^{r_1\,r_2}=
\begin{tikzpicture}
          [baseline={([yshift=-0.6ex]current bounding box.center)},scale=1.75]
          \foreach \t in {-1,...,-4}{
            \gridLine{\t}{0.25}{\t}{2.75};
          }
          \foreach \t in {1.5,...,4.5}{
            \gridLine{\t}{0.25}{\t}{2.75};
          }

          \foreach \x in {1,2}{
            \gridLine{-0.125}{\x}{-4.875}{\x};
            \gridLine{0.625}{\x}{5.375}{\x};
          }
          \foreach \x in {1}{
            \vmpsV{-0.125}{\x}{IcolVMPSgen}
            \sCircle{-1}{\x}{IcolU}
            \bCircle{-2}{\x}{IcolU}
            \sCircle{-3}{\x}{IcolU}
            \bCircle{-4}{\x}{IcolU}
            \vmpsV{0.625}{\x}{IcolVMPSgenc}
            \sCircle{1.5}{\x}{IcolUc}
            \bCircle{2.5}{\x}{IcolUc}
            \sCircle{3.5}{\x}{IcolUc}
            \bCircle{4.5}{\x}{IcolUc}

          }
          \foreach \x in {2}{
            \vmpsW{-0.125}{\x}{IcolVMPSgen}
            \bCircle{-1}{\x}{IcolU}
            \sCircle{-2}{\x}{IcolU}
            \bCircle{-3}{\x}{IcolU}
            \sCircle{-4}{\x}{IcolU}
            \vmpsW{0.625}{\x}{IcolVMPSgenc}
            \bCircle{1.5}{\x}{IcolUc}
            \sCircle{2.5}{\x}{IcolUc}
            \bCircle{3.5}{\x}{IcolUc}
            \sCircle{4.5}{\x}{IcolUc}
          }
          \node at ({\dx},{5.25*\dx}) {\scalebox{0.9}{$s_1$}};
          \node at ({2*\dx},{5.25*\dx}) {\scalebox{0.9}{$s_2$}};
          \node at ({\dx},{-5.75*\dx}) {\scalebox{0.9}{$r_1$}};
          \node at ({2*\dx},{-5.75*\dx}) {\scalebox{0.9}{$r_2$}};
        \end{tikzpicture},
        \qquad
   \tilde{\mathbb{W}}= \smashoperator{\sum_{s_1,s_2}}
   \tilde{\mathbb{W}}_{s_1\,s_2}^{s_1\,s_2}=
\begin{tikzpicture}
          [baseline={([yshift=-0.6ex]current bounding box.center)},scale=1.75]
          \foreach \t in {-1,...,-4}{
            \gridLine{\t}{0.25}{\t}{2.75};
          }
          \foreach \t in {1.5,...,4.5}{
            \gridLine{\t}{0.25}{\t}{2.75};
          }

          \foreach \x in {1,2}{
            \gridLine{-0.125}{\x}{-4.875}{\x};
            \gridLine{0.625}{\x}{5.375}{\x};
          }
          \ItopHook{-4.875}{1}{0.2}
          \ItopHook{-4.875}{2}{0.2}
          \IbottomHook{5.375}{1}{0.2}
          \IbottomHook{5.375}{2}{0.2}
          \griddashedLine{-4.875}{0.6}{5.375}{0.6};
          \griddashedLine{-4.875}{1.6}{5.375}{1.6};
          \foreach \x in {1}{
            \vmpsV{-0.125}{\x}{IcolVMPSgen}
            \sCircle{-1}{\x}{IcolU}
            \bCircle{-2}{\x}{IcolU}
            \sCircle{-3}{\x}{IcolU}
            \bCircle{-4}{\x}{IcolU}
            \vmpsV{0.625}{\x}{IcolVMPSgenc}
            \sCircle{1.5}{\x}{IcolUc}
            \bCircle{2.5}{\x}{IcolUc}
            \sCircle{3.5}{\x}{IcolUc}
            \bCircle{4.5}{\x}{IcolUc}

          }
          \foreach \x in {2}{
            \vmpsW{-0.125}{\x}{IcolVMPSgen}
            \bCircle{-1}{\x}{IcolU}
            \sCircle{-2}{\x}{IcolU}
            \bCircle{-3}{\x}{IcolU}
            \sCircle{-4}{\x}{IcolU}
            \vmpsW{0.625}{\x}{IcolVMPSgenc}
            \bCircle{1.5}{\x}{IcolUc}
            \sCircle{2.5}{\x}{IcolUc}
            \bCircle{3.5}{\x}{IcolUc}
            \sCircle{4.5}{\x}{IcolUc}
          }
\end{tikzpicture},
\ee
we can express the reduced density matrix~\eqref{eq:rhoAtn} as the following MPO
\be
\label{eq:MPSrhoA}
\rho_A(t)= \smashoperator{\sum_{s_j,r_j\in\{0,1\}}}
\tr\Big(
\tilde{\mathbb{W}}_{s_1 s_2}^{r_1 r_2}\cdots
\tilde{\mathbb{W}}_{s_{\abs{A}-1} s_{\abs{A}}}^{r_{\abs{A}-1} r_{\abs{A}}}
\tilde{\mathbb{W}}^{L-\abs{A}/2}
\Big)
\ketbra{s_1 s_2\ldots s_{\abs{A}}}{r_1 r_2\ldots r_{\abs{A}}},
\ee
where we conveniently consider the case of $\abs{A}$ even.

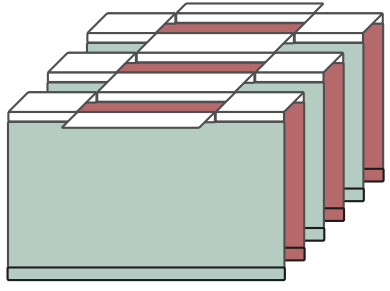
\begin{figure}
    \centering
\begin{tikzpicture}
  [baseline={([yshift=-0.6ex]current bounding box.center)},scale=1.75]
  \TECsheet{-2.5}{2.5};
  \TEsheet{-2}{2};
  \TECsheet{-1.5}{1.5};
  \TEsheet{-1}{1};
  \TECsheet{-0.5}{0.5};
  \TEsheet{0}{0};
  \bConnection{-2.5}{4.25}{-2.75}{4.5};
  \sconnection{-2}{2}{-2.5}{2.5};
  \sconnection{-2}{7.25}{-2.5}{7.75};
  \connection{-1.5}{3.25}{-2}{3.75};
  \sconnection{-1}{1}{-1.5}{1.5};
  \sconnection{-1}{6.25}{-1.5}{6.75};
  \connection{-0.5}{2.25}{-1}{2.75};
  \sconnection{0}{0}{-0.5}{0.5};
  \sconnection{0}{5.25}{-0.5}{5.75};
  \fConnection{0.4}{1.35}{0}{1.75};
\end{tikzpicture}\,,
\caption{Pictorial representation of $\tr[\rho_A(t)^3]$ in the folded picture. The portions corresponding to $A$ and $\bar A$ are connected together in a ``staggered" fashion: this staggering is implemented the operators $\mathcal{S}_{2n}$ and $\mathcal{S}_{2n}^\dag$ in \eqref{eq:SAn}.}
\label{fig:Renyipic}
\end{figure}

Inserting~\eqref{eq:MPSrhoA} in the definition \eqref{eq:Srenyi} of R\'enyi entropies and taking $\alpha=n$, with $n>1$ integer, we have 
\be
\label{eq:SAn}
S_A^{(n)}(t)=\frac{1}{1-n} 
\log\tr\left[(\tilde {\mathbb W}^{\ast\,\otimes n})^{|A|/2}
\mathcal S^\dag_{2n} (\tilde {\mathbb W}^{\otimes n})^{L-|A|/2}
\mathcal S_{2n}\right],
\ee
where the operator
$\mathcal{S}_{2n}$ denotes a periodic shift by one in the space of the $2n$
replicas (of which $n$ correspond to forward (green) and $n$ to backward (red) time-sheets).
More precisely, $\mathcal{S}_{2n}$ acts on the tensor product of $2n$ copies of the qudit chain in the $t$ direction as follows
\be
 \mathcal S_{2n}  \ket{\boldsymbol i_1}\otimes\ket{\boldsymbol i_2}\otimes \cdots 
 \otimes\ket{\boldsymbol i_{2n-1}}\otimes\ket{\boldsymbol i_{2n}} =
 \ket{\boldsymbol i_2}\otimes\ket{\boldsymbol i_3}\otimes \cdots
 \otimes\ket{\boldsymbol i_{2n}}\otimes\ket{\boldsymbol i_1}\,,\qquad
 \boldsymbol i_j \in \mathbb{Z}_d^{\times 2t},
\ee
where $\{\ket{\boldsymbol i}\}$ is a basis of $\mathbb C^{d^{2t}}$, which is
the Hilbert space of a qudit chain of length $2t$. The operator
$\mathcal{S}_{2n}$ appears because the sites of $A$ and $\bar A$ are contracted
to different replicas in the calculation of the R\'enyi entropies, see
Fig.~\ref{fig:Renyipic}. 

To simplify \eqref{eq:SAn} further we use a simple property of $\tilde {\mathbb W}$ (which is a special case of Property 1 in Paper I).
\begin{property}
\label{prop1}
    Whenever the initial state is normalised and the time evolution operator unitary, the spectrum of $\tilde  {\mathbb W}$ in \eqref{eq:Wmatrix} is given by $\{0,1\}$ and the algebraic and geometric multiplicity of the eigenvalue $1$ are equal to one. 
\end{property}
\begin{proof}
    The unitarity of time-evolution implies
    \be
    1=\braket{\Psi_t}=\tr\Big[\tilde{\mathbb W}^L\Big] = \sum_{j}\lambda_j^L,
    \ee
    where the second equality follows directly from the definition~\eqref{eq:Wmatrix}
    and $\lambda_j$ are eigenvalues of $\tilde{\mathbb W}$. This equality holds
    for any $L$, therefore $\lambda_j\in\{0,1\}$ and both the geometric and
    algebraic multiplicity of the eigenvalue $1$ have to be $1$.
\end{proof}

An immediate consequence of this is that we can write the thermodynamic limit of \eqref{eq:SAn} as 
\be
\label{eq:SAnth}
S_{A,\rm th}^{(n)}(t)\coloneqq \lim_{L \to \infty}S_{A}^{(n)}(t) =\frac{1}{1-n} \log\left[\frac{{}_n\!\mel{L}{\mathcal S_{2n} (\tilde {\mathbb W}^{\ast\,\otimes n})^{|A|/2} \mathcal S^\dag_{2n}}{R}_{n}}{{}_n\!\braket{L}{R}_n}\right]\,,
\ee
with 
\be
{}_n\!\bra{L} = {}^{n \otimes}\!\!\bra{L}, \qquad \ket{R}_{n}=\ket{R}\!\!^{\otimes n}\,, 
\ee
and $\bra{L}$ and $\ket{R}$ respectively denote the left and right \emph{fixed points} (i.e.\ eigenvectors corresponding to the eigenvalue $1$) of $\tilde  {\mathbb W}$. We see that, in the thermodynamic limit, the $n$-th R\'enyi entropy is expressed as a matrix element between $n$ copies of left and right fixed points. This expression further simplifies if one considers the entanglement of half of the system
\be
\label{eq:SAnthhalfsystem}
\lim_{|A| \to \infty}S_{A,\rm th}^{(n)}(t) =
\frac{2}{1-n} \log\Big[
    \abs{\frac{
        \tensor*[_n]{\!\mel{L}{\mathcal S_{2n}}{R^{\ast}}}{_n}
    }{ \tensor*[_n]{\braket{L}{R}}{_n}}}\Big].
\ee
Here we introduced the shorthand notation $\ket{R^{\ast}}=\ket{R}^{\ast}$ to denote the
complex conjugate of $\ket{R}$ (similarly, $\bra{L^{\ast}}=\bra{L}^{\ast}$) and we implicitly
used
\be
\tensor*[_n]{\!\mel{L^{\ast}}{\mathcal{S}_{2n}^{\dag}}{R}}{_n}
=
\Big(\tensor*[_n]{\!\mel{L}{\mathcal{S}_{2n}}{R^{\ast}}}{_n}\Big)^{\ast},
\ee
which follows directly from the permutation symmetry of $\tensor*[]{\ket{R}}{_n}$ and
$\tensor*[_n]{\bra{L}}{}$. Equation \eqref{eq:SAnthhalfsystem} implies that the information about the
asymptotic growth of entanglement is entirely encoded in the fixed points. Note that in systems with a strict maximal speed $v_{\rm max}$ for the propagation of signals --- as it is the case for local quantum circuits --- one does not need to consider the limit  $|A|\to\infty$ to obtain the simplified form~\eqref{eq:SAnthhalfsystem}: it is sufficient to take $|A|>2v_{\max} t$ so that the two boundaries are causally disconnected.

This approach can also be applied when the initial state is not homogeneous
(i.e.\ invariant under a small number of shifts), but is composed by the
junction of two \emph{different} homogeneous pieces. Namely 
\be
\ket{\Psi_0}  = 
\begin{tikzpicture}
    [baseline={([yshift=-2ex]current bounding box.center)},scale=1.75]
        \foreach \x in {6,...,11}{\gridLine{0}{\x}{-1}{\x}}
        \foreach \x in {6,8,...,10}{
            \vmpsV{0}{\x}{IcolVMPSgen}
             \node at ({(\x)*\dx},{-0.5*\dx}) {\scalebox{0.6}{R}};
            \vmpsW{0}{(\x+1)}{IcolVMPSgen}
             \node at ({(\x+1)*\dx},{-0.5*\dx}) {\scalebox{0.6}{R}};
        }
        \foreach \x in {0,...,5}{\gridLine{0}{\x}{-1}{\x}}
        \foreach \x in {0,2,...,4}{
            \vmpsV{0}{\x}{IcolVMPSgen}
            \node at ({(\x)*\dx},{-0.5*\dx}) {\scalebox{0.6}{L}};
            \vmpsW{0}{(\x+1)}{IcolVMPSgen}
             \node at ({(\x+1)*\dx},{-0.5*\dx}) {\scalebox{0.6}{L}};
        }
        \draw [decorate,decoration={brace,amplitude=5pt},xshift=0pt,yshift=0pt] 
    ({-.125*\dx},{1.*\dx})--({5.125*\dx},{1.*\dx}) node [midway,yshift=15pt]
    {$L$};
    \draw [decorate,decoration={brace,amplitude=5pt},xshift=0pt,yshift=0pt] 
    ({5.875*\dx},{1.*\dx})--({11.125*\dx},{1.*\dx}) node [midway,yshift=15pt]
    {$L$};
\end{tikzpicture}\,,
\label{eq:Psi0bipa}
\ee
where we took $L$ even. Quantum quenches from this kind of states are known as \emph{bipartitioning protocols}~\cite{bertini2021finitetemperature, bernard2016conformal, vasseur2016nonequilibrium, alba2021generalizedhydrodynamic} and can be thought of as the sudden junction of two homogeneous leads prepared in different states. In this case, the precise expression for the R\'enyi entropies depends on the position of $A$ with respect to the junction. For example, if the subsystem $A$ is starting at the site $x\geq0$ (i.e. on the right of the junction), we have
\be
\label{eq:SAnthbipa}
S_{x,A,\rm th}^{(n)}(t)=\frac{1}{1-n}
\log\Big[
    \frac{\tensor*[_n]{\mel{L_{\rm L}}{(\tilde {\mathbb W}_{\rm R}^{\otimes n})^{x/2}\mathcal S_{2n}
(\tilde {\mathbb W}_{\rm R}^{\ast\,\otimes n})^{|A|/2} \mathcal S^\dag_{2n}}{R_{\rm R}}}{_n}
}{\tensor*[_n]{\braket{L_{\rm L}}{R_{\rm R}}}{_n}}\Big]\,.
\ee
For the sake of simplicity in this paper we only consider the special case $x=0$, i.e.\ when $A$ starts right at the junction, i.e. 
\be
\label{eq:SAnthbipax0}
S_{A,\rm th}^{(n)}(t)=\frac{1}{1-n}
\log\Big[\frac{
    \tensor*[_n]{\mel{L_{\rm L}}{(\mathcal S_{2n}
(\tilde {\mathbb W}_{\rm R}^{\ast\,\otimes n})^{|A|/2} \mathcal S^\dag_{2n}}{R_{\rm R}}}{_n}
}{{\tensor*[_n]{\braket{L_{\rm L}}{R_{\rm R}}}{_n}}}\Big]\,.
\ee
If, in addition, the system has a strict maximal velocity, and the subsystem large enough, $|A|>2v_{\rm max}t$, the above expression reduces to
\be
\label{eq:SAnthhalfsystembipa}
S_{A,\rm th}^{(n)}(t)=\frac{1}{1-n}
\log\Big[
    \frac{\tensor*[_n]{\!\mel{L_{\rm L}}{\mathcal S_{2n}}{R_{\rm R}^{\ast}}}{_n}}{{\tensor*[_n]{\braket{L_{\rm L}}{R_{\rm R}}}{_n}}}
    \Big]
+\frac{1}{1-n} \log\Big[\Big(
   \frac{\tensor*[_n]{\!\mel{L_{\rm R}}{\mathcal S_{2n}}{R_{\rm R}^{\ast}}}{_n}}{{\tensor*[_n]{\braket{L_{\rm R}}{R_{\rm R}}}{_n}}}
   \Big)^{\ast} \Big].
\ee
Note that the two terms on the r.h.s.\ can be directly interpreted as the
entanglement produced at the two boundary points between $A$ and $\bar A$.
Indeed, the second term depends on the parameters of the right lead only, while
the first depends on the parameters of both left and right lead. Consistently,
repeating the same construction in the case of open boundary conditions and
taking $A$ to be semi-infinite one finds~\cite{klobas2021exact}
\be
\label{eq:SAnthhalfsystembipaopen}
\lim_{|A| \to \infty}S_{A,\rm th}^{(n)}(t)\Big|_{\mathrm{obc}}
=\frac{1}{1-n} \log\Big[
    \frac{\tensor[_n]{\!\mel{L_{\rm L}}{\mathcal S_{2n}}{R^{\ast}_{\rm R}}}{_n}}{{\tensor*[_n]{\braket{L_{\rm L}}{R_{\rm R}}}{_n}}}
    \Big]. 
\ee
Indeed, in this case there is a single boundary point between $A$ and $\bar A$. 

Our main goal will be to exploit the representations \eqref{eq:SAnthhalfsystem} and \eqref{eq:SAnthhalfsystembipa} to find the asymptotic behaviour of R\'enyi entropies for large times. Since \eqref{eq:SAnthhalfsystembipa} reduces to \eqref{eq:SAnth} for 
\be
\begin{tikzpicture}
    [baseline={([yshift=-0.6ex]current bounding box.center)},scale=1.75]
        \foreach \x in {0}{\gridLine{0}{\x}{-1}{\x}}
        \foreach \x in {0}{
            \vmpsV{0}{\x}{IcolVMPSgen}
             \node at ({(\x)*\dx},{-0.5*\dx}) {\scalebox{0.6}{R}};
        }
\end{tikzpicture}
=
\begin{tikzpicture}
    [baseline={([yshift=-0.6ex]current bounding box.center)},scale=1.75]
        \foreach \x in {0}{\gridLine{0}{\x}{-1}{\x}}
        \foreach \x in {0}{
            \vmpsV{0}{\x}{IcolVMPSgen}
             \node at ({(\x)*\dx},{-0.5*\dx}) {\scalebox{0.6}{L}};
        }
\end{tikzpicture}
=
\begin{tikzpicture}
    [baseline={([yshift=-1.6ex]current bounding box.center)},scale=1.75]
        \foreach \x in {0}{\gridLine{0}{\x}{-1}{\x}}
        \foreach \x in {0}{
            \vmpsV{0}{\x}{IcolVMPSgen}
        }
\end{tikzpicture}\, ,
\qquad  
\begin{tikzpicture}
    [baseline={([yshift=-0.6ex]current bounding box.center)},scale=1.75]
        \foreach \x in {0}{\gridLine{0}{\x}{-1}{\x}}
        \foreach \x in {0}{
            \vmpsW{0}{\x}{IcolVMPSgen}
             \node at ({(\x)*\dx},{-0.5*\dx}) {\scalebox{0.6}{R}};
        }
\end{tikzpicture}
=
\begin{tikzpicture}
    [baseline={([yshift=-0.6ex]current bounding box.center)},scale=1.75]
        \foreach \x in {0}{\gridLine{0}{\x}{-1}{\x}}
        \foreach \x in {0}{
            \vmpsW{0}{\x}{IcolVMPSgen}
             \node at ({(\x)*\dx},{-0.5*\dx}) {\scalebox{0.6}{L}};
        }
\end{tikzpicture}
=
\begin{tikzpicture}
    [baseline={([yshift=-1.6ex]current bounding box.center)},scale=1.75]
        \foreach \x in {0}{\gridLine{0}{\x}{-1}{\x}}
        \foreach \x in {0}{
            \vmpsW{0}{\x}{IcolVMPSgen}
        }
\end{tikzpicture}\,,
\ee
we can, without loss of generality, consider the inhomogeneous case~\eqref{eq:SAnthhalfsystembipa} only. 

\section{A solvable case: quantum cellular automaton Rule 54}
\label{eq:solvablestates}

The practical convenience of the representation~\eqref{eq:SAnthhalfsystembipa} depends on the form of the fixed points~$\bra{L}$ and~$\ket{R}$. For instance, they become extremely useful when the fixed points are written as matrix product states (MPS)s with a constant (i.e.\ time independent) bond dimension. This kind of simplification arises for some particular choices of the tensors~\eqref{eq:tensors}, i.e.\ for particular systems and initial states~\cite{bertini2019exact,piroli2020exact,klobas2021exact}. 

Here we focus on one of such choices. Specifically, we consider the quantum cellular automaton Rule 54, originally introduced in Ref.~\cite{bobenko1993two}, which has been recently shown to offer an exactly solvable benchmark for interacting integrable many-body dynamics, both in the classical~\cite{prosen2016integrability,prosen2017exact,inoue2018two,klobas2019time,buca2019exact,klobas2020matrix,klobas2020space}, and quantum~\cite{gopalakrishnan2018operator,gopalakrishnan2018hydrodynamics,friedman2019integrable,alba2019operator,alba2021diffusion,klobas2021exact} realm (see also the recent review~\cite{buca2021rule}). We can interpret it as a local quantum circuit where the time-evolution operator is written in the form \eqref{eq:UeUoMPO} with tensors~\cite{klobas2021exact, klobas2020space}
\be 
\begin{tikzpicture}
  [baseline={([yshift=-0.6ex]current bounding box.center)},scale=1.75]
  \gridLine{1}{0}{-1}{0};
  \gridLine{0}{1}{0}{-1};
  \bCircle{0}{0}{IcolU}
  \node at ({-1.3*\dx},{0}) {$\alpha$};
  \node at (0,{-1.3*\dx}) {$s$};
  \node at ({1.3*\dx},{0}) {$\beta$};
  \node at (0,{1.3*\dx}) {$r$};
\end{tikzpicture}=
  \delta_{\chi(s,\beta,r),\alpha}
  ,\qquad
 \begin{tikzpicture}
  [baseline={([yshift=-0.6ex]current bounding box.center)},scale=1.75]
  \gridLine{0.75}{0}{-0.75}{0};
  \gridLine{0}{0.75}{0}{-0.75};
  \sCircle{0}{0}{IcolU}
  \node at ({-1.25*\dx},{0}) {{$\alpha$}};
  \node at (0,{-\dx}) {{$s$}};
  \node at ({1.25*\dx},{0}) {{$\beta$}};
  \node at (0,{\dx}) {{$r$}};
\end{tikzpicture}= \delta_{s,\beta}\delta_{\beta,r}\delta_{r,\alpha},
\label{eq:balls}
\ee
where $d=\chi=2$ and $\chi(s,\beta,r)=(s+\beta+r+s r)\, {\rm mod}\, 2$. Note that, since Rule 54 can be represented as a local quantum circuit, is has a strict maximal velocity $v_{\rm max}$ for the propagation of signals. In particular, for our choice of units we have $v_{\rm max}=2$. 

Next, we consider initial-states 
\be
\ket{\Psi_{\vartheta, \vec{\varphi}}} =
\begin{tikzpicture}
    [baseline={([yshift=0.8ex]current bounding box.center)},scale=1.5]
    \foreach \x in {0,2,...,6}{
        \gridLine{0}{\x}{-1}{\x};
        \gridLine{0}{1+\x}{-1}{1+\x};
        \vmpsV{0}{\x}{colVMPS}
        \vmpsW{0}{(1+\x)}{colVMPS}
    }
    \draw [decorate,decoration={brace,amplitude=5pt},xshift=0pt,yshift=0pt] 
    ({7.25*\dx},{-0.25*\dx})--({-0.25*\dx},{-0.25*\dx}) node [midway,yshift=-12pt]
    {\scalebox{0.9}{$\vartheta$}};
\end{tikzpicture}
\label{eq:solvablefamily}
\ee
with tensors of the form 
\be 
\begin{tikzpicture}
    [baseline={([yshift=-2ex]current bounding box.center)},scale=1.5]
    \gridLine{0}{0}{-1}{0};
    \vmpsV{0}{0}{colVMPS}
    \node at (0,{1.5*\dx}) {{$s$}};
    % \node at (0,{-.75*\dx}) {$\vartheta$};
\end{tikzpicture}= \mathrm{e}^{\mathrm{i}\varphi_{1}} \delta_{s,0},\qquad
\begin{tikzpicture}
    [baseline={([yshift=-0.6ex]current bounding box.center)},scale=1.5]
    \gridLine{0}{0}{-1}{0};
    \vmpsW{0}{0}{colVMPS}
    \node at (0,{1.5*\dx}) {{$s$}};
    \node at (0,{-.75*\dx}) {\scalebox{0.9}{$\vartheta$}};
\end{tikzpicture}
=  \sqrt{1-\vartheta} \delta_{s,0} +
\sqrt{\vartheta}\mathrm{e}^{\mathrm{i}\varphi_{2}} \delta_{s,1}\,,
\label{eq:solvablefamily2}
\ee
where the parameter $\vartheta\in[0,1]$ will be referred to as the \emph{filling} while $\varphi_{1/2}\in[0,2\pi]$ as the \emph{phases}. 

In Paper I we prove that, choosing tensors of the form \eqref{eq:balls} and \eqref{eq:solvablefamily2}, the fixed points, $\bra{R}$ and $\ket{L}$, are MPSs of bond dimension 3. The latter depend on the filling but are independent of the phases. In fact, one can prove that fixed points are the same also when choosing different phases at each spatial point (as long as $\vartheta$ is the same everywhere). Explicitly, we have 
\be
\label{eq:fixedPointLR}
\bra{L_\vartheta}=
\begin{tikzpicture}
    [baseline={([yshift=-1.6ex]current bounding box.center)},scale=2]
    \draw[semithick,colLines,rounded corners=5] (0,{-4*\dx}) -- ({-0.75*\dx},{-4*\dx}) 
    -- ({-0.75*\dx},{-6*\dx}) -- ({0.75*\dx},{-6*\dx});
    \draw[semithick,colLines,rounded corners=5] (0,{-3*\dx}) -- ({-0.875*\dx},{-3*\dx}) 
    -- ({-0.875*\dx},{-6.25*\dx}) -- ({0.75*\dx},{-6.25*\dx});
    \draw[semithick,colLines,rounded corners=5] (0,{-2*\dx}) -- ({-\dx},{-2*\dx}) 
    -- ({-\dx},{-6.5*\dx}) -- ({0.75*\dx},{-6.5*\dx});
    \draw[semithick,colLines,rounded corners=5] (0,{-\dx}) -- ({-1.125*\dx},{-\dx}) 
    -- ({-1.125*\dx},{-6.75*\dx}) -- ({0.75*\dx},{-6.75*\dx});
    \foreach \t in {1,...,4}{
        \gridLine{\t}{0}{\t}{0.75};
    }
    \mpsWire{0.25}{0}{4.75}{0};
    \mpsBvec{0.25}{0};
    \mpsBvecW{4.75}{0}{colMPS};
    \mpsB{1}{0}{colMPS};
    \mpsA{2}{0}{colMPS};
    \mpsB{3}{0}{colMPS};
    \mpsA{4}{0}{colMPS};
    \node at (0,{-5.5*\dx}) {$\vartheta$};
\end{tikzpicture}\,,\qquad
\ket{R_\vartheta}=
\begin{tikzpicture}
    [baseline={([yshift=-1.6ex]current bounding box.center)},scale=2]
    \draw[semithick,colLines,rounded corners=5] (0,{-4*\dx}) -- ({0.75*\dx},{-4*\dx}) 
    -- ({0.75*\dx},{-6*\dx}) -- ({-0.75*\dx},{-6*\dx});
    \draw[semithick,colLines,rounded corners=5] (0,{-3*\dx}) -- ({0.875*\dx},{-3*\dx}) 
    -- ({0.875*\dx},{-6.25*\dx}) -- ({-0.75*\dx},{-6.25*\dx});
    \draw[semithick,colLines,rounded corners=5] (0,{-2*\dx}) -- ({\dx},{-2*\dx}) 
    -- ({\dx},{-6.5*\dx}) -- ({-0.75*\dx},{-6.5*\dx});
    \draw[semithick,colLines,rounded corners=5] (0,{-\dx}) -- ({1.125*\dx},{-\dx}) 
    -- ({1.125*\dx},{-6.75*\dx}) -- ({-0.75*\dx},{-6.75*\dx});
    \mpsBvec{0.25}{0};
    \mpsWire{0.25}{0}{4.75}{0};
    \foreach \t in {1,...,4}{
        \gridLine{\t}{-0.75}{\t}{0};
    }
    \foreach \t in {1,3}{
        \mpsA{\t}{0}{colMPS};
    }
    \foreach \t in {2,4}{
        \mpsB{\t}{0}{colMPS};
    }
    \mpsBvecV{4.75}{0}{colMPS};
    \node at (0,{-5.5*\dx}) {$\vartheta$};
\end{tikzpicture}\,,
\ee
where the ``bulk" tensors are given by
\begin{equation}
    \label{eq:boundarymatrices}
   \begin{aligned}
        \begin{tikzpicture}
            [baseline={([yshift=-0.6ex]current bounding box.center)},scale=1.7]
            \mpsWire{-0.625}{0}{.625}{0};
            \gridLine{0}{-0.625}{0}{0.625};
            \mpsA{0}{0}{colMPS}
            \node at ({0.875*\dx},0) {\scalebox{0.8}{$0$}};
            \node at ({-0.875*\dx},0) {\scalebox{0.8}{$0$}};
            \node at (0,{-1*\dx}) {\scalebox{0.8}{$\vartheta$}};
             \node at (0,{1.25*\dx}) {};
        \end{tikzpicture}\!\!= \!&
        \begin{bmatrix}
            1-\vartheta & 1-\vartheta & -(1-\vartheta) \\
            \vartheta & \vartheta & 1-\vartheta \\
            \vartheta & \displaystyle-\frac{\vartheta^2}{1-\vartheta} & -\vartheta
        \end{bmatrix},&
        \!\!
        \begin{tikzpicture}[baseline={([yshift=-0.6ex]current bounding box.center)},scale=1.7]
            \mpsWire{-0.625}{0}{.625}{0};
            \gridLine{0}{-0.625}{0}{0.625};
            \mpsA{0}{0}{colMPS}
            \node at ({-0.875*\dx},0) {\scalebox{0.8}{$0$}};
            \node at ({0.875*\dx},0) {\scalebox{0.8}{$1$}};
            \node at (0,{-1*\dx}) {\scalebox{0.8}{$\vartheta$}};
             \node at (0,{1.25*\dx}) {};
        \end{tikzpicture}\!\!\!=\!\!
        \begin{tikzpicture}[baseline={([yshift=-0.6ex]current bounding box.center)},scale=1.7]
            \mpsWire{-0.625}{0}{.625}{0};
            \gridLine{0}{-0.625}{0}{0.625};
            \mpsA{0}{0}{colMPS}
            \node at ({-0.875*\dx},0) {\scalebox{0.8}{$0$}};
            \node at ({0.875*\dx},0) {\scalebox{0.8}{$1$}};
            \node at (0,{-1*\dx}) {\scalebox{0.8}{$\vartheta$}};
            \node at (0,{1.25*\dx}) {};
        \end{tikzpicture}\!\!=
        \!&
        \begin{bmatrix}
            0 & 1-\vartheta & -(1-\vartheta) \\
            \vartheta & 0 & 0 \\
            \vartheta & 0 & 0 
        \end{bmatrix}\!\!,\\
        \begin{tikzpicture}[baseline={([yshift=-0.6ex]current bounding box.center)},scale=1.7]
            \mpsWire{-0.625}{0}{.625}{0};
            \gridLine{0}{-0.625}{0}{0.625};
            \mpsB{0}{0}{colMPS};
            \node at ({-0.875*\dx},0) {\scalebox{0.8}{$s$}};
            \node at ({0.875*\dx},0) {\scalebox{0.8}{$r$}};
        \end{tikzpicture}\!\!=\!
        &\begin{bmatrix}
            \delta_{r,0}\delta_{s,0} & 0 & 0 \\ 
            0 & \delta_{r,1}\delta_{s,1} & 0 \\
            0 & 0 & \delta_{r,1}\delta_{s,1}
        \end{bmatrix},&
        \!\begin{tikzpicture}[baseline={([yshift=-0.6ex]current bounding box.center)},scale=1.7]
            \mpsWire{-0.625}{0}{.625}{0};
            \gridLine{0}{-0.625}{0}{0.625};
            \mpsA{0}{0}{colMPS}
            \node at ({-0.875*\dx},0) {\scalebox{0.8}{$1$}};
            \node at ({0.875*\dx},0) {\scalebox{0.8}{$1$}};
        \end{tikzpicture}\!\!=\!
        &\begin{bmatrix}
            0 & 1 & 0 \\ 1 & 0 & 0 \\ 0 & 0 & 0
        \end{bmatrix}\!\!,
    \end{aligned}
\end{equation}
and boundary vectors are
\be
\label{eq:boundaryvectors}
\begin{tikzpicture}
    [baseline={([yshift=-0.6ex]current bounding box.center)},scale=1.8]
    \mpsWire{0}{0}{0.625}{0};
    \mpsBvecW{0.625}{0}{colMPS};
\end{tikzpicture}=
\begin{bmatrix}
    1 \\ 0 \\ 0 
\end{bmatrix},\qquad
\begin{tikzpicture}
    [baseline={([yshift=0.8ex]current bounding box.center)},scale=1.8]
    \mpsWire{0}{0}{0.625}{0};
    \mpsBvecV{0.625}{0}{colMPS};
    \node at (0,{-1.25*\dx}) {\scalebox{0.9}{$\vartheta$}};
\end{tikzpicture}=
-\frac{1}{1-\vartheta}
\begin{bmatrix}
    (1-\vartheta)^2 \\ \vartheta(1-\vartheta) \\ -{\vartheta^2} 
\end{bmatrix},\qquad
\begin{tikzpicture}
    [baseline={([yshift=-0.6ex]current bounding box.center)},scale=1.8]
    \mpsBvec{0}{0};
    \mpsWire{0}{0}{-0.625}{0};
\end{tikzpicture} = \frac{1}{\sqrt{2}}
\begin{bmatrix}1\\1\\0
\end{bmatrix}.
\ee
These choices give left and right fixed point fulfilling 
\be
\braket{L_{\vartheta_1}}{R_{\vartheta_2}}=
\begin{tikzpicture}
  [baseline={([yshift=0.6ex]current bounding box.center)},scale=2]
  \mpsWire{0}{0}{0.75}{0};
  \mpsBvec{0}{0};
  \mpsBvecV{0.75}{0}{colMPS};
       \mpsWire{0}{0.75}{0.75}{0.75};
       \mpsBvec{0}{0.75};
       \mpsBvecW{0.75}{0.75}{colMPS};
       \node at (0,{-1.35*\dx}) {\scalebox{0.8}{$\vartheta_1$}};
       \node at ({0.75*\dx},{-1.35*\dx}) {\scalebox{0.8}{$\vartheta_2$}};
\end{tikzpicture} =1, \qquad \forall \vartheta_1,\vartheta_2.
\label{eq:LRoverlap}
\ee
In the above diagrams we explicitly reported $\vartheta$ to signal the dependence on the filling. In the following, however, whenever the choice of $\vartheta$ is unambiguous we will ease the notation by removing it.

As proven in Paper I the state $\ket{\Psi_{\vartheta, \vec{\varphi}}}$ relaxes (locally) to a family of Gibbs states. In particular considering density matrix reduced to a finite subsystem $A$ we have 
\be
      \rho_{A}(t)\simeq \rho_{\mathrm{GE},A} = \frac{{\rm tr}_{\bar A}(\mathrm{e}^{-\mu(\vartheta) N})}{{\rm tr}(\mathrm{e}^{-\mu(\vartheta) N})}\,,
      \qquad N=N_{+}+N_{-},
\label{eq:GE}
\ee
where $\simeq$ denotes the leading contribution for large times, $N_{\pm}$ are the number of left and right-moving quasiparticles (solitons) explicitly given by~\cite{klobas2020exactPhD, gopalakrishnan2018hydrodynamics}
\be
      \begin{aligned}
          N_{\rm +} &= \sum_{x\in\mathbb Z_{L}}  P_{2x}^{-} P^{-}_{2x+1}  
          +  \sum_{x\in\mathbb Z_{2L}} P^{+}_{x} P^-_{x+1} P^+_{x+2}\,, \\
          N_{\rm -} &= \sum_{x\in\mathbb Z_{L}}  P_{2x-1}^{-} P^{-}_{2x}  
          +\sum_{x\in\mathbb Z_{2L}} P^{+}_{x} P^-_{x+1} P^+_{x+2}\,,
      \end{aligned}\qquad
P^{\pm}\coloneqq \frac{\1\pm\sigma_3}{2}\,,
\ee
and the chemical potential $\mu(\theta)$ reads as 
\be
\label{eq:muvartheta}
e^{-\mu(\vartheta)} = \frac{\vartheta}{1-\vartheta}\qquad\Rightarrow\qquad \vartheta =  \frac{1}{1+e^{\mu(\vartheta)}}\,.
\ee
This shows that $\vartheta$ in \eqref{eq:solvablefamily} sets the density of quasiparticles in the stationary state.

In fact, the states \eqref{eq:solvablefamily} can also be used to design solvable bipartitioning protocols. Indeed, as proven in Paper I, considering initial states of the form 
\be
\label{eq:solvablebipartitioning}
\ket{\Psi_{\vartheta_{\rm L}, \vec{\varphi}_{\rm L}, \vartheta_{\rm R}, \vec{\varphi}_{\rm R}}} = 
      \ket{\Psi_{\vartheta_{\rm L}, \vec{\varphi}_{\rm L}}}  
      \otimes  \ket{\Psi_{\vartheta_{\rm R}, \vec{\varphi}_{\rm R}}},
\ee
one finds $\bra{L_{\rm L}}=\bra{L_{\vartheta_{\rm L}}}$ and $\ket{R_{\rm R}}=\ket{R_{\vartheta_{\rm R}}}$ (with both $\bra{L_{\vartheta}}$ and $\ket{R_{\vartheta}}$ of the form \eqref{eq:fixedPointLR}).
In this case any finite subsystem $A$ at finite distance from the junction relaxes to a family of \emph{generalised} Gibbs states. Namely  
\be
 \rho_{A}(t)\simeq \rho_{\mathrm{GGE},A} = \frac{{\rm tr}_{\bar A}(\mathrm{e}^{-\mu_{\rm L}(\vartheta_{\rm L},\vartheta_{\rm R}) N_{\rm -}-\mu_{\rm R}(\vartheta_{\rm L},\vartheta_{\rm R}) N_{\rm +}})}{{\rm tr}(\mathrm{e}^{-\mu_{\rm L}(\vartheta_{\rm L},\vartheta_{\rm R}) N_{\rm -}-\mu_{\rm R}(\vartheta_{\rm L},\vartheta_{\rm R}) N_{\rm +}})}\,,
\label{eq:GGE}
\ee
where $\mu_{\mathrm{R/L}}(\vartheta_{\rm L},\vartheta_{\rm R})$ is given by 
\be
e^{-\mu_{\rm R/L}(\vartheta_{\rm L},\vartheta_{\rm R})} = \frac{\vartheta_{\rm R/L}(1-\vartheta_{\rm L/R})}
{(1-\vartheta_{\rm R/L})^2}\,.
\label{eq:muvarthetaLR}
\ee
Importantly, in Rule 54 the relaxation happens with finite rate~\cite{klobas2021exact} (see also Paper I). In particular, the finite-time corrections to~\eqref{eq:GE} and~\eqref{eq:GGE} are exponentially small in $t-3|A|/2$.  

Finally, we recall (see e.g.\ Paper I) that $\rho_{\mathrm{GGE},A}$ in Eq.~\eqref{eq:GGE} (and hence also its particular case \eqref{eq:GE}) is conveniently expressed in terms of the following MPO
\be\label{eq:GGEmpo}
\rho_{\mathrm{GGE},A}=
\frac{1}{Z_A}
  \begin{tikzpicture}
    [baseline={([yshift=-3ex]current bounding box.center)},scale=2]
    \gridLine{0.75}{1}{-0.75}{1};
    \gridLine{0.75}{2}{-0.75}{2};
    \gridLine{0.75}{3}{-0.75}{3};
    \gridLine{0.75}{4}{-0.75}{4};
    \mpsWire{0}{0}{-0.625}{0};
    \mpsWire{0}{5}{-0.625}{5};
    \vmpsWire{0}{0}{0}{5};
    \mpsBvec{-0.625}{0};
    \mpsBvec{-0.625}{5};
    \mpsBvecW{0}{0}{colSMPS};
    \mpsBvecV{0}{5}{colSMPS};
    \vmpsV{0}{1}{colSMPS};
    \vmpsW{0}{2}{colSMPS};
    \vmpsV{0}{3}{colSMPS};
    \vmpsW{0}{4}{colSMPS};
    \draw [decorate,decoration={brace,amplitude=5pt},xshift=0pt,yshift=0pt] 
    ({0.75*\dx},{0.875*\dx})--({4.25*\dx},{0.875*\dx}) node [midway,yshift=12pt]
     {$\abs{A}$};
    %\draw [decorate,decoration={brace,amplitude=5pt},xshift=0pt,yshift=0pt] 
    %({5.125*\dx},{-1.125*\dx})--({-0.125*\dx},{-1.125*\dx}) node [midway,yshift=-12pt]
      %{\scalebox{0.9}{$\vartheta_1,\vartheta_2$}};
  \end{tikzpicture},\qquad
  Z_A= {1+\vartheta_{\rm L}+\vartheta_{\rm R}}.
\ee
The bulk tensors
$\begin{tikzpicture}[baseline={([yshift=-0.6ex]current bounding box.center)},scale=1.5]
    \vmpsWire{0}{-0.5}{0}{0.5}
    \gridLine{0.5}{0}{-0.5}{0};
    \vmpsV{0}{0}{colSMPS}
\end{tikzpicture}$, 
$\begin{tikzpicture}[baseline={([yshift=-0.6ex]current bounding box.center)},scale=1.5]
    \vmpsWire{0}{-0.5}{0}{0.5}
    \ngridLine{0.5}{0}{-0.5}{0};
    \vmpsW{0}{0}{colSMPS}
\end{tikzpicture}$
are diagonal in the two copies of the physical space, and the auxiliary space is $3$-dimensional,
\be
\begin{aligned}
\begin{tikzpicture}[baseline={([yshift=-0.6ex]current bounding box.center)},scale=2]
    \vmpsWire{0}{-0.5}{0}{0.5}
    \gridLine{0.75}{0}{-0.75}{0}
    \vmpsV{0}{0}{colSMPS}
    \node at (0,{1.125*\dx}) {\scalebox{0.8}{$0$}};
    \node at (0,{-1.125*\dx}) {\scalebox{0.8}{$0$}};
    %\node at (0,{-1.5*\dx}) {\scalebox{0.8}{$\vartheta_1,\vartheta_2$}};
\end{tikzpicture}&= %\mkern-8mu&=
\begin{bmatrix}
    1 & 0 & 0 \\
    \mathrm{e}^{-\mu_{\rm R}(\vartheta_{\rm L},\vartheta_{\rm R})} & 0 & 0 \\
    1 & 0 & 0 
\end{bmatrix},&\quad
\begin{tikzpicture}[baseline={([yshift=-0.6ex]current bounding box.center)},scale=2]
    \vmpsWire{0}{-0.5}{0}{0.5}
    \gridLine{0.75}{0}{-0.75}{0}
    \vmpsV{0}{0}{colSMPS}
    \node at (0,{1.125*\dx}) {\scalebox{0.8}{$0$}};
    \node at (0,{-1.125*\dx}) {\scalebox{0.8}{$1$}};
    %\node at (0,{-0.5*\dx}) {\scalebox{0.8}{$\vartheta_1,\vartheta_2$}};
\end{tikzpicture}&= %\mkern-8mu&=\mkern-8mu
\begin{tikzpicture}[baseline={([yshift=-0.6ex]current bounding box.center)},scale=2]
    \vmpsWire{0}{-0.5}{0}{0.5}
    \gridLine{0.75}{0}{-0.75}{0}
    \vmpsV{0}{0}{colSMPS}
    \node at (0,{1.125*\dx}) {\scalebox{0.8}{$1$}};
    \node at (0,{-1.125*\dx}) {\scalebox{0.8}{$0$}};
    %\node at (0,{-0.5*\dx}) {\scalebox{0.8}{$\vartheta_1,\vartheta_2$}};
\end{tikzpicture}=0,\\%\mkern-8mu=0,\\
\begin{tikzpicture}[baseline={([yshift=-0.6ex]current bounding box.center)},scale=2]
    \vmpsWire{0}{-0.5}{0}{0.5}
    \gridLine{0.75}{0}{-0.75}{0}
    \vmpsV{0}{0}{colSMPS}
    \node at (0,{1.125*\dx}) {\scalebox{0.8}{$1$}};
    \node at (0,{-1.125*\dx}) {\scalebox{0.8}{$1$}};
    %\node at (0,{-0.5*\dx}) {\scalebox{0.8}{$\vartheta_1,\vartheta_2$}};
\end{tikzpicture}&= %\mkern-8mu&=
\begin{bmatrix}
    0 & \mathrm{e}^{-\mu_{\rm R}(\vartheta_{\rm L},\vartheta_{\rm R})} & 0 \\
    0 & 0 & 1 \\
    0 & 0 & \mathrm{e}^{-\mu_{\rm L}(\vartheta_{\rm L},\vartheta_{\rm R})} 
\end{bmatrix},&
\begin{tikzpicture}[baseline={([yshift=-0.6ex]current bounding box.center)},scale=2]
    \vmpsWire{0}{-0.5}{0}{0.5}
    \gridLine{0.75}{0}{-0.75}{0}
    \vmpsW{0}{0}{colSMPS}
    %\node at (0,{-0.5*\dx}) {\scalebox{0.8}{$\vartheta_1,\vartheta_2$}};
\end{tikzpicture} &= %\mkern-8mu&=\mkern-8mu
    (1-\vartheta_{\rm L})(1-\vartheta_{\rm R}) 
\left.
\begin{tikzpicture}[baseline={([yshift=-0.6ex]current bounding box.center)},scale=2]
    \vmpsWire{0}{-0.5}{0}{0.5}
    \gridLine{0.75}{0}{-0.75}{0}
    \vmpsV{0}{0}{colSMPS}
    %\node at (0,{-0.5*\dx}) {\scalebox{0.8}{$\vartheta_2,\vartheta_1$}};
\end{tikzpicture}
    \right|_{\mu_{\rm L}\leftrightarrow\mu_{\rm R}}.
\end{aligned}
\ee
The boundary tensors
$\begin{tikzpicture}[baseline={([yshift=-0.6ex]current bounding box.center)},scale=1.5]
    \vmpsWire{0}{0}{0}{0.625}
    \mpsWire{0}{0}{-0.625}{0};
    \mpsBvec{-0.625}{0}
    \mpsBvecW{0}{0}{colSMPS}
\end{tikzpicture}$, 
$\begin{tikzpicture}[baseline={([yshift=-0.6ex]current bounding box.center)},scale=1.5]
    \vmpsWire{0}{0}{0}{-0.625}
    \mpsWire{0}{0}{-0.625}{0};
    \mpsBvec{-0.625}{0}
    \mpsBvecV{0}{0}{colSMPS}
\end{tikzpicture}$
are $3$-dimensional (row and column) vectors and their explicit expression is reported in Appendix A of Paper I.

\section{Exact results for R\'enyi entropies}
\label{sec:mainresult}

In this section we show that combining the representations \eqref{eq:SAnthhalfsystembipa} and \eqref{eq:SAnthhalfsystembipaopen} with the exact expressions \eqref{eq:fixedPointLR} the problem of computing the growth of R\'enyi entropies is mapped into that of contracting a certain tensor network. This can be done exactly in the asymptotic limit $1\ll t \leq |A|/4$. Moreover, using \eqref{eq:GGEmpo}, we also show that also the stationary value reached by the entropies for large times is characterised by a tensor network. As we shall see, the latter is contracted exactly in the limit of large $|A|$. Let us begin by proving the latter statement.

\subsection{Stationary values}

As a consequence of \eqref{eq:GGE} we have that for ${t>3|A|/2}$ the R\'enyi entropies fulfil 
\be
S^{(\alpha)}_A(t) \simeq S^{(\alpha)}_{\text{GGE},A} = \frac{1}{1-\alpha}
\log\left[\tr\Big(\rho_{\text{GGE},A}^{\alpha}\Big)\right],
\ee
where we recall that $\simeq$ denotes equality up to exponential corrections and $\rho_{\text{GGE},A}$ the GGE reduced to the subsystem $A$. Using the MPO representation of $\rho_{\text{GGE},A}$ (cf. Eq.~\eqref{eq:GGEmpo}) we can express R\'enyi entropies with index $n$ (integer and larger than one) in terms of the following tensor network
\be 
\tr\rho_{\mathrm{GGE},A}^n=
\frac{1}{Z_A^n}
\begin{tikzpicture}[baseline={([yshift=0ex]current bounding box.center)},scale=2]
    \foreach \x in{1,...,4}{
    \gridLine{0.25}{\x}{5.75}{\x};
    \griddashedLine{0.25}{(\x-0.25)}{5.75}{(\x-0.25)};
    \ItopHook{0.25}{\x}{0.125}
    \IbottomHook{5.75}{\x}{0.125}
    }
    \foreach \t in{1,...,5}{
        \vmpsWire{\t}{0}{\t}{5}
        \vmpsV{\t}{1}{colSMPS}
        \vmpsW{\t}{2}{colSMPS}
        \vmpsV{\t}{3}{colSMPS}
        \vmpsW{\t}{4}{colSMPS}
        \mpsWire{\t}{0}{\t-0.5}{0}
        \mpsWire{\t}{5}{\t-0.5}{5}
        \mpsBvec{(\t-0.5)}{0}
        \mpsBvec{(\t-0.5)}{5}
        \mpsBvecW{\t}{0}{colSMPS}
        \mpsBvecV{\t}{5}{colSMPS}
    }
    \draw [decorate,decoration={brace,amplitude=5pt},xshift=0pt,yshift=0pt] 
    ({4.125*\dx},{-6.125*\dx})--({0.75*\dx},{-6.125*\dx}) node [midway,yshift=-10pt]
    {$\abs{A}$};
    \draw [decorate,decoration={brace,amplitude=5pt},xshift=0pt,yshift=0pt] 
    ({5.25*\dx},{-0.75*\dx})--({5.25*\dx},{-5.25*\dx}) node [midway,xshift=10pt]
    {$n$};
    \draw [gray,very thick,rounded corners=2]
    ({2.625*\dx},{-6*\dx}) rectangle ({4.375*\dx},0);
    \node at ({3.5*\dx},{0.5*\dx}) {$T_n$};
\end{tikzpicture}=
\frac{1}{Z_A^n}\ 
\tensor*[^{\otimes n}]{\!\bra*{
\begin{tikzpicture}[baseline={([yshift=-0.6ex]current bounding box.center)},scale=2]
        \mpsWire{0}{0}{-0.5}{0}
        \mpsBvec{-0.5}{0}
        \mpsBvecW{0}{0}{colSMPS}
\end{tikzpicture}}}{}
    T_n^{\abs{A}/2}
    \tensor*{\ket*{
\begin{tikzpicture}[baseline={([yshift=-0.6ex]current bounding box.center)},scale=2]
        \mpsWire{0}{0}{-0.5}{0}
        \mpsBvec{-0.5}{0}
        \mpsBvecV{0}{0}{colSMPS}
\end{tikzpicture}}}{^{\otimes n}}.
\ee
This immediately implies that for large $\abs{A}$ the R\'enyi entropy is dominated by the leading eigenvalue $\Lambda_n$ of the transfer matrix $T_n$ (defined in the above diagram). Namely 
\be
\label{eq:formalStationaryEntropy}
S^{(n)}_{\mathrm{GGE},A}
=
\frac{\abs{A}}{2(1-n)} \log\Big(\Lambda_n(\vartheta_1,\vartheta_2)\Big)
+\mathcal{O}\big({\abs{A}^{0}}\big),
\ee
where we used the fact that $Z_A$ does not grow with $\abs{A}$ (cf. Eq.~\eqref{eq:GGEmpo}). 

To find $\Lambda_n$ we make use of the following relations
\be
\label{eq:projectorRelationsTn}
\begin{tikzpicture}[baseline={([yshift=-0.6ex]current bounding box.center)},scale=2]
    \vmpsWire{0}{0.25}{0}{2.75}
    \vmpsWire{1}{0.25}{1}{2.75}
    \gridLine{-0.5}{1}{1.5}{1}
    \gridLine{-0.5}{2}{1.5}{2}
    \vmpsV{0}{1}{colSMPS}
    \vmpsV{1}{1}{colSMPS}
    \vmpsW{0}{2}{colSMPS}
    \vmpsW{1}{2}{colSMPS}
\end{tikzpicture}=
\begin{tikzpicture}[baseline={([yshift=-0.6ex]current bounding box.center)},scale=2]
    \vmpsWire{0}{0.5}{0}{3.25}
    \vmpsWire{1}{0.5}{1}{3.25}
    \gridLine{-0.5}{1}{1.5}{1}
    \gridLine{-0.5}{2}{1.5}{2}
    \vmpsV{0}{1}{colSMPS}
    \vmpsV{1}{1}{colSMPS}
    \vmpsW{0}{2}{colSMPS}
    \vmpsW{1}{2}{colSMPS}
    \proj{0}{1}{2.75}{colTensor}
\end{tikzpicture},\qquad
\begin{tikzpicture}[baseline={([yshift=-0.6ex]current bounding box.center)},scale=2]
    \vmpsWire{0}{-0.25}{0}{2.75}
    \vmpsWire{1}{-0.25}{1}{2.75}
    \gridLine{-0.5}{1}{1.5}{1}
    \gridLine{-0.5}{2}{1.5}{2}
    \vmpsV{0}{1}{colSMPS}
    \vmpsV{1}{1}{colSMPS}
    \vmpsW{0}{2}{colSMPS}
    \vmpsW{1}{2}{colSMPS}
    \proj{0}{1}{0.25}{colTensor}
\end{tikzpicture}=
\begin{tikzpicture}[baseline={([yshift=-0.6ex]current bounding box.center)},scale=2]
    \vmpsWire{0}{-0.25}{0}{3.25}
    \vmpsWire{1}{-0.25}{1}{3.25}
    \gridLine{-0.5}{1}{1.5}{1}
    \gridLine{-0.5}{2.5}{1.5}{2.5}
    \vmpsV{0}{1}{colSMPS}
    \vmpsV{1}{1}{colSMPS}
    \vmpsW{0}{2.5}{colSMPS}
    \vmpsW{1}{2.5}{colSMPS}
    \proj{0}{1}{1.75}{colTensor}
    \proj{0}{1}{0.25}{colTensor}
\end{tikzpicture},
\ee
where we introduced the projector defined by 
\be
\label{eq:projTn}
\begin{tikzpicture}[baseline={([yshift=-0.6ex]current bounding box.center)},scale=2]
    \vmpsWire{0}{-0.5}{0}{0.5}
    \vmpsWire{1}{-0.5}{1}{0.5}
    \proj{0}{1}{0}{colTensor}
    \node at ({-0.75*\dx},0) {\scalebox{0.9}{$x$}};
    \node at ({-0.75*\dx},{-\dx}) {\scalebox{0.9}{$y$}};
    \node at ({0.75*\dx},0) {\scalebox{0.9}{$z$}};
    \node at ({0.75*\dx},{-\dx}) {\scalebox{0.9}{$w$}};
\end{tikzpicture}=\delta_{x,y}\delta_{y,z}\delta_{z,w}.
\ee
The identities~\eqref{eq:projectorRelationsTn} imply that the eigenvalues of $T_n$ coincide
with the spectrum of the reduced transfer matrix $\tilde{T}_n$ defined by applying the 
projector~\eqref{eq:projTn} on all the pairs of auxiliary legs
\be
\tilde{T}_n = 
\begin{tikzpicture}[baseline={([yshift=-0.6ex]current bounding box.center)},scale=2]
    \foreach \x in {1,3}{
    \gridLine{0.25}{\x}{5.75}{\x};
    \griddashedLine{0.25}{(\x-0.25)}{5.75}{(\x-0.25)};
    \ItopHook{0.25}{\x}{0.125}
    \IbottomHook{5.75}{\x}{0.125}}
    \foreach \t in {1,...,5}{
        \vmpsWire{\t}{-0.625}{\t}{4.625}
        \vmpsV{\t}{1}{colSMPS}
        \vmpsW{\t}{3}{colSMPS}
    }
    \foreach \t in {1,3}{
        \proj{\t}{(\t+1)}{-0.1875}{colTensor}
        \proj{\t}{(\t+1)}{3.8125}{colTensor}
        \proj{\t}{(\t+1)}{1.8125}{colTensor}
    }
    \foreach \t in {2,4}{
        \proj{\t}{(\t+1)}{0.1875}{colTensor}
        \proj{\t}{(\t+1)}{4.1875}{colTensor}
        \proj{\t}{(\t+1)}{2.1875}{colTensor}
    }
\end{tikzpicture}.
\ee
Therefore the non-zero eigenvalues of $T_n$ are given by the spectrum of a $3\times 3$ matrix
\be
\Sp(T_n) = \Sp(\tilde{T}_n) = \{0\} \cup
\Sp\left(
\bar{\vartheta}^n_{1}\bar{\vartheta}^n_{2}
\begin{bmatrix}
    1+\mathrm{e}^{-n(\mu_1+\mu_2)} & \mathrm{e}^{-n \mu_1} & \mathrm{e}^{-n \mu_2}\\
    1+\mathrm{e}^{-n\mu_2} & \mathrm{e}^{-n(\mu_1+\mu_2)} & \mathrm{e}^{-n \mu_2}\\
    1+\mathrm{e}^{-n\mu_1} & \mathrm{e}^{-n \mu_1} & \mathrm{e}^{-n (\mu_1+\mu_2)}
\end{bmatrix}
\right),
\ee
where we use the shorthand notation $\bar{\vartheta}_{1/2}=1-\vartheta_{1/2}$.
In particular, the leading eigenvalue
$\Lambda_n(\vartheta_1,\vartheta_2)$ can be expressed as
\be
\Lambda_n(\vartheta_1,\vartheta_2)=
\frac{(1-\vartheta_{1})^n(1-\vartheta_{2})^n}{(1-\vartheta^{(n)}_1)(1-\vartheta^{(n)}_2)},
\label{eq:Lambdan}
\ee
where we defined $\vartheta_{1,2}^{(n)}$ fulfilling 
\be
\frac{\vartheta_{1,2}^{(n)}(1-\vartheta_{2,1}^{(n)})}
{(1-\vartheta_{1,2}^{(n)})^2}=
\left(\frac{\vartheta_{1,2}(1-\vartheta_{2,1})}
{(1-\vartheta_{1,2})^2}\right)^n.
\ee
Note that $\vartheta_{1,2}^{(n)}$ can be understood as generalisations
of the filling functions $\vartheta_{1,2}$ to the case where chemical potentials
$\mu_{1,2}$ are replaced  by $ n \mu_{1,2}$ (see Eq.~\eqref{eq:muvarthetaLR}). We also remark that the result obtained by substituting \eqref{eq:Lambdan} in \eqref{eq:formalStationaryEntropy} agrees with the TBA prediction of Ref.~\cite{alba2017quench}.

The result \eqref{eq:formalStationaryEntropy} can be analytically continued to 
$\mathcal D=\{z\in\mathbb C:\,\, {\rm Re}[z]>0\}$. Indeed, the function $\Lambda_z(\vartheta_1,\vartheta_2)$ --- obtained by replacing $n$ in Eq.~\eqref{eq:Lambdan} with $z\in\mathbb C$ --- is holomorphic and bounded in
$\mathcal D$. Therefore, Carlson's Theorem~\cite{rubel1956necessary} ensures that it is the only analytic continuation of $\{\Lambda_n(\vartheta_1,\vartheta_2)\}_{n=1,2,3,\ldots}$ fulfilling 
\be
\begin{aligned}
    | \Lambda_z(\vartheta_1,\vartheta_2) |&\leq C e^{\tau |z|},& z&\in\mathcal D,&\qquad
    | \Lambda_{1+iy}(\vartheta_1,\vartheta_2)|&\leq C e^{ c |y|},& y&\in \mathbb R\,,
\end{aligned}
\ee
with $C,\tau\in\mathbb R$ and $c<\pi$. As this is a requirement that we expect from physical grounds, we choose $\Lambda_z(\vartheta_1,\vartheta_2)$ as the relevant analytic continuation. In particular, in the limit $z\to 1$ we find 
\be
S_{\mathrm{GGE},A}=-\frac{\abs{A}}{2}
\sum_{j=1}^2
\frac{1+2\vartheta_{3-j}}{1+\vartheta_1+\vartheta_2}
\left(
\vartheta_j \log\vartheta_j+
(1-\vartheta_j) \log(1-\vartheta_j)
\right) + \mathcal{O}(\abs{A}^0),
\ee
which coincides with the expression of the Yang-Yang entropy in the state \eqref{eq:GGE}~\cite{friedman2019integrable}. In the homogeneous case, when $\vartheta_1=\vartheta_2=\vartheta$, R\'enyi entropies take a free-fermionic form (see e.g.\ \cite{alba2017quench})
\be\label{eq:GEstationaryRenyi}
S^{(n)}_{\mathrm{GE},A}=-\frac{\abs{A}}{1-n}
\log\Big((1-\vartheta)^n+\vartheta^n\Big)+ \mathcal{O}(\abs{A}^0).
\ee

\subsection{Asymptotic slopes}
\label{sec:Renyi}
Let us consider the elementary building blocks 
\be
b_n(\vartheta_1,\vartheta_2)\coloneqq \frac{1}{1-n} {\log\Big[ \frac{\tensor*[_n]{\!\mel*{L_{\vartheta_1}}{\mathcal{S}_{2n}}{R_{\vartheta_2}^{\ast}}}{_n}}{\tensor*[_n]{\!\braket*{L_{\vartheta_1}}{R_{\vartheta_2}}}{_n}} \Big]}, \label{eq:goal}
\ee 
where $\bra{L_{\vartheta_1}}$, $\ket{R_{\vartheta_2}}$ are both of the form \eqref{eq:fixedPointLR}. Recalling \eqref{eq:SAnthhalfsystembipa} we see that $b_n(\vartheta_1,\vartheta_2)$ can be interpreted as the $n$-th R\'enyi entropy generated at one of the boundaries of the subsystem $A$. This means that, upon analytic continuation, evaluating \eqref{eq:goal} gives direct access to \emph{all} R\'enyi entropies (including von Neumann) for $t\leq |A|/4$.

Considering the graphical representation \eqref{eq:fixedPointLR} of the fixed points, we can express the matrix element in~\eqref{eq:goal} in terms of the following tensor network
\be
\tensor[_n]{\!\mel{L_{\vartheta_1}}{\mathcal{S}_{2n}}{R_{\vartheta_2}^{\ast}}}{_n}
 =   \begin{tikzpicture}[baseline={([yshift=-2.6ex]current bounding box.center)},scale=1.95]
    \foreach \t in {0,...,5}{
        \gridLine{\t}{0.5}{\t}{6.5}
        \leftHook{\t}{0.5};
        \rightHook{\t}{6.5};
        \griddashedLine{(\t+0.3)}{0.5}{(\t+0.3)}{6.5};
    }
    \foreach \x in {1,3,5}{
      \mpsWire{-1}{\x}{6}{\x};
      \foreach \t in {0,2,4}{\mpsB{\t}{\x}{colMPS}};
      \foreach \t in {1,3,5}{\mpsA{\t}{\x}{colMPS}};
      \mpsBvec{-1}{\x}
      \mpsBvecW{6}{\x}{colMPS};
    }
    \foreach \x in {2,4,6}{
      \mpsWire{-1}{\x}{6}{\x};
      \foreach \t in {0,2,4}{\mpsA{\t}{\x}{colMPS}};
      \foreach \t in {1,3,5}{\mpsB{\t}{\x}{colMPS}};
      \mpsBvec{-1}{\x}
      \mpsBvecV{6}{\x}{colMPS};
    }
    \draw [decorate,decoration={brace,amplitude=5pt},xshift=0pt,yshift=0pt] 
    ({0.75*\dx},{1.125*\dx})--({6.25*\dx},{1.125*\dx}) node [midway,yshift=10pt]
    {$2n$};
    \draw [decorate,decoration={brace,amplitude=5pt},xshift=0pt,yshift=0pt] 
    ({6.75*\dx},{0.125*\dx})--({6.75*\dx},{-5.125*\dx}) node [midway,xshift=10pt]
    {$2t$};
    \foreach \x in {1,3,5}{
        \node at ({\x*\dx},{-6.75*\dx}) {\scalebox{0.9}{$\vartheta_1$}};}
     \foreach \x in {2,4,6}{
         \node at ({\x*\dx},{-6.75*\dx}) {\scalebox{0.9}{$\vartheta_2$}};}
   %\node at ({0*\dx},{0.18}) {};
  \end{tikzpicture}\mkern-14mu=
  \bra{\mathcal U_{n}} \mathcal T^t_{n} \ket{\mathcal D_{n}},
\label{eq:transfermatrep}
\ee
where we introduced
\be
{\mathcal{T}_{n}} \!=\!
  \begin{tikzpicture}[baseline={([yshift=-2.6ex]current bounding box.center)},scale=1.95]
    \foreach \t in {0,1}{
        \gridLine{\t}{0.5}{\t}{6.5}
        \leftHook{\t}{0.5};
        \rightHook{\t}{6.5};
        \griddashedLine{(\t+0.3)}{0.5}{(\t+0.3)}{6.5};
    }
    \foreach \x in {1,3,5}{
      \mpsWire{-1}{\x}{2}{\x};
      \foreach \t in {0}{\mpsB{\t}{\x}{colMPS}};
      \foreach \t in {1}{\mpsA{\t}{\x}{colMPS}};
      %\mpsBvec{-1}{\x}
      %\mpsBvecW{6}{\x}{colMPS};
    }
    \foreach \x in {2,4,6}{
      \mpsWire{-1}{\x}{2}{\x};
      \foreach \t in {0}{\mpsA{\t}{\x}{colMPS}};
      \foreach \t in {1}{\mpsB{\t}{\x}{colMPS}};
     % \mpsBvec{-1}{\x}
     % \mpsBvecV{6}{\x}{colMPS};
    }
    \draw [decorate,decoration={brace,amplitude=5pt},xshift=0pt,yshift=0pt] 
    ({0.875*\dx},{1.125*\dx})--({6.125*\dx},{1.125*\dx}) node [midway,yshift=10pt]
    {$2n$};
    \foreach \x in {1,3,5}{
        \node at ({\x*\dx},{-2.5*\dx}) {\scalebox{0.9}{$\vartheta_1$}};}
     \foreach \x in {2,4,6}{
         \node at ({\x*\dx},{-2.5*\dx}) {\scalebox{0.9}{$\vartheta_2$}};}
  \end{tikzpicture},\quad
 \ket{\mathcal D_n} \!=\mkern-8mu \begin{tikzpicture}
    [baseline={([yshift=-2.6ex]current bounding box.center)},scale=1.95]
    \foreach \x in {0,...,5}{
    \mpsWire{0}{\x*0.85}{-0.75}{\x*0.85};
  }
    %\vmpsWire{0}{0}{0}{5};
    \mpsBvecW{0}{0*0.85}{colMPS};
    \mpsBvecV{0}{1*0.85}{colMPS};
    \mpsBvecW{0}{2*0.85}{colMPS};
    \mpsBvecV{0}{3*0.85}{colMPS};
    \mpsBvecW{0}{4*0.85}{colMPS};
    \mpsBvecV{0}{5*0.85}{colMPS};
     \draw [decorate,decoration={brace,amplitude=5pt},xshift=0pt,yshift=0pt] 
    ({0.875*\dx-1*\dx},{0.85*\dx})--({5.125*\dx*0.85},{0.8*\dx}) node [midway,yshift=10pt]
    {$2n$};
     \foreach \x in {0,2,4}{
         \node at ({\x*\dx*0.85},{-0.2}) {\scalebox{0.9}{$\vartheta_1$}};}
     \foreach \x in {1,3,5}{
         \node at ({\x*\dx*0.85},{-0.2}) {\scalebox{0.9}{$\vartheta_2$}};}
      \node at ({0*\dx},{0.75}) {};
 \end{tikzpicture},\quad
 \bra{\mathcal U_n} \!=\! \begin{tikzpicture}
    [baseline={([yshift=1.2ex]current bounding box.center)},scale=1.95]
    \foreach \x in {0,...,5}{
      \mpsWire{0}{0.75*\x}{0.75}{0.75*\x};
      \mpsBvec{0}{0.75*\x};
    }
      \draw [decorate,decoration={brace,amplitude=5pt},xshift=0pt,yshift=0pt] 
    ({5.125*0.75*\dx},{-0.8*\dx}) -- ({0.875*\dx-1*\dx},{-0.8*\dx}) node [midway,yshift=-10pt]
    {$2n$};
  \end{tikzpicture}.
\ee
From the representation \eqref{eq:transfermatrep} we see that the asymptotic behaviour of \eqref{eq:goal} is determined by the largest eigenvalue of the transfer matrix $\mathcal T_n$. Therefore, we proceed by identifying its spectrum. To this aim it is convenient to merge together the tensors~$\begin{tikzpicture}[baseline={([yshift=-0.6ex]current bounding box.center)},scale=1.5]
    \mpsWire{-0.5}{0}{0.5}{0};
    \gridLine{0}{-0.5}{0}{0.5};
    \mpsA{0}{0}{colMPS}
  \end{tikzpicture}$,
  $\begin{tikzpicture}
    [baseline={([yshift=-0.6ex]current bounding box.center)},scale=1.5]
    \mpsWire{-0.5}{0}{0.5}{0};
    \gridLine{0}{-0.5}{0}{0.5};
    \mpsB{0}{0}{colMPS}
  \end{tikzpicture}$ on two consecutive rows and columns, i.e. 
\be \label{eq:defM}
M_{a b}^{x y}=\mkern-6mu
\begin{tikzpicture}
    [baseline={([yshift=-0.6ex]current bounding box.center)},scale=1.75]
    \dtensorLine{-1}{0}{1}{0};
    \dtensorLine{0}{-1}{0}{1};
    \tensorM{0}{0}{colITensor};
    \node at ({-1.3*\dx},0) {$a$};
    \node at ({1.3*\dx},0) {$b$};
    \node at (0,{1.3*\dx}) {$x$};
    \node at (0,{-1.3*\dx}) {$y$};
\end{tikzpicture}
\mkern-6mu\equiv\mkern-6mu
\begin{tikzpicture}
    [baseline={([yshift=-0.6ex]current bounding box.center)},scale=1.75]
    \mpsWire{-0.75}{0}{1.75}{0};
    \mpsWire{-0.75}{1}{1.75}{1};
    \gridLine{0}{-0.75}{0}{1.75};
    \gridLine{1}{-0.75}{1}{1.75};
    \mpsB{0}{0}{colMPS}
    \mpsA{0}{1}{colMPS}
    \mpsA{1}{0}{colMPS}
    \mpsB{1}{1}{colMPS}
    \node at ({-1.3*\dx},0) {$r_1$};
    \node at ({2.3*\dx},0) {$s_1$};
    \node at ({-1.3*\dx},{-\dx}) {$r_2$};
    \node at ({2.3*\dx},{-\dx}) {$s_2$};
    \node at (0,{1.3*\dx}) {$z_1$};
    \node at ({\dx},{1.3*\dx}) {$z_2$};
    \node at (0,{-2.3*\dx}) {$w_1$};
    \node at ({\dx},{-2.3*\dx}) {$w_2$};
\end{tikzpicture},\qquad
\begin{aligned}
    x&=3 z_1 + z_2,& 
    y&=3 w_1 + w_2,& x,y&\in\mathbb{Z}_9,\\
    a&=2 r_1 + r_2,&
    b&=2 s_1 + s_2,& a,b&\in\mathbb{Z}_4,
\end{aligned}
\ee
so that $\mathcal{T}_n$ is rewritten in terms of $n$ horizontally connected tensors~$M$ with
periodic boundaries
\be
\mathcal{T}_n =
  \begin{tikzpicture}
    [baseline={([yshift=-2.6ex]current bounding box.center)},scale=1.75]
    \TleftHook{0}{-1};
    \TrightHook{0}{4};
    \dtensorLine{0}{-1}{0}{1.75};
     \dtensorLine{0}{2.25}{0}{4}
     \dtensordottedLine{0}{1.75}{0}{2.25}
    \dtensorLine{1}{0}{-1}{0};
    %\dtensorLine{1}{1}{-1}{1};
    \dtensorLine{1}{1}{-1}{1};
    \dtensorLine{1}{3}{-1}{3};
    \tensorM{0}{0}{colITensor};
    \tensorM{0}{1}{colITensor};
    %\tensorM{0}{1}{colITensor};
    \tensorM{0}{3}{colITensor};
  \draw [decorate,decoration={brace,amplitude=5pt},xshift=0pt,yshift=0pt] 
    ({-0.125*\dx},{\dx})--({3.125*\dx},{\dx}) node [midway,yshift=10pt]
    {$n$};
  \end{tikzpicture}.
  \label{eq:Tn}
\ee
Moreover, we also make a convenient local basis transformation
\be \label{eq:defMt}
\tilde{M}=
\begin{tikzpicture}
    [baseline={([yshift=-0.6ex]current bounding box.center)},scale=1.75]
    \dtensorLine{-1}{0}{1}{0};
    \dtensorLine{0}{-1}{0}{1};
    \tensorM{0}{0}{colTensor};
\end{tikzpicture}=
\begin{tikzpicture}
    [baseline={([yshift=-0.6ex]current bounding box.center)},scale=1.75]
    \dtensorLine{-1.75}{0}{1.75}{0};
    \dtensorLine{0}{-1}{0}{1};
    \tensorM{0}{0}{colITensor};
    \bCircle{1}{0}{white};
    \bCircle{-1}{0}{white};
    %\node at (0,{\dx}) {\scalebox{0.8}{$P$}};
    \node at (0,{\dx}) {\scalebox{0.8}{$-$}};
\end{tikzpicture},\qquad
\begin{tikzpicture}
    [baseline={([yshift=-0.6ex]current bounding box.center)},scale=1.75]
    \dtensorLine{0.25}{0}{1.75}{0};
    \bCircle{1}{0}{white};
\end{tikzpicture}=P,\qquad
\begin{tikzpicture}
    [baseline={([yshift=-0.6ex]current bounding box.center)},scale=1.75]
    \dtensorLine{0.25}{0}{1.75}{0};
    \bCircle{1}{0}{white};
    \node at (0,{-\dx}) {\scalebox{0.8}{$-$}};
\end{tikzpicture}=P^{-1},
\ee
where we defined 
\be
P=\!\begin{bmatrix}
  1 & & & & & & & & \\
    &1& 1& & & & & & \\
    &1&\displaystyle-\frac{\vartheta_2}{1-\vartheta_2}& & & & & & \\
    & &  &1 & & &1& & \\
    & &  &  &1&1& &1& 1\\
    & &  &  &1&\displaystyle-\frac{\vartheta_2}{1-\vartheta_2}& &1&\displaystyle-\frac{\vartheta_2}{1-\vartheta_2}\\
    & &  &1 & & &\displaystyle-\frac{\vartheta_1}{1-\vartheta_1}& & \\
    & &  &  &\displaystyle-\frac{\vartheta_1}{1-\vartheta_1} &\displaystyle-\frac{\vartheta_1}{1-\vartheta_1}& &1&1\\
    & &  &  &\displaystyle-\frac{\vartheta_1}{1-\vartheta_1}
    &\displaystyle\frac{\vartheta_1 \vartheta_2}{(1-\vartheta_1)(1-\vartheta_2)}&
    &1&\displaystyle-\frac{\vartheta_2}{1-\vartheta_2} \\
\end{bmatrix}\!.
\ee
We denote by $\tilde{\mathcal{T}}_n$ the transfer matrix in the new basis (defined as \eqref{eq:Tn}, with~$M$ replaced by~$\tilde{M}$).

Since $\tilde{\mathcal{T}}_n$ and $\mathcal{T}_n$ are related by a similarity transformation, their spectra coincide. To determine them we make use of the following Lemma (proven in Appendix~\ref{sec:prooflemma1}).
\begin{lemma}
\label{lemma1}
For any~$k\ge 1$
\be\label{eq:lemma1eq}
  \begin{tikzpicture}
    [baseline={([yshift=-0.6ex]current bounding box.center)},scale=1.75]
    \dtensorLine{-0.75}{0}{4.75}{0};
    \dtensorLine{6.25}{0}{9.75}{0};
    \dtensorLine{-0.75}{1}{4.75}{1};
    \dtensorLine{6.25}{1}{9.75}{1};
    \dtensorLine{0}{-1}{0}{2};
    \dtensorLine{2}{-1}{2}{2};
    \dtensorLine{4}{-1}{4}{2};
    \dtensorLine{7}{-1}{7}{2};
    \dtensorLine{9}{-1}{9}{2};
    \tensorM{0}{0}{colTensor};
    \tensorM{0}{1}{colTensor};
    \tensorM{2}{0}{colTensor};
    \tensorM{2}{1}{colTensor};
    \tensorM{4}{0}{colTensor};
    \tensorM{4}{1}{colTensor};
    \tensorM{7}{0}{colTensor};
    \tensorM{7}{1}{colTensor};
    \tensorM{9}{0}{colTensor};
    \tensorM{9}{1}{colTensor};
    \node[draw=white,fill=white,inner sep=1pt] at (0,{\dx}) {$x_1$};
    \node[draw=white,fill=white,inner sep=1pt] at (0,{-\dx}) {$x_2$};
    \node[draw=white,fill=white,inner sep=1pt] at (0,{-3*\dx}) {$x_3$};
    \node at (0,{-5.25*\dx}) {$\vdots$};
    \node[draw=white,fill=white,inner sep=1pt] at (0,{-8*\dx}) {$x_k$};
    \node[draw=white,fill=white,inner sep=1pt] at (0,{-10*\dx}) {$x_1$};
    \node[draw=white,fill=white,inner sep=1pt] at ({\dx},{\dx}) {$y_1$};
    \node[draw=white,fill=white,inner sep=1pt] at ({\dx},{-\dx}) {$y_2$};
    \node[draw=white,fill=white,inner sep=1pt] at ({\dx},{-3*\dx}) {$y_3$};
    \node at ({\dx},{-5.25*\dx}) {$\vdots$};
    \node[draw=white,fill=white,inner sep=1pt] at ({\dx},{-8*\dx}) {$y_k$};
    \node[draw=white,fill=white,inner sep=1pt] at ({\dx},{-10*\dx}) {$y_1$};
  \end{tikzpicture}=
  \prod_{j=1}^k\delta_{y_k,x_k}
  \begin{tikzpicture}
    [baseline={([yshift=-0.6ex]current bounding box.center)},scale=1.75]
    \dtensorLine{-0.75}{0}{4.75}{0};
    \dtensorLine{6.25}{0}{9.75}{0};
    \dtensorLine{-0.75}{1}{4.75}{1};
    \dtensorLine{6.25}{1}{9.75}{1};
    \dtensorLine{0}{-1}{0}{2};
    \dtensorLine{2}{-1}{2}{2};
    \dtensorLine{4}{-1}{4}{2};
    \dtensorLine{7}{-1}{7}{2};
    \dtensorLine{9}{-1}{9}{2};
    \tensorM{0}{0}{colTensor};
    \tensorM{0}{1}{colTensor};
    \tensorM{2}{0}{colTensor};
    \tensorM{2}{1}{colTensor};
    \tensorM{4}{0}{colTensor};
    \tensorM{4}{1}{colTensor};
    \tensorM{7}{0}{colTensor};
    \tensorM{7}{1}{colTensor};
    \tensorM{9}{0}{colTensor};
    \tensorM{9}{1}{colTensor};
    \node[draw=white,fill=white,inner sep=1pt] at (0,{\dx}) {$x_1$};
    \node[draw=white,fill=white,inner sep=1pt] at (0,{-\dx}) {$x_2$};
    \node[draw=white,fill=white,inner sep=1pt] at (0,{-3*\dx}) {$x_3$};
    \node at (0,{-5.25*\dx}) {$\vdots$};
    \node[draw=white,fill=white,inner sep=1pt] at (0,{-8*\dx}) {$x_k$};
    \node[draw=white,fill=white,inner sep=1pt] at (0,{-10*\dx}) {$x_1$};
    \node[draw=white,fill=white,inner sep=1pt] at ({\dx},{\dx}) {$x_1$};
    \node[draw=white,fill=white,inner sep=1pt] at ({\dx},{-\dx}) {$x_2$};
    \node[draw=white,fill=white,inner sep=1pt] at ({\dx},{-3*\dx}) {$x_3$};
    \node at ({\dx},{-5.25*\dx}) {$\vdots$};
    \node[draw=white,fill=white,inner sep=1pt] at ({\dx},{-8*\dx}) {$x_k$};
    \node[draw=white,fill=white,inner sep=1pt] at ({\dx},{-10*\dx}) {$x_1$};
  \end{tikzpicture}.
  \ee
\end{lemma}
This lemma has the remarkable consequence that traces of powers of~${\mathcal{T}}_n$ can be obtained by considering a simple~$9\times9$ matrix. Namely we have 
\be\label{eq:tracePowReduced}
{\rm tr}\big(\mathcal{T}_n^k\big)=
{\rm tr}\big(\tilde{\mathcal{T}}_n^k\big)=
\sum_{x_1,x_2,\ldots,x_k}
  \begin{tikzpicture}
    [baseline={([yshift=-2.4ex]current bounding box.center)},scale=1.75]
      \draw [decorate,decoration={brace,amplitude=5pt},xshift=0pt,yshift=0pt] 
    ({-0.25*\dx},{1.25*\dx})--({5.75*\dx},{1.25*\dx}) node [midway,yshift=10pt]
    {$n$};
    \foreach \t in {0,2,4,7,9}{
      \TleftHook{\t}{-0.75};
      \TrightHook{\t}{6.25};
      \dtensorLine{\t}{-0.75}{\t}{6.25};
      \node[draw=white,fill=white,inner sep=0pt] at ({4.25*\dx},{-\t*\dx}) {$\,\cdots$};
    }
    \foreach \x in {0,1,2,3,5.5}{
      \dtensorLine{-0.75}{\x}{4.75}{\x};
      \dtensorLine{6.25}{\x}{9.75}{\x};
    }
    \foreach \t in {0,2,4,7,9}{
      \foreach \x in {0,1,2,3,5.5}{
        \tensorM{\t}{\x}{colTensor};
      }
    }
    \foreach \x in {0,1,2,3,5.5}{
      \node[draw=white,fill=white,inner sep=1pt] at ({\x*\dx},{\dx}) {$x_1$};
      \node[draw=white,fill=white,inner sep=1pt] at ({\x*\dx},{-\dx}) {$x_2$};
      \node[draw=white,fill=white,inner sep=1pt] at ({\x*\dx},{-3*\dx}) {$x_3$};
      \node at ({\x*\dx},{-5.25*\dx}) {$\vdots$};
      \node[draw=white,fill=white,inner sep=1pt] at ({\x*\dx},{-8*\dx}) {$x_k$};
      \node[draw=white,fill=white,inner sep=1pt] at ({\x*\dx},{-10*\dx}) {$x_1$};
    }
\end{tikzpicture}=
\begin{tikzpicture}
  [baseline={([yshift=-0.6ex]current bounding box.center)},scale=1.75]
  \draw [decorate,decoration={brace,amplitude=5pt},xshift=0pt,yshift=0pt] 
    ({1*\dx},{0.25*\dx})--({1*\dx},{-9.25*\dx}) node [midway,xshift=10pt]
    {$k$};
  \TtopHook{-0.75}{0};
  \TbottomHook{9.75}{0};
  \dtensorLine{-0.75}{0}{4.75}{0};
  \dtensorLine{6.25}{0}{9.75}{0};
  \foreach \t in {0,2,4,7,9}{
    \TleftHook{\t}{-0.75};
    \TrightHook{\t}{0.75};
    \dtensorLine{\t}{-0.75}{\t}{0.75};
    \tensorM{\t}{0}{colTensor};
    \node at (0,{-\t*\dx}) {\scalebox{0.8}{$n$}};
  }
  \node at ({0},{-5.25*\dx}) {$\vdots$};
\end{tikzpicture} = {\rm tr} \big(\tilde{\tau}_n^k\big),
\ee
where we introduced the tensor 
\be
\begin{tikzpicture}
  [baseline={([yshift=-0.6ex]current bounding box.center)},scale=1.75]
  \dtensorLine{0}{-1}{0}{1};
  \dtensorLine{-1}{0}{1}{0};
  \tensorM{0}{0}{colTensor};
  \node at (0,0) {\scalebox{0.8}{$n$}};
  \node[draw=white,fill=white,inner sep=1pt] at (0,{1.25*\dx}) {$x$};
  \node[draw=white,fill=white,inner sep=1pt] at (0,{-1.25*\dx}) {$y$};
    \node[draw=white,fill=white,inner sep=1pt] at ({1.25*\dx},0) {$a$};
  \node[draw=white,fill=white,inner sep=1pt] at ({-1.25*\dx},0) {$b$};
\end{tikzpicture}=\big(\tilde{M}^{x y}_{ab}\big)^n.
\ee
Explicitly, the matrix elements of~$\tilde{\tau}_n$ are expressed as
\be
\big[\tilde{\tau}_n\big]_{x_1,x_2}=
\sum_{a_1,\ldots,a_n=1}^4 \tilde{M}^{x_1 x_2}_{a_1a_2}\cdots  \tilde{M}^{x_1 x_2}_{a_n a_1},
\ee
which yields 
\be
\mkern-4mu\tilde{\tau}_n\mkern-4mu=\mkern-6mu
\begin{bmatrix}
  (1-\vartheta_1)^n (1-\vartheta_2)^n& \mkern-26mu 0 & 0 & 0 & 0 & 1 & (1-\vartheta_2)^n & 0 & 1 \\
  \vartheta_2^n(1-\vartheta_1)^n & \mkern-26mu 0 & 0 & 0 & 0 & 0 & \vartheta_2^n & 0 & 0 \\
  0 & \mkern-26mu 0 & 0 & 0 & 1 & 0 & 0 & 1 & 0 \\
  0 & \mkern-26mu \vartheta_1^n & \vartheta_1^n & 0 & 0 & 0 &  0 & 0 & 0 \\
  0 & \mkern-26mu 0 & 0 & \vartheta_2^n & 0 & 0 &  0 & 0 & 0 \\
  0 & \mkern-26mu 0 & 0 & (1-\vartheta_2)^n & 0 & 0 &  0 & 0 & 0 \\
  0 & \mkern-26mu (1-\vartheta_1)^n & (1-\vartheta_1)^n & 0 & 0 & 0 &  0 & 0 & 0 \\
  \vartheta_1^n \vartheta_2^n & \mkern-26mu 0 & 0 & 0 & 0 & 0 &  0 & 0 & 0 \\
  \vartheta_1^n (1-\vartheta_2)^n & \mkern-26mu 0 & 0 & 0 & 0 & 0 &  0 & 0 & 0 \\
\end{bmatrix}\!.
\ee
Since the relation~\eqref{eq:tracePowReduced} holds for any~$k$, the non-zero
eigenvalues of~$\mathcal{T}_n$ and~$\tilde{\tau}_n$ coincide. The eigenvalues
of the latter are easily obtained. In particular, it is straightforward to see
that the only three non-zero eigenvalues of $\tilde{\tau}_n$ are the solutions
to the following cubic equation
\be
\lambda^3 = 
\left((1-\vartheta_1)^n \lambda+\vartheta_1^n\right)
\left((1-\vartheta_2)^n\lambda+ \vartheta_2^n\right).
\label{eq:cubicEq}
\ee
The main properties of this equation are studied in Appendix~\ref{sec:propertiescubiceq} and can be summarised as follows 
\begin{lemma}
\label{lemma2} 
For $\vartheta_1,\vartheta_2\in(0,1)$ the solution of Eq.~\eqref{eq:cubicEq} with strictly larger magnitude is real and positive. Its explicit expression reads as  
\be
\mkern-3mu
\lambda_n(\vartheta_1, \vartheta_2)
    \mkern-4mu=\mkern-4mu\frac{(1-\vartheta_1)^n(1-\vartheta_2)^n}{3}
    \mkern-2mu+\mkern-2mu
    \sqrt[3]{\Delta_{n,1} 
    \mkern-2mu+\mkern-2mu\sqrt{\Delta_{n,1}^2-\Delta_{n,2}^3}}+\frac{\Delta_{n,2}}{ \sqrt[3]{\Delta_{n,1} \!+\!\sqrt{\Delta_{n,1}^2-\Delta_{n,2}^3}}},
\label{eq:lambdan}
\ee
where 
\be
\begin{aligned}
    \Delta_{n,1} &= \frac{1}{6} (1-\vartheta_1)^{2n} (1-\vartheta_2)^{2n}
    \sum_{j=1}^2 \left(\frac{\vartheta_j}{1-\vartheta_j}\right)^n
    \!+\frac{1}{2} \vartheta_1^n \vartheta_2^n
    \!+ \frac{1}{27}(1-\vartheta_1)^{3n}  (1-\vartheta_2)^{3 n}\,,\\
    \Delta_{n,2} &= \frac{1}{3} \vartheta_1^n (1-\vartheta_2)^n
    +\frac{1}{3} \vartheta_2^n (1-\vartheta_1)^n +\frac{1}{9}(1-\vartheta_1)^{2n}(1-\vartheta_2)^{2n}\,.
\end{aligned}
\ee
\end{lemma}
Putting all together and noting that $\tensor*[_n]{\!\braket*{L_{\vartheta_1}}{R_{\vartheta_2}}}{_n}=1$ (cf.~\eqref{eq:LRoverlap}) we find that for $t\gg1$, the building block \eqref{eq:goal} displays a linear growth with slope given by 
\be
r_n(\vartheta_1, \vartheta_2) \coloneqq \lim_{t\to\infty} \frac{b_n(\vartheta_1, \vartheta_2)}{t}=  \frac{1}{1-n}\log[\lambda_n(\vartheta_1, \vartheta_2)].
\label{eq:result}
\ee
For example, in the case of the min-entropy (i.e.\ $n\to\infty$) we find the following explicit result  
\be
r_\infty(\vartheta_1, \vartheta_2)=\begin{cases}
-\log[(1-\vartheta_1)  (1-\vartheta_2)],
    \qquad &
    \begin{aligned}
        \tfrac{\vartheta_1}{1-\vartheta_1}&\leq (1-\vartheta_1)  (1-\vartheta_2),\\
        \tfrac{\vartheta_2}{1-\vartheta_2}&\leq (1-\vartheta_1)  (1-\vartheta_2),
    \end{aligned}\\[2em]
-\log[\vartheta_1^{1/2}(1-\vartheta_2)^{1/2}],
    & \begin{aligned}
        \tfrac{\vartheta_1}{1-\vartheta_1} &> \vartheta_1^{1/2}(1-\vartheta_2)^{1/2},\\
        \tfrac{\vartheta_2}{1-\vartheta_2} &\leq \vartheta_1^{1/2}(1-\vartheta_2)^{1/2},
    \end{aligned}\\[2em]
-\log[\vartheta_2^{1/2}(1-\vartheta_1)^{1/2}],
    & \begin{aligned}
        \tfrac{\vartheta_2}{1-\vartheta_2} &> \vartheta_2^{1/2}(1-\vartheta_1)^{1/2},\\
        \tfrac{\vartheta_1}{1-\vartheta_1} &\leq \vartheta_2^{1/2}(1-\vartheta_1)^{1/2},
    \end{aligned}\\[2em]
-\log[\vartheta_1^{1/3}  \vartheta_2^{1/3}],
    & \begin{aligned}
        \tfrac{\vartheta_1}{1-\vartheta_1}&\geq \vartheta_1^{1/3}  \vartheta_2^{1/3}, \\
        \tfrac{\vartheta_2}{1-\vartheta_2}&\geq \vartheta_1^{1/3}  \vartheta_2^{1/3}.
    \end{aligned}
\end{cases}
\ee
Note that the above characterisation of the spectrum of $\mathcal{T}_n$ can also be used to find the leading corrections to \eqref{eq:result}. To do that, however, one would also need to find the eigenvector associated to $\lambda_n(\vartheta_1, \vartheta_2)$.

\begin{figure}
    \centering
    \includegraphics[width=0.75\textwidth]{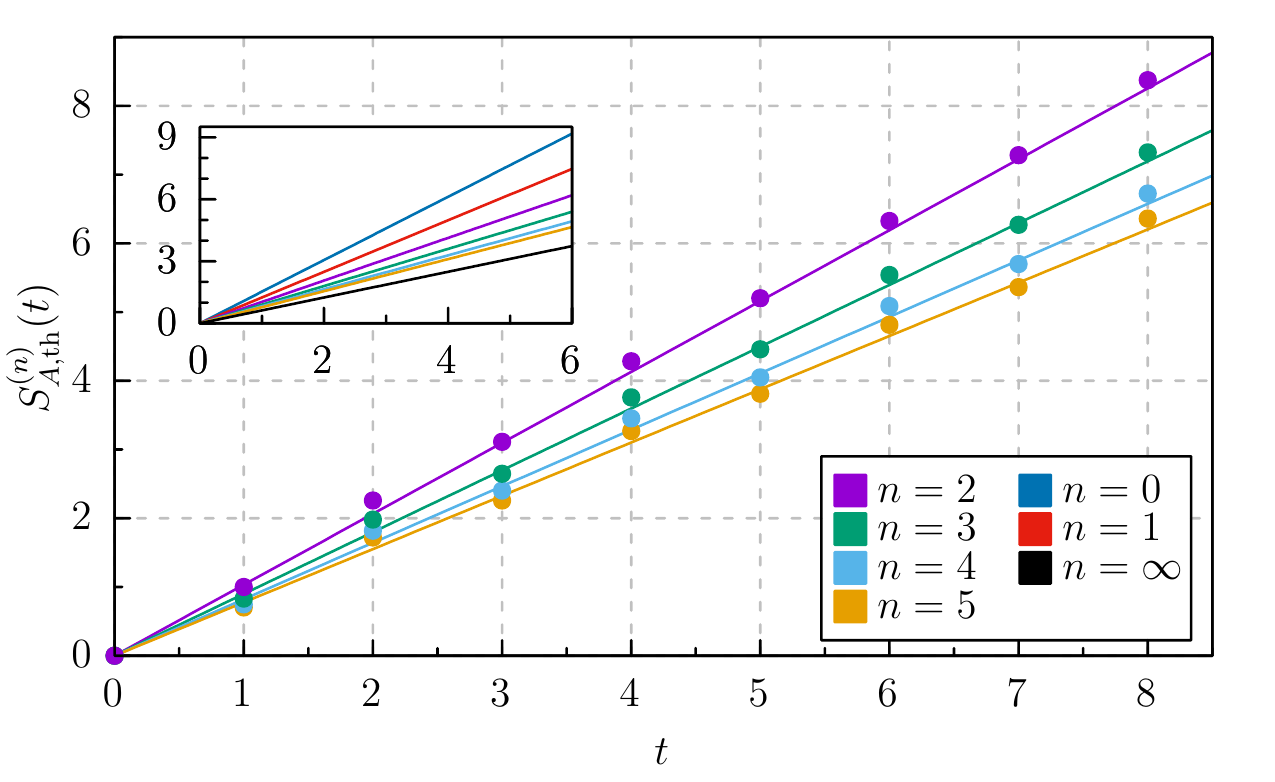}
    \caption{\label{fig:renyiExample1} Growth of the R\'enyi
    $n$-entropy after a homogeneous quench from the state characterised by
    $(\vartheta_{\rm L},\vartheta_{\rm R})=(0.65,0.15)$, for typical values of $n$. Solid lines are the asymptotic prediction obtained by
    neglecting the subleading corrections and taking into account only the
    largest eigenvalue of the tensor network~\eqref{eq:transfermatrep}, while
    the dots correspond to exact finite-time values in the $L\to\infty$ limit.
    }
\end{figure}
Substituting \eqref{eq:result} in \eqref{eq:SAnthhalfsystembipa} we arrive at the main result of this paper 
\be
\smashoperator{\lim_{\substack{t\to\infty \\[0.25em] {|A|}/{t}=\zeta \geq 4}}}
    \frac{S^{(n)}_{A,\rm th}(t)}{t} =
    r_n(\vartheta_{\rm L},\vartheta_{\rm R})+
    r_n(\vartheta_{\rm R},
\vartheta_{\rm R}).
\label{eq:mainresult}
\ee
    This equation provides a rigorous proof of the fact that R\'enyi entropies of large subsystems grow linearly in the asymptotic regime and gives an exact expression for their slope. A comparison between the asymptotic result~\eqref{eq:mainresult} and the exact numerical evaluation of~\eqref{eq:SAnthhalfsystembipa} for finite times is reported in Fig.~\ref{fig:renyiExample1}.
    
\begin{figure}
    \centering
    \includegraphics[width=0.75\textwidth]{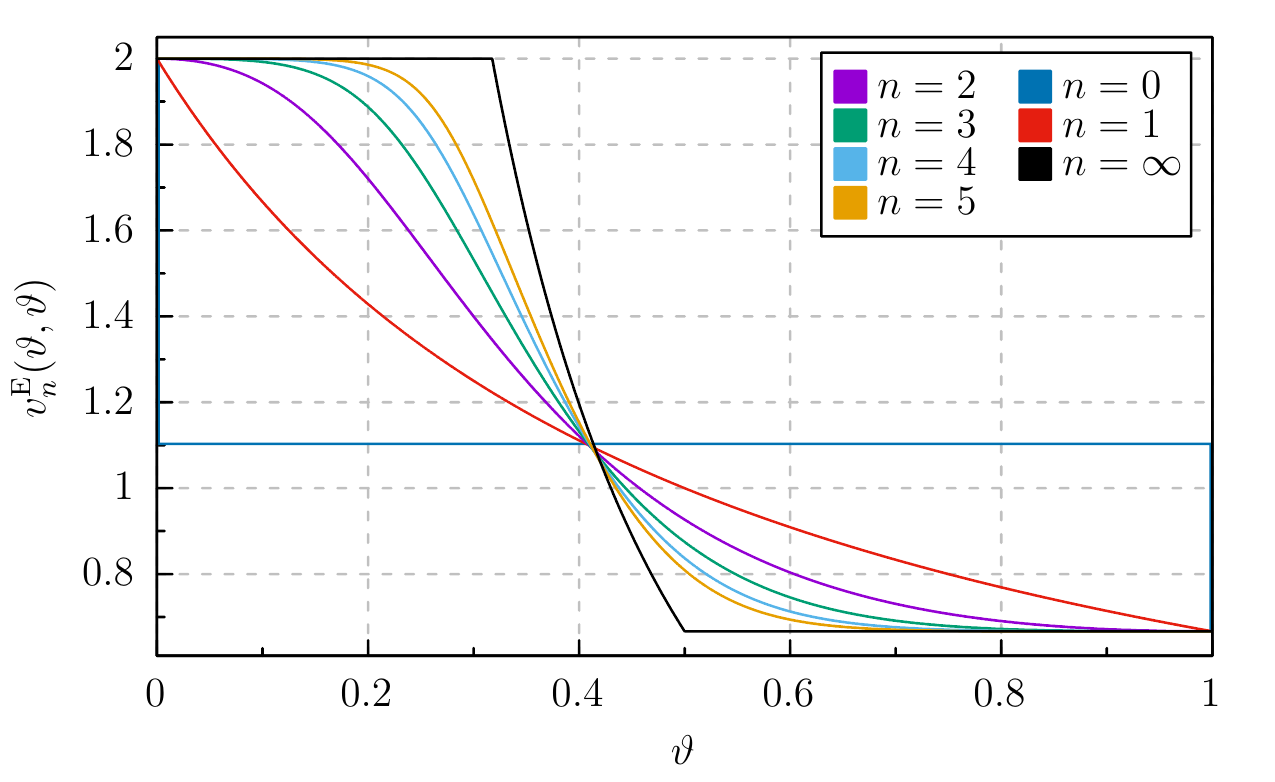}
    \caption{Entanglement velocity (cf. \eqref{eq:entanglementvelocity}) as a function of the filling $\vartheta$ (note
    that here we are considering a homogeneous case $\vartheta_{\rm
    L}=\vartheta_{\rm R}=\vartheta$), for different R\'enyi indices $n$.
    }
    \label{fig:renyiExample2}
\end{figure}

    We recall that \eqref{eq:mainresult} applies to the case of a bipartitioning protocol with two leads initially prepared in different solvable states~\eqref{eq:solvablefamily} and where the subsystem $A$ starts at the junction. The special case $\vartheta_{\rm L}= \vartheta_{\rm R}=\vartheta$ describes the growth of entanglement after a homogeneous quench from a solvable state. A remarkable consequence of \eqref{eq:mainresult} is that the entanglement velocity 
\be
v^{\rm E}_{n}(\vartheta_{\rm L},\vartheta_{\rm R}) \coloneqq \smashoperator{\lim_{\substack{t\to\infty \\[0.25em] {|A|}/{t}=\zeta \geq 4}}}
    \quad\frac{S^{(n)}_{A,\rm th}(t)}{t s^{(n)}_{\rm GGE}}=\frac{r_n(\vartheta_{\rm L},\vartheta_{\rm R})}{ s^{(n)}_{\rm GGE}}+\frac{r_n(\vartheta_{\rm R},\vartheta_{\rm R})}{ s^{(n)}_{\rm GGE}}\,, 
\label{eq:entanglementvelocity}
\ee
where $s^{(n)}_{\rm GGE}$ is the entropy density of the GGE (cf.~\eqref{eq:formalStationaryEntropy}), depends non-trivially on $n$. See Fig.~\ref{fig:renyiExample2} for a representative example.
 
The result \eqref{eq:result} can again be analytically continued to 
$\mathcal D=\{z\in\mathbb C:\,\, {\rm Re}[z]>0\}$. Indeed, the function
$\lambda_z(\vartheta_1, \vartheta_2)$ --- obtained by replacing $n$ in
\eqref{eq:lambdan} with $z\in\mathbb C$ --- is holomorphic and bounded in
$\mathcal D$. Specifically (see Appendix~\ref{sec:propertiescubiceq})
\be
|\lambda_z(\vartheta_1, \vartheta_2)|\leq \lambda_{{\rm Re}[z]}(\vartheta_1, \vartheta_2) < 3\,. 
\label{eq:inequalitylambda}
\ee
Applying again Carlson's Theorem~\cite{rubel1956necessary} we then have that $\lambda_z(\vartheta_{1}, \vartheta_{2})$ is the only analytic continuation of $\{\lambda_n(\vartheta_{1},\vartheta_{2})\}_{n=1,2,3,\dots}$
which fulfils the physically sensible bounds  
\be
\begin{aligned}
    | \lambda_z (\vartheta_{1}, \vartheta_{2})|&\leq C e^{\tau |z|},& z&\in\mathcal D,&\qquad
    | \lambda_{1+iy}(\vartheta_{1}, \vartheta_{2})|&\leq C e^{ c |y|},& y&\in \mathbb R\,,
\end{aligned}
\ee
with $C,\tau\in\mathbb R$ and $c<\pi$.  

Considering now $z=1+\delta$ with $\delta\ll1$ from \eqref{eq:lambdan} we find  
\be
\lambda_{1+\delta}(\vartheta_1, \vartheta_2) =
1+\frac{\delta}{1+\vartheta_1+\vartheta_2}\sum_{j=1}^2\big(\vartheta_j \log \vartheta_j
+(1-\vartheta_j) \log (1-\vartheta_j)\big)
+\mathcal O(\delta^2),
\ee
which gives the following result for the slope of the von Neumann entropy
\be
 r(\vartheta_1, \vartheta_2) \coloneqq\! \lim_{z\to1}  r_z(\vartheta_1, \vartheta_2)\! = -\frac{1}{1+\vartheta_1+\vartheta_2}
 \sum_{j=1}^2 \big(\vartheta_j \log \vartheta_j + (1-\vartheta_j)\log(1-\vartheta_j) \big)
 ,
\ee
or, equivalently
\be
\smashoperator{\lim_{\substack{t\to\infty\\[0.25em]{|A|}/{t}=\zeta \geq 4}}}
\frac{S_{A,\rm th}(t)}{t} =r(\vartheta_{\rm L},\vartheta_{\rm R})
+r(\vartheta_{\rm R},\vartheta_{\rm R}).
\ee

\section{The quasiparticle picture}
\label{sec:qp}

In the famous work~\cite{calabrese2005evolution}, Calabrese and Cardy
proposed a simple picture that explains the growth of entanglement in terms of
correlated quasiparticles created by the quench. In the simplest formulation
one imagines that at $t=0$ the quench produces pairs of quasiparticles at every
point in space and for $t>0$ they begin to propagate with opposite velocities $\pm v$.
Quasiparticles forming each pair are \emph{correlated} or \emph{entangled}, while those in
different pairs are uncorrelated. Then, one postulates that, for any time $t$,
the entanglement between a given subsystem $A$ and its complement $\bar A$ is
proportional to the number of correlated pairs shared between $A$ and $\bar A$. 

Considering a \emph{homogeneous} quench this picture gives the following
expression for the Von Neumann entropy 
\be
S_{A,\rm th} = \min(4vt,2|A|) s\,,
\label{eq:quasiparticleS}
\ee
where by $s$ we denoted the contribution to the entanglement of a pair multiplied
by the density of pairs. This expression can be immediately generalised to the case of
$N_s$ different species of quasiparticles with a dispersion relation
parametrised by $\lambda\in [-\Lambda,\Lambda]$ 
\be
    S_{A,\rm th} = \sum_{n=1}^{N_s} \int_{-\Lambda}^{\Lambda} {\rm d}\lambda 
    \min((v_{n,\lambda}-v_{n,-\lambda})t,|A|) s_{n,\lambda}\,.
\label{eq:quasiparticlemultiS}
\ee
Here we took correlated pairs formed by particles with the same $n$ and opposite
$\lambda$s, $v_{n,\lambda}$ is the velocity of the quasiparticle labelled by
$(n,\lambda)$ and $s_{n,\lambda}$ the contribution of the pairs labelled by
$(n,\lambda)$ to the density of entanglement entropy. 

This picture can be generalised to describe inhomogeneous quenches by allowing
the contribution to the entanglement of a given pair to depend on the emission
point~\cite{bertini2018entanglement} and the quasiparticles to have a curved
trajectory~\cite{alba2019entanglement}, namely
\be
\label{eq:quasiparticlemultiSinhomo}
S_{A,\rm th} 
= \sum_{n=1}^{N_s} \int_{-\Lambda}^{\Lambda} \!\!\!{\rm d}\lambda\!\!\int \!\!{\rm d}x\,\,\,
\chi_A(X_{n,\lambda}(t,x))(1-\chi_A(X_{n,-\lambda}(t,x)))s_{n,\lambda}(x)\,,
\ee
where $X_{n,\lambda}(t,x)$ is the position at time $t$ of the quasiparticle
$(n,\lambda)$ emitted in $x$ at time~$0$. Additional refinements accounting for
initial states producing $n$-plets of correlated excitations~\cite{bertini2018entanglementand, bastianello2018spreading}, and non-unitary non-interacting dynamics~\cite{cao2019entanglement, alba2021spreading, carollo2021emergent} have also been
developed. 

Interestingly, due to the simplicity of Rule 54 we can explicitly show that our solvable initial states~\eqref{eq:solvablefamily} consist precisely of pairs of oppositely-moving quasiparticles. This can be seen by expressing them in the computational basis
\be \label{eq:initialStateCompBasis}
\ket{\Psi_{\vartheta,\vec{\varphi}}}=
\sum_{s_1,s_3,\ldots,s_L}
\left(\mathrm{e}^{\mathrm{i}\varphi_1}\sqrt{1-\vartheta}\right)^L
\left(\mathrm{e}^{\mathrm{i}\varphi_2}\sqrt{\frac{\vartheta}{1-\vartheta}}
\right)^{s_1+s_2+\ldots+s_L}
%\sum_{j=1}^L s_j}
\ket{0s_10s_20s_3\cdots 0 s_L}.
\ee
Each of the basis states that enter the above sum at position $2j$ is either $\ket{0}$ or $\ket{1}$. This means that the local configuration around $2j$ is either $\ket{000}$ or $\ket{010}$. We now claim that the first option implies no quasiparticle at position $2j$, while the second one corresponds to a pair of oppositely-moving quasiparticles that have temporarily merged into one site. 

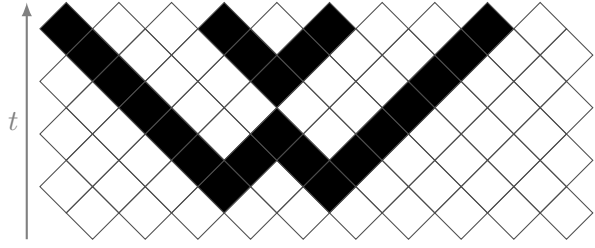
\begin{figure}
    \center
    \begin{tikzpicture}[scale=0.35]
    \draw[-latex,thick,gray] ({-3.5},{-1}) -- ({-3.5},{8}) node [midway,xshift=-5] {$t$};
    \rectangle{-2}{1}{0};
    \rectangle{-1}{0}{0};
    \rectangle{0}{1}{0};
    \rectangle{1}{0}{0};
    \rectangle{2}{1}{0};
    \rectangle{3}{0}{0};
    \rectangle{4}{1}{1};
    \rectangle{5}{0}{0};
    \rectangle{6}{1}{0};
    \rectangle{7}{0}{0};
    \rectangle{8}{1}{1};
    \rectangle{9}{0}{0};
    \rectangle{10}{1}{0};
    \rectangle{11}{0}{0};
    \rectangle{12}{1}{0};
    \rectangle{13}{0}{0};
    \rectangle{14}{1}{0};
    \rectangle{15}{0}{0};
    \rectangle{16}{1}{0};
    \rectangle{17}{0}{0};

    \rectangle{-2}{3}{0};
    \rectangle{-1}{2}{0};
    \rectangle{0}{3}{0};
    \rectangle{1}{2}{0};
    \rectangle{2}{3}{1};
    \rectangle{3}{2}{1};
    \rectangle{4}{3}{0};
    \rectangle{5}{2}{1};
    \rectangle{6}{3}{1};
    \rectangle{7}{2}{1};
    \rectangle{8}{3}{0};
    \rectangle{9}{2}{1};
    \rectangle{10}{3}{1};
    \rectangle{11}{2}{0};
    \rectangle{12}{3}{0};
    \rectangle{13}{2}{0};
    \rectangle{14}{3}{0};
    \rectangle{15}{2}{0};
    \rectangle{16}{3}{0};
    \rectangle{17}{2}{0};

    \rectangle{-2}{5}{0};
    \rectangle{-1}{4}{0};
    \rectangle{0}{5}{1};
    \rectangle{1}{4}{1};
    \rectangle{2}{5}{0};
    \rectangle{3}{4}{0};
    \rectangle{4}{5}{0};
    \rectangle{5}{4}{0};
    \rectangle{6}{5}{1};
    \rectangle{7}{4}{0};
    \rectangle{8}{5}{0};
    \rectangle{9}{4}{0};
    \rectangle{10}{5}{0};
    \rectangle{11}{4}{1};
    \rectangle{12}{5}{1};
    \rectangle{13}{4}{0};
    \rectangle{14}{5}{0};
    \rectangle{15}{4}{0};
    \rectangle{16}{5}{0};
    \rectangle{17}{4}{0};

    \rectangle{-2}{7}{1};
    \rectangle{-1}{6}{1};
    \rectangle{0}{7}{0};
    \rectangle{1}{6}{0};
    \rectangle{2}{7}{0};
    \rectangle{3}{6}{0};
    \rectangle{4}{7}{1};
    \rectangle{5}{6}{1};
    \rectangle{6}{7}{0};
    \rectangle{7}{6}{1};
    \rectangle{8}{7}{1};
    \rectangle{9}{6}{0};
    \rectangle{10}{7}{0};
    \rectangle{11}{6}{0};
    \rectangle{12}{7}{0};
    \rectangle{13}{6}{1};
    \rectangle{14}{7}{1};
    \rectangle{15}{6}{0};
    \rectangle{16}{7}{0};
    \rectangle{17}{6}{0};
\end{tikzpicture}
    \caption{\label{fig:pairstructure}
    Time-evolution of $\ket{0}^{\otimes 5} \otimes \ket{010} \otimes \ket{0} \otimes \ket{010} \otimes \ket{0}^{\otimes 6}$. 
    Up to the even-odd staggering, the horizontal coordinate of a diamond corresponds to the physical site, the vertical position indicates the time-step (increasing upwards). Black diamonds correspond to $\ket{1}$s and white ones to $\ket{0}$s.}
\end{figure}

This can be appreciated by noting that the time-evolution operator $\mathbb{U}$ is deterministic in the computational basis, therefore each basis state is mapped into exactly one other. In the sequence of bit-strings representing basis states, the freely-moving quasiparticles are given by pairs of consecutive $\ket{1}$ on top of the background of $\ket{0}$. This is conveniently depicted by introducing a staggered zig-zag lattice where even sites are displaced upwards with respect to odd sites, and black and white diamonds correspond to $\ket{1}$ and $\ket{0}$ respectively. Due to the staggering a freely-moving quasiparticle in a row is graphically represented by one (and not two) black fields. Figure~\ref{fig:pairstructure} contains an illustration of the time-evolution of a representative basis state included in the sum~\eqref{eq:initialStateCompBasis} with $2L=18$. One observes that each of the local states $\ket{010}$ indeed corresponds to a pair of oppositely moving quasiparticles, and $\ket{000}$ behaves as the empty space. Since in the initial state there cannot be more than one consecutive $\ket{1}$, all quasiparticles appear as pairs.

\subsection{Von Neumann entropy: exact confirmation of the quasiparticle picture}

The quasiparticle picture is believed to apply whenever the system possesses
stable quasiparticles~\cite{calabrese2017lecture}. In particular, this is the
case for \emph{integrable models} where the``entangling" quasiparticles have
been conjectured to coincide with the stable excitations on the stationary
state reached after the quench~\cite{alba2017entanglement}. Using this
identification one can make \eqref{eq:quasiparticlemultiS} and
\eqref{eq:quasiparticlemultiSinhomo} predictive by computing all the --- yet
unknown --- functions featured in those equations by means of TBA~\cite{yang1969thermodynamics,
takahashi1999thermodynamics}. 

More specifically, let us consider~\eqref{eq:quasiparticlemultiS} which
depends on two unknown functions.  The first, $v_n(\lambda)$, is naturally
identified with the velocity of the excitations --- accessible in
TBA~\cite{bonnes2014light} --- while the second, $s_n(\lambda)$, can be fixed
by imposing the equality between the entanglement and the thermodynamic entropy
in the stationary state~\cite{alba2017entanglement}.
All this is particularly simple for excitations on the stationary states
\eqref{eq:GGE} in Rule 54. Indeed, on these states there is only one species of
excitations and $\lambda$ can take only two values ($\lambda\in\{\pm\}$) (for
further details see Paper I, the Supplemental Material of
Ref.~\cite{friedman2019integrable}, and the review~\cite{buca2021rule}). Therefore,
the prediction~\eqref{eq:quasiparticlemultiS} is effectively of the
form~\eqref{eq:quasiparticleS} with 
\be
v=v_\vartheta =\frac{2}{1+2 \vartheta}\,,\qquad
s = s_\vartheta  = - \vartheta \log \vartheta - \left(1- \vartheta\right)
\log \left(1-\vartheta\right),
\label{eq:fixhomo}
\ee
where $\vartheta$ is precisely the filling characterising the Gibbs state~\eqref{eq:GE}. Namely, it is written in terms of the chemical potential as in Eq.~\eqref{eq:muvartheta}.  

We then see that the limits 
\be
\smashoperator[r]{\lim_{\substack{t\to\infty \\[0.25em] {|A|}/{t}=\zeta \geq 4}}}
\frac{S_{A,\rm th}}{t} = 4 v s,\qquad 
\smashoperator[r]{\lim_{\substack{t\to\infty \\[0.25em] {|A|}/{t}=\zeta \leq {2}/{3}}}}
\frac{S_{A,\rm th}}{t} = 2 s \zeta, 
\ee
computed with the quasiparticle picture agree with our exact results for all values of $\vartheta$. To the best of our knowledge this result, together with the special case ($\vartheta=1/2$) presented in Ref.~\cite{klobas2021exact}, provides the first rigorous confirmation of the quasiparticle picture in the presence of interactions. 

The same check can be performed in the case of bipartitioning protocols. In this case,
following~\cite{bertini2018entanglement, alba2019entanglement}, we impose 
\be
\dot{X}_{\pm}(x,t)=v_\pm(x,t),
\qquad s_{+}(x) = s_{-}(x) = s_{\rm L} \Theta(-x)+s_{\rm R} \Theta(x), 
\label{eq:fixinhomo}
\ee
where 
\be
s_{\rm L/R} = - \vartheta_{\rm L/R} \log \vartheta_{\rm L/R} 
- \left(1- \vartheta_{\rm L/R}\right) \log \left(1-\vartheta_{\rm L/R}\right),
\ee  
and $v_\pm(x,t)$ is the velocity of excitations on the locally quasistationary state at point
$(x,t)$ as computed by Generalized
Hydrodynamics~\cite{bertini2016transport, castroalvaredo2016emergent}. In particular,
using the explicit result for $v_\pm(x,t)$ reported in Paper I we have 
\be
X_{-}(x,t)=\begin{cases}
x - v_{\rm L} t, & x< 0, \\
\frac{x}{2}(1+\frac{v_{\rm L}}{v_{\rm R}}) - v_{\rm L} t, \quad& 0< x\leq 2 v_{\rm R} t,\\
x-v_{\rm R} t,  & x \geq 2 v_{\rm R} t, 
\end{cases}
\ee
and 
\be
X_{+}=\begin{cases}
x + v_{\rm L} t, & x< - 2 v_{\rm L} t, \\
    \frac{x}{2}(1+\frac{v_{\rm R}}{v_{\rm L}}) + v_{\rm R} t, \quad& - 2 v_{\rm L} t\leq x\leq 0, \\
x+v_{\rm R}t,  & x>0,
\end{cases}
\ee
where we introduced 
\be
v_{\mathrm{L/R}}=\frac{2}{1+2 \vartheta_{\mathrm{L/R}}}\,.
\ee
Plugging it into~\eqref{eq:quasiparticlemultiSinhomo}, a simple (but tedious) calculation gives 
\be
S_{A,\rm th} = \begin{cases}
\displaystyle
\frac{2 v_{\rm R}v_{\rm L}}{v_{\rm R}+v_{\rm L}}
    t (s_{\rm R}+s_{\rm L})+ 2 v_{\rm R} t s_{\rm R},
    & \displaystyle  \frac{|A|}{t} \geq \frac{v_{\rm R}
    ( v_{\rm R}+3 v_{\rm L})}{(v_{\rm R}+v_{\rm L})},\\
\\
\displaystyle \frac{2 v_{\rm R}v_{\rm L}}{v_{\rm R}+v_{\rm L}}
    t (s_{\rm L}-s_{\rm R}) + 2 |A| s_{\rm R},
    & \displaystyle v_{\rm R} \leq \frac{|A|}{t}
    \leq \frac{v_{\rm R} ( v_{\rm R}+3 v_{\rm L})}{(v_{\rm R}+v_{\rm L})}, \\
\\
\displaystyle \frac{2 v_{\rm L} }{v_{\rm R}+v_{\rm L}} |A| s_{\rm L}+
    \frac{2 v_{\rm R} }{v_{\rm R}+v_{\rm L}} |A| s_{\rm R},\qquad
    & \displaystyle \frac{|A|}{t}\leq v_{\rm R},
\end{cases}
\ee
where we took $A=[0,|A|]$. Noting that 
\be
\frac{v_{\rm R} ( v_{\rm R}+3 v_{\rm L})}{(v_{\rm R}+v_{\rm L})}< 4,
\qquad v_{\rm R}\geq\frac{2}{3},
\ee
we have 
\be
\smashoperator[r]{\lim_{\substack{t\to\infty \\[0.25em] {|A|}/{t}
=\zeta \geq 4}}}\frac{S_{A,\rm th}}{t} 
\!=\!\frac{2 v_{\rm R}v_{\rm L}}{v_{\rm R}
\!+\! v_{\rm L}} (s_{\rm R}\mkern-1mu +\mkern-1mu s_{\rm L})
+ 2 v_{\rm R} s_{\rm R},\mkern8mu
\smashoperator[r]{\lim_{\substack{t\to\infty \\[0.25em] {|A|}/{t}
=\zeta \leq \frac{2}{3}}}}\frac{S_{A,\rm th}}{t} 
\mkern-4mu=\mkern-4mu\frac{2 v_{\rm L} }{v_{\rm R}\!+\! v_{\rm L}} s_{\rm L}\zeta 
+ \frac{2 v_{\rm R} }{v_{\rm R}\!+\!v_{\rm L}} s_{\rm R} \zeta, 
\ee
which, once again, agree with our exact results for all possible values of
$\vartheta_{\rm L/R}\in[0,1]$. To the best of our knowledge, this is the first rigorous confirmation of the quasiparticle picture for inhomogeneous quenches.

\subsection{R\'enyi Entropies: no consistent quasiparticle description}
In non-interacting systems the quasiparticle picture can be directly extended
to R\'enyi entropies with $\alpha\neq1$. As pointed out in
Refs.~\cite{alba2017quench, alba2017renyi}, however, in the presence of interactions
this extension becomes far less straightforward. The reason appears to
be connected to the fact that $S^{(\alpha)}_{A}$ have a stronger non-linear
dependence on the state compared to the Von Neumann entanglement entropy. This
makes it harder to understand which excitations --- or better the excitations over
which stationary state --- are relevant for the quasiparticle picture. As a result,
a consistent extension of the quasiparticle picture for R\'enyi entropies in
interacting systems has not yet been found.
Here we use our exact results to show that insisting on a quasiparticle
description for higher R\'enyi entropies one has to take excitations over a
stationary state with unclear physical meaning. For simplicity, we focus on the
homogeneous quench~\eqref{eq:solvablefamily} as it contains all the basic
elements of our reasoning. 

\begin{figure}
    \centering
    \includegraphics[width=0.75\textwidth]{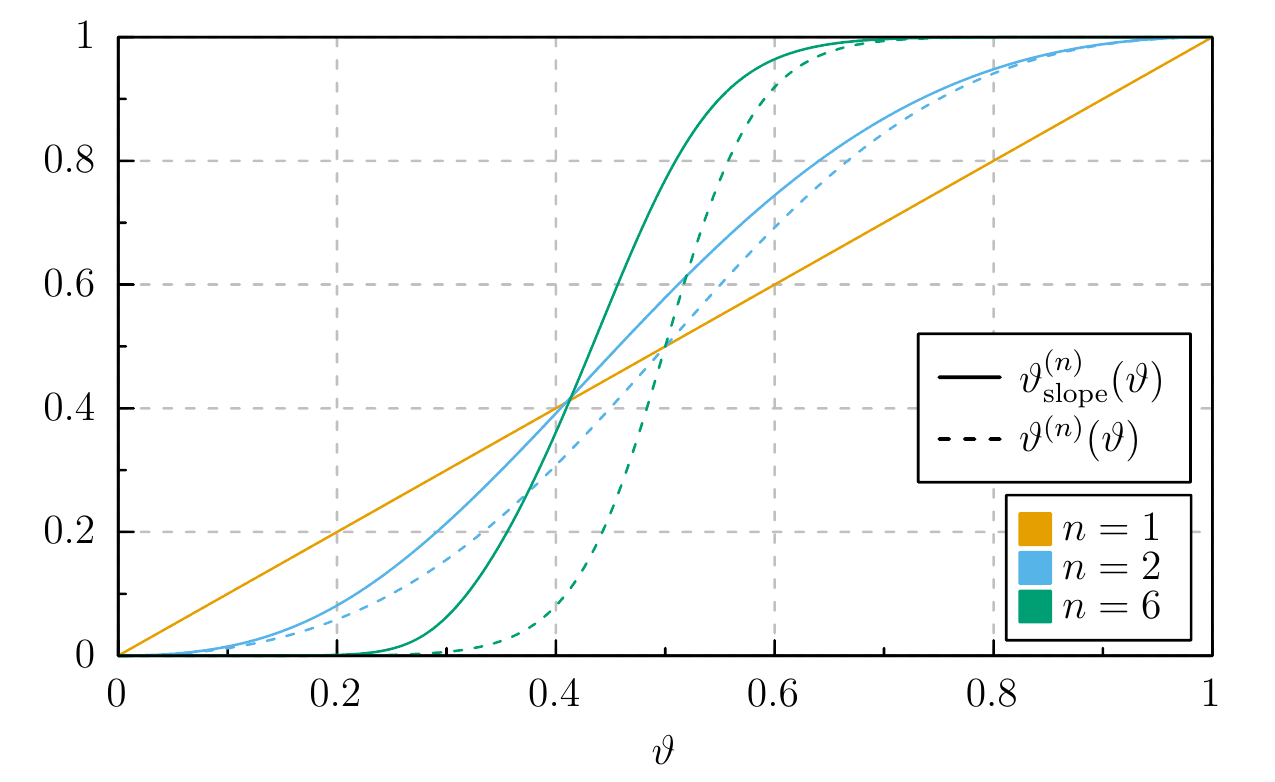}
    \caption{\label{fig:difGplot} Comparison between the two effective $n$-dependent
    filling fractions~\eqref{eq:stateRenyi} and~\eqref{eq:alphapowerGGE}, for a few
    different choices of $n$. In the case of von Neumann entropy (i.e.\ for $n=1$),
    the effective filling fraction is just~$\vartheta$, while for $n\neq 1$ we get
    two different predictions: $\vartheta_{\rm slope}^{(n)}$ coming from
    the renormalised quasiparticle velocity and $\vartheta^{(n)}$ extracted from the
    stationary state. The two quantities agree for small $\vartheta$ or $1-\vartheta$,
    while for the intermediate $\vartheta$, the difference between the two increases with
    $\abs{n-1}$.
    }
\end{figure}

Requiring the validity of the quasiparticle picture we find the following
asymptotic formula for the R\'enyi entropies
\be
S^{(\alpha)}_{A,\rm th} = \min(4 v_\alpha(\vartheta) t,2|A|) s^{(\alpha)}(\vartheta)\,,
\label{eq:quasiparticleSren}
\ee
where now $v_\alpha, s^{(\alpha)}$ are unknown functions. The density of R\'enyi
entropy ``carried" by a quasiparticle pair can be fixed using the expression for the
stationary-state R\'enyi entropy~\eqref{eq:GEstationaryRenyi}, namely
\be
s^{(\alpha)}(\vartheta)
=\smashoperator{\lim_{\substack{t\to\infty \\[0.25em] {|A|}/{t}=\zeta \leq {2}/{3}}}}
\frac{S^{(\alpha)}_{A,\rm th}}{2 |A|} =
\frac{1}{1-\alpha} \log[\vartheta^\alpha+(1-\vartheta)^{\alpha}]. 
\ee
Using now the exact expression for the rate of entanglement spreading~\eqref{eq:result}
we have that the quasiparticle velocity must be given by 
\be
v_\alpha(\vartheta) =
\ \smashoperator{\lim_{\substack{t\to\infty\\[0.25em]{|A|}/{t}=\zeta \geq 4}}}\ 
\frac{S^{(\alpha)}_{A,\rm th}}{4 t s^{(\alpha)}}=v^{\rm E}_{\alpha}(\vartheta,\vartheta) = \frac{\log[\lambda_\alpha(\vartheta, \vartheta)]}
{\log[\vartheta^\alpha+(1-\vartheta)^{\alpha}]}\,.
\ee
Now we note that 
\be
v_{\alpha\neq1}(\vartheta) \neq \frac{2}{1+2\vartheta}\,,
\ee
which means that the quasiparticles \emph{cannot} be though of as excitations on the
stationary state~\eqref{eq:GGE}, i.e.\ the state describing the expectation
values of local observables at infinite times after the quench. Nevertheless,
one can interpret $v_{\alpha}(\vartheta)$ as the velocity excitations over the
stationary state with filling 
\be
\vartheta^{(\alpha)}_{\rm slope}(\vartheta)=\frac{2-v_\alpha(\vartheta)}{2v_\alpha(\vartheta)}\,.
\label{eq:stateRenyi}
\ee
Indeed, since $v_\alpha(\vartheta)\in[2/3,2]$, Eq.~\eqref{eq:stateRenyi} is always in $[0,1]$ and hence describes a legitimate filling.  

The physical meaning of \eqref{eq:stateRenyi} is, however, unclear. In particular, for~$\alpha\neq 1$ the filling~\eqref{eq:stateRenyi} does \emph{not} coincide with that of the macrostate that describes the stationary value of the $\alpha$ R\'enyi entropy in TBA~\cite{alba2017quench, alba2017renyi}. Indeed, in our case the latter has filling (cf.~\eqref{eq:Lambdan})
\be
\label{eq:alphapowerGGE}
\vartheta^{(\alpha)}(\vartheta)=\frac{\vartheta^\alpha}{\vartheta^\alpha+(1-\vartheta)^\alpha}.
\ee
Even though \eqref{eq:stateRenyi} and \eqref{eq:alphapowerGGE} are close for small and large fillings they are different functions of $\vartheta$. See the representative example in Fig.~\ref{fig:difGplot}.

\section{Conclusions}
\label{sec:conclusions}
In the paper we used a time-channel approach to find exact results for the entanglement dynamics in the quantum cellular automaton Rule 54, which is arguably the simplest example of interacting integrable model. We showed that the entanglement dynamics from a class of solvable initial states is characterised by a certain tensor network and that, remarkably, the latter can be contracted exactly. We used our results to test the quasiparticle picture for the entanglement spreading in Rule 54. In particular, we confirmed that the quasiparticle picture provides quantitatively accurate predictions for the evolution of the von Neumann entanglement entropy in the presence of interactions, both in homogeneous and inhomogeneous situations. Therefore validating the predictions of both Ref.~\cite{alba2017entanglement} and Ref.~\cite{alba2019entanglement}. We also argued that our results seem to exclude a consistent quasiparticle interpretation for the evolution of other R\'enyi entropies. Indeed, we showed that the potential quasiparticles responsible for the spreading of R\'enyi entropies cannot be interpreted as excitations on a physically meaningful stationary state.   

An interesting direction for future research is to extend the techniques presented here to the study of the various kinds of operator space entanglement~\cite{prosen2007operator, zanardi2001entanglement}. These include the entanglement of local operators, of the reduced density matrix, and of the time evolving operator. The latter is particularly relevant for the questions considered in this paper because it gives access to the ``line tension", which is the function needed to obtain quantitative predictions from the membrane picture (see e.g.~\cite{zhou2020entanglement}). It would be interesting to also test these predictions against our results, especially those concerning the R\'enyi entropies that do not seem to be described by the quasiparticle picture. 

\section*{Acknowledgements}
We thank Fabian Essler and Maurizio Fagotti for useful discussions and Lorenzo Piroli for collaboration on closely related projects. 

\paragraph{Funding information}
This work has been supported by the EPSRC through the grant EP/S020527/1 (KK) and by the Royal Society through the University Research Fellowship No. 201101 (BB).

\begin{appendix}
\section{Proof of Lemma~\ref{lemma1}}
\label{sec:prooflemma1}
We begin by introducing the following shorthand notation. We call
$\{\tilde{M}^{x y}\}_{x,y=1}^9$ the set of~$4\times 4$ matrices with matrix
elements given by the tensor~$\tilde{M}$, i.e.\
\be
\left[\tilde{M}^{x y}\right]_{a b} \coloneqq \tilde{M}^{x y}_{a b}.
\ee
In this new notation the statement of Lemma~\ref{lemma1} reads as 
\be
\begin{aligned}
&\big(\tilde{M}^{x_1 x_2}\tilde{M}^{y_1 y_2}\big)\otimes
\big(\tilde{M}^{x_2 x_3}\tilde{M}^{y_2 y_3}\big)\otimes
\cdots\otimes\big(\tilde{M}^{x_k x_1}\tilde{M}^{y_k y_1}\big) \\
&= \left[\prod_{j=1}^k \delta_{x_j,y_j}\right] \big(\tilde{M}^{x_1 x_2})^2\otimes
\big(\tilde{M}^{x_2 x_3}\big)^2\otimes\cdots\otimes\big(\tilde{M}^{x_1 x_2}\big)^2\,.
\end{aligned}
\ee
Before proving this statement in full generality, let us first consider $k=1$.
In this case one can explicitly evaluate all the products of pairs of
matrices~$\tilde{M}^{x_1 x_1}\tilde{M}^{y_1 y_1}$ and realise that the only
non-zero combination comes from~$x_1=y_1=1$, i.e. 
\be
  %\begin{tikzpicture}
    %[baseline={([yshift=-0.6ex]current bounding box.center)},scale=1.75]
    %\dtensorLine{-0.75}{0}{0.75}{0};
    %\dtensorLine{-0.75}{1}{0.75}{1};
    %\dtensorLine{0}{-0.75}{0}{1.75};
    %\tensorM{0}{0}{colTensor};
    %\tensorM{0}{1}{colTensor};
    %\node[draw=white,fill=white,inner sep=1pt] at (0,{\dx}) {$x_1$};
    %\node[draw=white,fill=white,inner sep=1pt] at (0,{-\dx}) {$x_1$};
   % \node[draw=white,fill=white,inner sep=1pt] at ({\dx},{\dx}) {$y_1$};
    %\node[draw=white,fill=white,inner sep=1pt] at ({\dx},{-\dx}) {$y_1$};
  %\end{tikzpicture}=
\tilde{M}^{x_1 x_1}\tilde{M}^{y_1 y_1}=\delta_{x_1,0}\delta_{y_1,0}
\,\frac{1}{16}
\begin{bmatrix}
  1&0&0&0\\
  0&0&0&0\\
  0&0&0&0\\
  0&0&0&0
\end{bmatrix},
\ee
and the property~\eqref{eq:lemma1eq} holds. For~$k=2$ one can similarly check that the
product of two tensors
\be
%  \begin{tikzpicture}
 %   [baseline={([yshift=-0.6ex]current bounding box.center)},scale=1.75]
  %  \dtensorLine{-0.75}{0}{2.75}{0};
  %  \dtensorLine{-0.75}{1}{2.75}{1};
  %  \dtensorLine{0}{-0.75}{0}{1.75};
  %  \dtensorLine{2}{-0.75}{2}{1.75};
  %  \tensorM{0}{0}{colTensor};
  %  \tensorM{0}{1}{colTensor};
   % \tensorM{2}{0}{colTensor};
   % \tensorM{2}{1}{colTensor};
   % \node[draw=white,fill=white,inner sep=1pt] at (0,{\dx}) {$x_1$};
   % \node[draw=white,fill=white,inner sep=1pt] at (0,{-\dx}) {$x_2$};
   % \node[draw=white,fill=white,inner sep=1pt] at (0,{-3*\dx}) {$x_1$};
   % \node[draw=white,fill=white,inner sep=1pt] at ({\dx},{\dx}) {$y_1$};
   % \node[draw=white,fill=white,inner sep=1pt] at ({\dx},{-\dx}) {$y_2$};
    %\node[draw=white,fill=white,inner sep=1pt] at ({\dx},{-3*\dx}) {$y_1$};
  %\end{tikzpicture}=
\big(\tilde{M}^{x_1 x_2}\tilde{M}^{y_1 y_2}\big)\otimes
\big(\tilde{M}^{x_2 x_1}\tilde{M}^{y_2 y_1}\big)
\ee
is nonzero only for the following $5$ combinations of indices
\be
\begin{tabular}{c | c | c | c}
  $x_1$ & $y_1$ & $x_2$ & $y_2$ \\
  \hline
  $0$ & $0$ & $0$ & $0$ \\
  $0$ & $0$ & $8$ & $8$ \\
  $8$ & $8$ & $0$ & $0$ \\
  $1$ & $1$ & $6$ & $6$ \\
  $6$ & $6$ & $1$ & $1$
\end{tabular},
\ee
which proves~\eqref{eq:lemma1eq} for~$k=2$.

To prove the lemma for general~$k$ we show that for any two \emph{different}
cycles of indices~$(x_1,x_2,x_3,\ldots,x_k,x_1)$ and~$(y_1,y_2,\ldots,y_k,y_1)$
at least one of the matrix products $\tilde{M}^{x_j x_{j+1}}\cdot
\tilde{M}^{y_j y_{j+1}}$ is~$0$. This can be demonstrated by defining the
$81\times 81$ adjacency matrix~$A$ with elements that are $1$ if the two pairs
$(x_j,y_j)$ and $(x_{j+1},y_{j+1})$ are connected by a nonzero matrix product,
and $0$ otherwise,
\be
A_{(x_1 y_1),(x_2 y_2)} : =
\begin{cases}
  1, &  \tilde{M}^{x_1 x_2}\cdot\tilde{M}^{y_1 y_2} = 0,\\
  0, &  \text{otherwise}.
\end{cases}
\ee
If a pair of cycles~$(x_1,x_2,\ldots,x_k,x_1)$ and $(y_1,y_2,\ldots,y_k,y_1)$ gives
a nonzero value to the l.h.s.\ of~\eqref{eq:lemma1eq}, all the matrix elements of~$A$
appearing in the following product have to be $1$,
\be
A_{(x_1 y_1),(x_2 y_2)} A_{(x_2 y_2),(x_3 y_3)}\cdots A_{(x_k y_k),(x_1 y_1)}=1,
\ee
which is one of the contributions to the diagonal matrix element
$[A^k]_{(x_1,y_1),(x_1,y_1)}$.  

Next, let us define a $9\times 9$ \emph{reduced} adjacency matrix $\tilde{A}$
that contains only the elements where $x_j$ and $y_j$ are the same
\be
\tilde{A}_{x_1,x_2} : = A_{(x_1,x_1),(x_2,x_2)}.
\ee
By explicit diagonalisation of $A$ and $\tilde A$ we find that they have the
same non-zero eigenvalues, which are the solutions of the following cubic
equation 
\be
x^3=(x+1)^2.
\ee
This implies
\be
{\rm tr} A^k = {\rm tr} \tilde{A}^k, \qquad \forall k, 
\ee
or, in other words, that only non-zero contributions to diagonal
elements~$[A^k]_{(x_1,y_1),(x_1,y_1)}$ come from elements with the same values
of~$x_j$ and~$y_j$
\be
A_{(x_1 y_1),(x_2 y_2)} A_{(x_2 y_2),(x_3 y_3)}\cdots
A_{(x_k y_k),(x_1 y_1)}
=\delta_{x_1,y_1}\cdots \delta_{x_k,y_k}
\tilde{A}_{x_1,x_2} \tilde{A}_{x_2,x_3}\cdots \tilde{A}_{x_k,x_1}.
\ee
This completes the proof of the lemma.

\section{Proof of Lemma~\ref{lemma2}}
\label{sec:propertiescubiceq}

To prove Lemma~\ref{lemma2} we rewrite Eq.~\eqref{eq:cubicEq} as 
\be
p(\lambda,n)=0,
\label{eq:cubicEqpoly}
\ee
where we defined the polynomial 
\be
p(\lambda,n)=\lambda^3 +a_{2,n} \lambda^2 +a_{1,n} \lambda +a_{0,n},
\ee
with 
\be
\begin{aligned}
a_{0,n}&=- \vartheta_1^n\vartheta_2^n, & a_{1,n}&=- \left(\vartheta_1^n (1-\vartheta_2)^n +\vartheta_2^n (1-\vartheta_1)^n \right),\\
a_{2,n}&=-  (1-\vartheta_1)^n(1-\vartheta_1)^n,\qquad &  a_{3,n}&=1.
\end{aligned}
\ee
Since $a_{3,n}$ is positive and all $\{a_{j,n}\}_{j=0}^2$ are negative for $\vartheta_1, \vartheta_2\in(0,1)$, Descartes' rule of signs (see e.g.\ Ref~\cite{descartes1954geometry}) implies that Eq.~\eqref{eq:cubicEqpoly} has only one real positive solution which we denote by $\lambda_n(\vartheta_1,\vartheta_2)$. Moreover, since $a_{3,n}$ is positive, we also have 
\be
p(\lambda,n)>0,\quad \forall \lambda > \lambda_n(\vartheta_1,\vartheta_2)\,, 
\ee
which implies 
\be
\lambda^3 > |a_{0,n}|+|a_{1,n}| \lambda +|a_{2,n}| \lambda^2 ,\quad \forall \lambda > \lambda_n(\vartheta_1,\vartheta_2)\,. 
\label{eq:inequality}
\ee
Next, we recall that Rouch\'e's Theorem (see e.g.\ Ref.~\cite{rudin1987real}) implies that whenever a polynomial
\be
g(\lambda) = \sum_{k=0}^m b_k \lambda^k,
\ee
has coefficients $b_k\in\mathbb C$ fulfilling 
\be
|b_m| R^m \leq \sum_{k=0}^{m-1} |b_k| R^k, 
\ee
for some $R\in\mathbb R$, all the solutions to $g(\lambda)=0$ are contained in the circle of radius $R$. Applying this to \eqref{eq:inequality} we find that all solutions to Eq.~\eqref{eq:cubicEqpoly} are contained in the circle of radius $\lambda_n(\vartheta_1,\vartheta_2)$. To conclude we should prove that the absolute values of the other two solutions to Eq.~\eqref{eq:cubicEq} are strictly smaller than $\lambda_n(\vartheta_1,\vartheta_2)$. 

We proceed by contradiction. Let us assume that all solutions to Eq.~\eqref{eq:cubicEqpoly} have the same absolute value. Since the polynomial has real coefficients this means that the solutions are 
\be
\{\lambda_n(\vartheta_1,\vartheta_2),\ \lambda_n(\vartheta_1,\vartheta_2) e^{i \theta},
\ \lambda_n(\vartheta_1,\vartheta_2) e^{-i \theta}\},
\ee
for some $\theta\in\mathbb R$. This in turn implies that $p(\lambda,n)$ must coincide with 
\be
\begin{aligned}
    &(\lambda-\lambda_n(\vartheta_1,\vartheta_2))(\lambda-\lambda_n(\vartheta_1,\vartheta_2)e^{i \theta})(\lambda-\lambda_n(\vartheta_1,\vartheta_2)e^{-i \theta})\\
    &\quad
    = \lambda^3 -  (1+2\cos\theta) \lambda_n(\vartheta_1,\vartheta_2) \lambda^2  +  (1+2\cos\theta) \lambda_n(\vartheta_1,\vartheta_2)^2 \lambda -  \lambda_n(\vartheta_1,\vartheta_2)^3.
\end{aligned}
\ee
We see that this cannot happen for any $\theta$ because either the coefficient of $\lambda^2$ or that of $\lambda$ are positive, while both $a_{2,n}$ and $a_{1,n}$ are negative. We now assume that there is a single additional solution to Eq.~\eqref{eq:cubicEqpoly} with absolute value equal to $\lambda_n(\vartheta_1,\vartheta_2)$. This implies that the set of solutions reads as 
\be
\{\lambda_n(\vartheta_1,\vartheta_2),\quad -\lambda_n(\vartheta_1,\vartheta_2),\quad -\lambda_n(\vartheta_1,\vartheta_2)+c \},
\ee
with some $c\in(0,\lambda_n(\vartheta_1,\vartheta_2)]$. Therefore, $p(\lambda,n)$ must coincide with 
\be
\begin{aligned}
&(\lambda-\lambda_n(\vartheta_1,\vartheta_2))(\lambda+\lambda_n(\vartheta_1,\vartheta_2))(\lambda+\lambda_n(\vartheta_1,\vartheta_2)-c)\\
&\quad = \lambda^3 +  (\lambda_n(\vartheta_1,\vartheta_2)-c) \lambda^2  -  \lambda_n(\vartheta_1,\vartheta_2)^2 \lambda -  \lambda_n(\vartheta_1,\vartheta_2)^2 (\lambda_n(\vartheta_1,\vartheta_2)-c)\,.
\end{aligned}
\ee
This is once again impossible because the coefficient of $\lambda^2$ is positive, while $a_{2,n}$ is negative. Therefore the only possibility is that all other solutions to Eq.~\eqref{eq:cubicEqpoly} have absolute value strictly smaller than $\lambda_n(\vartheta_1,\vartheta_2)$. Finally, the explicit expression~\eqref{eq:lambdan} is found using the general solution of the cubic equation (one can immediately verify that Eq.~\eqref{eq:lambdan} is indeed real, positive, and fulfils Eq.~\eqref{eq:cubicEq} for all $\vartheta_1,\vartheta_2\in[0,1]$). 

Let us now move on and prove \eqref{eq:inequalitylambda}. We begin recalling that $\lambda_z(\vartheta_1,\vartheta_2)$ is obtained replacing $n$ in Eq.~\eqref{eq:lambdan} with $z\in\mathcal D=\{z\in\mathbb C\!:\,\, {\rm Re}[z]>0\}$ and solves Eq.~\eqref{eq:cubicEqpoly} with $n$ replaced by $z\in\mathcal D=\{z\in\mathbb C\!:\,\, {\rm Re}[z]>0\}$. Next, we observe that 
\be
1\geq a_{2,{\rm Re}[z]} \geq |a_{2,z}|,\qquad 2\geq a_{1,{\rm Re}[z]} \geq |a_{1,z}|,\qquad 1\geq a_{0,{\rm Re}[z]} \geq |a_{0,z}|\,.  
\ee
Combining these two facts we find  
\be
\begin{aligned}
\lambda_{{\rm Re}[z]}(\vartheta_1,\vartheta_2)^3 & = a_{0,{\rm Re}[z]}+ a_{1,{\rm Re}[z]}\lambda_{{\rm Re}[z]}(\vartheta_1,\vartheta_2) + a_{2,{\rm Re}[z]}  \lambda_{{\rm Re}[z]}(\vartheta_1,\vartheta_2)^2\\
& \geq |a_{0,z}|+ |a_{1,z}|\lambda_{{\rm Re}[z]}(\vartheta_1,\vartheta_2) + |a_{2,z}|  \lambda_{{\rm Re}[z]}(\vartheta_1,\vartheta_2)^2\,.
\end{aligned}
\ee 
Using again Rouch\'e's Theorem we then have 
\be
|\lambda_{z}(\vartheta_1,\vartheta_2)|\leq \lambda_{{\rm Re}[z]}(\vartheta_1,\vartheta_2)\,.
\ee
Finally we observe that 
\be
3^3  > 1+ 2\times 3 + 1\times  3^2 \geq a_{0,{\rm Re}[z]} + a_{1,{\rm Re}[z]} 3 + a_{2,{\rm Re}[z]}  3^2\,,
\ee
which implies 
\be
3>\lambda_{{\rm Re}[z]}(\vartheta_1,\vartheta_2)\,.
\ee
\end{appendix}

\bibliography{bibliography}

\begin{thebibliography}{10}
\providecommand{\url}[1]{\texttt{#1}}
\providecommand{\urlprefix}{URL }
\expandafter\ifx\csname urlstyle\endcsname\relax
  \providecommand{\doi}[1]{doi:\discretionary{}{}{}#1}\else
  \providecommand{\doi}{doi:\discretionary{}{}{}\begingroup
  \urlstyle{rm}\Url}\fi
\providecommand{\eprint}[2][]{\url{#2}}

\bibitem{calabrese2005evolution}
P.~Calabrese and J.~Cardy,
\newblock \emph{Evolution of entanglement entropy in one-dimensional systems},
\newblock J. Stat. Mech. \textbf{2005}(04), P04010 (2005),
\newblock \doi{10.1088/1742-5468/2005/04/p04010}.

\bibitem{liu2014entanglement}
H.~Liu and S.~J. Suh,
\newblock \emph{Entanglement tsunami: Universal scaling in holographic
  thermalization},
\newblock Phys. Rev. Lett. \textbf{112}, 011601 (2014),
\newblock \doi{10.1103/PhysRevLett.112.011601}.

\bibitem{fagotti2008evolution}
M.~Fagotti and P.~Calabrese,
\newblock \emph{Evolution of entanglement entropy following a quantum quench:
  Analytic results for the {XY} chain in a transverse magnetic field},
\newblock Phys. Rev. A \textbf{78}, 010306 (2008),
\newblock \doi{10.1103/PhysRevA.78.010306}.

\bibitem{alba2018entanglement}
V.~Alba and P.~Calabrese,
\newblock \emph{{Entanglement dynamics after quantum quenches in generic
  integrable systems}},
\newblock SciPost Phys. \textbf{4}, 17 (2018),
\newblock \doi{10.21468/SciPostPhys.4.3.017}.

\bibitem{alba2017entanglement}
V.~Alba and P.~Calabrese,
\newblock \emph{Entanglement and thermodynamics after a quantum quench in
  integrable systems},
\newblock PNAS \textbf{114}(30), 7947 (2017),
\newblock \doi{10.1073/pnas.1703516114}.

\bibitem{laeuchli2008spreading}
A.~M. L\"auchli and C.~Kollath,
\newblock \emph{Spreading of correlations and entanglement after a quench in
  the one-dimensional {Bose--Hubbard} model},
\newblock J. Stat. Mech. \textbf{2008}(05), P05018 (2008),
\newblock \doi{10.1088/1742-5468/2008/05/p05018}.

\bibitem{kim2013ballistic}
H.~Kim and D.~A. Huse,
\newblock \emph{Ballistic spreading of entanglement in a diffusive
  nonintegrable system},
\newblock Phys. Rev. Lett. \textbf{111}, 127205 (2013),
\newblock \doi{10.1103/PhysRevLett.111.127205}.

\bibitem{pal2018entangling}
R.~Pal and A.~Lakshminarayan,
\newblock \emph{Entangling power of time-evolution operators in integrable and
  nonintegrable many-body systems},
\newblock Phys. Rev. B \textbf{98}, 174304 (2018),
\newblock \doi{10.1103/PhysRevB.98.174304}.

\bibitem{bertini2019entanglement}
B.~Bertini, P.~Kos and T.~Prosen,
\newblock \emph{Entanglement spreading in a minimal model of maximal many-body
  quantum chaos},
\newblock Phys. Rev. X \textbf{9}(2), 021033 (2019),
\newblock \doi{10.1103/PhysRevX.9.021033}.

\bibitem{piroli2020exact}
L.~Piroli, B.~Bertini, J.~I. Cirac and T.~Prosen,
\newblock \emph{Exact dynamics in dual-unitary quantum circuits},
\newblock Phys. Rev. B \textbf{101}, 094304 (2020),
\newblock \doi{10.1103/PhysRevB.101.094304}.

\bibitem{gopalakrishnan2019unitary}
S.~Gopalakrishnan and A.~Lamacraft,
\newblock \emph{Unitary circuits of finite depth and infinite width from
  quantum channels},
\newblock Phys. Rev. B \textbf{100}, 064309 (2019),
\newblock \doi{10.1103/PhysRevB.100.064309}.

\bibitem{Greiner2015}
R.~Islam, R.~Ma, P.~M. Preiss, M.~E. Tai, A.~Lukin, M.~Rispoli and M.~Greiner,
\newblock \emph{Measuring entanglement entropy in a quantum many-body system},
\newblock Nature \textbf{528}, 77 (2015),
\newblock \doi{10.1038/nature15750}.

\bibitem{Greiner2016}
A.~M. Kaufman, M.~E. Tai, A.~Lukin, M.~Rispoli, R.~Schittko, P.~M. Preiss and
  M.~Greiner,
\newblock \emph{Quantum thermalization through entanglement in an isolated
  many-body system},
\newblock Science \textbf{353}(6301), 794 (2016),
\newblock \doi{10.1126/science.aaf6725}.

\bibitem{Greiner2019}
A.~Lukin, M.~Rispoli, R.~Schittko, M.~E. Tai, A.~M. Kaufman, S.~Choi,
  V.~Khemani, J.~L{\'e}onard and M.~Greiner,
\newblock \emph{Probing entanglement in a many-body{\textendash}localized
  system},
\newblock Science \textbf{364}(6437), 256 (2019),
\newblock \doi{10.1126/science.aau0818}.

\bibitem{dechiara2006entanglement}
G.~D. Chiara, S.~Montangero, P.~Calabrese and R.~Fazio,
\newblock \emph{Entanglement entropy dynamics of {Heisenberg} chains},
\newblock J. Stat. Mech. \textbf{2006}(03), P03001 (2006),
\newblock \doi{10.1088/1742-5468/2006/03/p03001}.

\bibitem{znidaric2008many}
M.~{\v Z}nidari{\v c}, T.~Prosen and P.~Prelov{\v s}ek,
\newblock \emph{Many-body localization in the {Heisenberg} {$XXZ$} magnet in a
  random field},
\newblock Phys. Rev. B \textbf{77}, 064426 (2008),
\newblock \doi{10.1103/PhysRevB.77.064426}.

\bibitem{nandkishore2015many}
R.~Nandkishore and D.~A. Huse,
\newblock \emph{Many-body localization and thermalization in quantum
  statistical mechanics},
\newblock Annu. Rev. Condens. Matter Phys. \textbf{6}(1), 15 (2015),
\newblock \doi{10.1146/annurev-conmatphys-031214-014726}.

\bibitem{kormos2017real}
M.~Kormos, M.~Collura, G.~Tak{\'a}cs and P.~Calabrese,
\newblock \emph{Real-time confinement following a quantum quench to a
  non-integrable model},
\newblock Nat. Phys. \textbf{13}(3), 246 (2017),
\newblock \doi{10.1038/nphys3934}.

\bibitem{li2019measurement}
Y.~Li, X.~Chen and M.~P.~A. Fisher,
\newblock \emph{Measurement-driven entanglement transition in hybrid quantum
  circuits},
\newblock Phys. Rev. B \textbf{100}, 134306 (2019),
\newblock \doi{10.1103/PhysRevB.100.134306}.

\bibitem{skinner2019measurement}
B.~Skinner, J.~Ruhman and A.~Nahum,
\newblock \emph{Measurement-induced phase transitions in the dynamics of
  entanglement},
\newblock Phys. Rev. X \textbf{9}, 031009 (2019),
\newblock \doi{10.1103/PhysRevX.9.031009}.

\bibitem{vasseur2019entanglement}
R.~Vasseur, A.~C. Potter, Y.-Z. You and A.~W.~W. Ludwig,
\newblock \emph{Entanglement transitions from holographic random tensor
  networks},
\newblock Phys. Rev. B \textbf{100}, 134203 (2019),
\newblock \doi{10.1103/PhysRevB.100.134203}.

\bibitem{nahum2017quantum}
A.~Nahum, J.~Ruhman, S.~Vijay and J.~Haah,
\newblock \emph{Quantum entanglement growth under random unitary dynamics},
\newblock Phys. Rev. X \textbf{7}, 031016 (2017),
\newblock \doi{10.1103/PhysRevX.7.031016}.

\bibitem{schollwock2011density}
U.~Schollw{\"o}ck,
\newblock \emph{The density-matrix renormalization group in the age of matrix
  product states},
\newblock Ann. Phys. \textbf{326}(1), 96 (2011),
\newblock \doi{10.1016/j.aop.2010.09.012}.

\bibitem{bertini2019exact}
B.~Bertini, P.~Kos and T.~Prosen,
\newblock \emph{Exact correlation functions for dual-unitary lattice models in
  $1+1$ dimensions},
\newblock Phys. Rev. Lett. \textbf{123}, 210601 (2019),
\newblock \doi{10.1103/PhysRevLett.123.210601}.

\bibitem{claeys2021ergodic}
P.~W. Claeys and A.~Lamacraft,
\newblock \emph{Ergodic and non-ergodic dual-unitary quantum circuits with
  arbitrary local {H}ilbert space dimension},
\newblock Phys. Rev. Lett. \textbf{126}, 100603 (2021),
\newblock \doi{10.1103/PhysRevLett.126.100603}.

\bibitem{suzuki2021computational}
R.~Suzuki, K.~Mitarai and K.~Fujii,
\newblock \emph{Computational power of one- and two-dimensional dual-unitary
  quantum circuits},
\newblock {a}rXiv:2103.09211 (2021), \eprint{https://arxiv.org/abs/2103.09211}.

\bibitem{bertini2020operatorI}
B.~Bertini, P.~Kos and T.~Prosen,
\newblock \emph{Operator entanglement in local quantum circuits {I}: Chaotic
  dual-unitary circuits},
\newblock SciPost Phys. \textbf{8}(4), 067 (2020),
\newblock \doi{10.21468/SciPostPhys.8.4.067}.

\bibitem{bertini2020operatorII}
B.~Bertini, P.~Kos and T.~Prosen,
\newblock \emph{Operator entanglement in local quantum circuits {II}: Solitons
  in chains of qubits},
\newblock SciPost Phys. \textbf{8}(4), 068 (2020),
\newblock \doi{10.21468/SciPostPhys.8.4.068}.

\bibitem{claeys2020maximum}
P.~W. Claeys and A.~Lamacraft,
\newblock \emph{Maximum velocity quantum circuits},
\newblock Phys. Rev. Research \textbf{2}, 033032 (2020),
\newblock \doi{10.1103/PhysRevResearch.2.033032}.

\bibitem{reid2021entanglement}
I.~Reid and B.~Bertini,
\newblock \emph{Entanglement barriers in dual-unitary circuits},
\newblock Phys. Rev. B \textbf{104}, 014301 (2021),
\newblock \doi{10.1103/PhysRevB.104.014301}.

\bibitem{klobas2021exact}
K.~Klobas, B.~Bertini and L.~Piroli,
\newblock \emph{Exact thermalization dynamics in the ``{R}ule 54'' quantum
  cellular automaton},
\newblock Phys. Rev. Lett. \textbf{126}, 160602 (2021),
\newblock \doi{10.1103/PhysRevLett.126.160602}.

\bibitem{pozsgay2014quantum}
B.~Pozsgay,
\newblock \emph{Quantum quenches and generalized {G}ibbs ensemble in a {B}ethe
  ansatz solvable lattice model of interacting bosons},
\newblock J. Stat. Mech. \textbf{2014}(10), P10045 (2014),
\newblock \doi{10.1088/1742-5468/2014/10/p10045}.

\bibitem{pozsgay2016real}
B.~Pozsgay and V.~Eisler,
\newblock \emph{Real-time dynamics in a strongly interacting bosonic hopping
  model: Global quenches and mapping to the {XX} chain},
\newblock J. Stat. Mech. \textbf{2016}(5), 053107 (2016),
\newblock \doi{10.1088/1742-5468/2016/05/053107}.

\bibitem{zadnik2021foldedI}
L.~Zadnik and M.~Fagotti,
\newblock \emph{The folded spin-1/2 {XXZ} model: {I.} diagonalisation, jamming,
  and ground state properties},
\newblock SciPost Phys. Core \textbf{4}, 10 (2021),
\newblock \doi{10.21468/SciPostPhysCore.4.2.010}.

\bibitem{zadnik2021foldedII}
L.~Zadnik, K.~Bidzhiev and M.~Fagotti,
\newblock \emph{The folded spin-1/2 {XXZ} model: {II.} thermodynamics and
  hydrodynamics with a minimal set of charges},
\newblock SciPost Phys. \textbf{10}, 99 (2021),
\newblock \doi{10.21468/SciPostPhys.10.5.099}.

\bibitem{pozsgay2021integrable}
B.~Pozsgay, T.~Gombor, A.~Hutsalyuk, Y.~Jiang, L.~Pristy{\'a}k and E.~Vernier,
\newblock \emph{An integrable spin chain with {Hilbert} space fragmentation and
  solvable real time dynamics},
\newblock {a}rXiv:2105.02252 (2021), \eprint{https://arxiv.org/abs/2105.02252}.

\bibitem{pozsgay2021yang}
B.~Pozsgay,
\newblock \emph{A {Yang}-{Baxter} integrable cellular automaton with a four
  site update rule},
\newblock J. Phys. A: Math. Theor. \textbf{54}(38), 384001 (2021),
\newblock \doi{10.1088/1751-8121/ac1dbf}.

\bibitem{gombor2021integrable}
T.~Gombor and B.~Pozsgay,
\newblock \emph{Integrable spin chains and cellular automata with medium range
  interaction},
\newblock {a}rXiv:2108.02053 (2021), \eprint{https://arxiv.org/abs/2108.02053}.

\bibitem{klobas2021exactI}
K.~Klobas and B.~Bertini,
\newblock \emph{Exact relaxation to {G}ibbs and non-equilibrium steady states
  in the quantum cellular automaton {R}ule 54},
\newblock {a}rXiv:2104.04511 (2021), \eprint{https://arxiv.org/abs/2104.04511}.

\bibitem{bertini2016transport}
B.~Bertini, M.~Collura, J.~De~Nardis and M.~Fagotti,
\newblock \emph{Transport in out-of-equilibrium {XXZ} chains: Exact profiles of
  charges and currents},
\newblock Phys. Rev. Lett. \textbf{117}, 207201 (2016),
\newblock \doi{10.1103/PhysRevLett.117.207201}.

\bibitem{castroalvaredo2016emergent}
O.~A. Castro-Alvaredo, B.~Doyon and T.~Yoshimura,
\newblock \emph{Emergent hydrodynamics in integrable quantum systems out of
  equilibrium},
\newblock Phys. Rev. X \textbf{6}, 041065 (2016),
\newblock \doi{10.1103/PhysRevX.6.041065}.

\bibitem{bertini2018entanglement}
B.~Bertini, M.~Fagotti, L.~Piroli and P.~Calabrese,
\newblock \emph{Entanglement evolution and generalised hydrodynamics:
  {N}oninteracting systems},
\newblock J. Phys. A: Math. Theor. \textbf{51}(39), 39LT01 (2018),
\newblock \doi{10.1088/1751-8121/aad82e}.

\bibitem{alba2019entanglement}
V.~Alba, B.~Bertini and M.~Fagotti,
\newblock \emph{{Entanglement evolution and generalised hydrodynamics:
  {I}nteracting integrable systems}},
\newblock SciPost Phys. \textbf{7}, 5 (2019),
\newblock \doi{10.21468/SciPostPhys.7.1.005}.

\bibitem{banuls2009matrix}
M.~C. Ba\~nuls, M.~B. Hastings, F.~Verstraete and J.~I. Cirac,
\newblock \emph{Matrix product states for dynamical simulation of infinite
  chains},
\newblock Phys. Rev. Lett. \textbf{102}, 240603 (2009),
\newblock \doi{10.1103/PhysRevLett.102.240603}.

\bibitem{muller2012tensor}
A.~M{\"u}ller-Hermes, J.~I. Cirac and M.~C. Ba{\~n}uls,
\newblock \emph{Tensor network techniques for the computation of dynamical
  observables in one-dimensional quantum spin systems},
\newblock New J. Phys. \textbf{14}(7), 075003 (2012),
\newblock \doi{10.1088/1367-2630/14/7/075003}.

\bibitem{osborne2006efficient}
T.~J. Osborne,
\newblock \emph{Efficient approximation of the dynamics of one-dimensional
  quantum spin systems},
\newblock Phys. Rev. Lett. \textbf{97}, 157202 (2006),
\newblock \doi{10.1103/PhysRevLett.97.157202}.

\bibitem{cirac2020matrix}
I.~Cirac, D.~Perez-Garcia, N.~Schuch and F.~Verstraete,
\newblock \emph{Matrix product states and projected entangled pair states:
  Concepts, symmetries, and theorems},
\newblock {a}rXiv:2011.12127 (2020), \eprint{https://arxiv.org/abs/2011.12127}.

\bibitem{calabrese2006time}
P.~Calabrese and J.~Cardy,
\newblock \emph{Time dependence of correlation functions following a quantum
  quench},
\newblock Phys. Rev. Lett. \textbf{96}, 136801 (2006),
\newblock \doi{10.1103/PhysRevLett.96.136801}.

\bibitem{calabrese2007quantum}
P.~Calabrese and J.~Cardy,
\newblock \emph{Quantum quenches in extended systems},
\newblock J. Stat. Mech. \textbf{2007}(06), P06008 (2007),
\newblock \doi{10.1088/1742-5468/2007/06/P06008}.

\bibitem{amico2008entanglement}
L.~Amico, R.~Fazio, A.~Osterloh and V.~Vedral,
\newblock \emph{Entanglement in many-body systems},
\newblock Rev. Mod. Phys. \textbf{80}, 517 (2008),
\newblock \doi{10.1103/RevModPhys.80.517}.

\bibitem{laflorencie2016quantum}
N.~Laflorencie,
\newblock \emph{Quantum entanglement in condensed matter systems},
\newblock Phys. Rep. \textbf{646}, 1 (2016),
\newblock \doi{10.1016/j.physrep.2016.06.008}.

\bibitem{li2008entanglement}
H.~Li and F.~D.~M. Haldane,
\newblock \emph{Entanglement spectrum as a generalization of entanglement
  entropy: Identification of topological order in non-abelian fractional
  quantum {H}all effect states},
\newblock Phys. Rev. Lett. \textbf{101}, 010504 (2008),
\newblock \doi{10.1103/PhysRevLett.101.010504}.

\bibitem{Linke2018}
N.~M. Linke, S.~Johri, C.~Figgatt, K.~A. Landsman, A.~Y. Matsuura and
  C.~Monroe,
\newblock \emph{Measuring the {R}\'enyi entropy of a two-site {F}ermi-{H}ubbard
  model on a trapped ion quantum computer},
\newblock Phys. Rev. A \textbf{98}, 052334 (2018),
\newblock \doi{10.1103/PhysRevA.98.052334}.

\bibitem{Elben2020mixed}
A.~Elben, R.~Kueng, H.~Y.~R. Huang, R.~van Bijnen, C.~Kokail, M.~Dalmonte,
  P.~Calabrese, B.~Kraus, J.~Preskill, P.~Zoller and B.~Vermersch,
\newblock \emph{Mixed-state entanglement from local randomized measurements},
\newblock Phys. Rev. Lett. \textbf{125}, 200501 (2020),
\newblock \doi{10.1103/PhysRevLett.125.200501}.

\bibitem{zhou2020single}
Y.~Zhou, P.~Zeng and Z.~Liu,
\newblock \emph{Single-copies estimation of entanglement negativity},
\newblock Phys. Rev. Lett. \textbf{125}, 200502 (2020),
\newblock \doi{10.1103/PhysRevLett.125.200502}.

\bibitem{alba2017quench}
V.~Alba and P.~Calabrese,
\newblock \emph{Quench action and {R}\'enyi entropies in integrable systems},
\newblock Phys. Rev. B \textbf{96}, 115421 (2017),
\newblock \doi{10.1103/PhysRevB.96.115421}.

\bibitem{yang1969thermodynamics}
C.~N. Yang and C.~P. Yang,
\newblock \emph{Thermodynamics of a one‐dimensional system of bosons with
  repulsive delta‐function interaction},
\newblock J. Math. Phys. \textbf{10}(7), 1115 (1969),
\newblock \doi{10.1063/1.1664947}.

\bibitem{takahashi1999thermodynamics}
M.~Takahashi,
\newblock \emph{Thermodynamics of one-dimensional solvable models},
\newblock Cambridge University Press,
\newblock \doi{10.1017/CBO9780511524332} (1999).

\bibitem{bertini2021finitetemperature}
B.~Bertini, F.~Heidrich-Meisner, C.~Karrasch, T.~Prosen, R.~Steinigeweg and
  M.~{\v{Z}}nidari{\v{c}},
\newblock \emph{Finite-temperature transport in one-dimensional quantum lattice
  models},
\newblock Rev. Mod. Phys. \textbf{93}(2), 025003 (2021),
\newblock \doi{10.1103/RevModPhys.93.025003}.

\bibitem{bernard2016conformal}
D.~Bernard and B.~Doyon,
\newblock \emph{Conformal field theory out of equilibrium: A review},
\newblock J. Stat. Mech. \textbf{2016}(6), 064005 (2016),
\newblock \doi{10.1088/1742-5468/2016/06/064005}.

\bibitem{vasseur2016nonequilibrium}
R.~Vasseur and J.~E. Moore,
\newblock \emph{Nonequilibrium quantum dynamics and transport: From
  integrability to many-body localization},
\newblock J. Stat. Mech. \textbf{2016}(6), 064010 (2016),
\newblock \doi{10.1088/1742-5468/2016/06/064010}.

\bibitem{alba2021generalizedhydrodynamic}
V.~Alba, B.~Bertini, M.~Fagotti, L.~Piroli and P.~Ruggiero,
\newblock \emph{Generalized-hydrodynamic approach to inhomogeneous quenches:
  Correlations, entanglement and quantum effects},
\newblock {a}rXiv:2104.00656 (2021), \eprint{https://arxiv.org/abs/2104.00656}.

\bibitem{bobenko1993two}
A.~Bobenko, M.~Bordemann, C.~Gunn and U.~Pinkall,
\newblock \emph{On two integrable cellular automata},
\newblock Commun. Math. Phys. \textbf{158}(1), 127 (1993),
\newblock \doi{10.1103/PhysRevLett.122.250603}.

\bibitem{prosen2016integrability}
T.~Prosen and C.~Mej{\' i}a-Monasterio,
\newblock \emph{Integrability of a deterministic cellular automaton driven by
  stochastic boundaries},
\newblock J. Phys. A: Math. Theor. \textbf{49}(18), 185003 (2016),
\newblock \doi{10.1088/1751-8113/49/18/185003}.

\bibitem{prosen2017exact}
T.~Prosen and B.~Bu{\v{c}}a,
\newblock \emph{Exact matrix product decay modes of a boundary driven cellular
  automaton},
\newblock J. Phys. A: Math. Theor. \textbf{50}(39), 395002 (2017),
\newblock \doi{10.1088/1751-8121/aa85a3}.

\bibitem{inoue2018two}
A.~Inoue and S.~Takesue,
\newblock \emph{Two extensions of exact nonequilibrium steady states of a
  boundary-driven cellular automaton},
\newblock J. Phys. A: Math. Theor. \textbf{51}(42), 425001 (2018),
\newblock \doi{10.1088/1751-8121/aadc29}.

\bibitem{klobas2019time}
K.~Klobas, M.~Medenjak, T.~Prosen and M.~Vanicat,
\newblock \emph{Time-dependent matrix product ansatz for interacting reversible
  dynamics},
\newblock Commun. Math. Phys. \textbf{371}(2), 651 (2019),
\newblock \doi{10.1007/s00220-019-03494-5}.

\bibitem{buca2019exact}
B.~Bu{\v c}a, J.~P. Garrahan, T.~Prosen and M.~Vanicat,
\newblock \emph{Exact large deviation statistics and trajectory phase
  transition of a deterministic boundary driven cellular automaton},
\newblock Phys. Rev. E \textbf{100}, 020103(R) (2019),
\newblock \doi{10.1103/PhysRevE.100.020103}.

\bibitem{klobas2020matrix}
K.~Klobas, M.~Vanicat, J.~P. Garrahan and T.~Prosen,
\newblock \emph{Matrix product state of multi-time correlations},
\newblock J. Phys. A: Math. Theor. \textbf{53}(33), 335001 (2020),
\newblock \doi{10.1088/1751-8121/ab8c62}.

\bibitem{klobas2020space}
K.~Klobas and T.~Prosen,
\newblock \emph{{Space-like dynamics in a reversible cellular automaton}},
\newblock SciPost Phys. Core \textbf{2}, 10 (2020),
\newblock \doi{10.21468/SciPostPhysCore.2.2.010}.

\bibitem{gopalakrishnan2018operator}
S.~Gopalakrishnan,
\newblock \emph{Operator growth and eigenstate entanglement in an interacting
  integrable {F}loquet system},
\newblock Phys. Rev. B \textbf{98}, 060302(R) (2018),
\newblock \doi{10.1103/PhysRevB.98.060302}.

\bibitem{gopalakrishnan2018hydrodynamics}
S.~Gopalakrishnan, D.~A. Huse, V.~Khemani and R.~Vasseur,
\newblock \emph{Hydrodynamics of operator spreading and quasiparticle diffusion
  in interacting integrable systems},
\newblock Phys. Rev. B \textbf{98}, 220303(R) (2018),
\newblock \doi{10.1103/PhysRevB.98.220303}.

\bibitem{friedman2019integrable}
A.~J. Friedman, S.~Gopalakrishnan and R.~Vasseur,
\newblock \emph{Integrable many-body quantum {F}loquet-{T}houless pumps},
\newblock Phys. Rev. Lett. \textbf{123}, 170603 (2019),
\newblock \doi{10.1103/PhysRevLett.123.170603}.

\bibitem{alba2019operator}
V.~Alba, J.~Dubail and M.~Medenjak,
\newblock \emph{Operator entanglement in interacting integrable quantum
  systems: The case of the {R}ule 54 chain},
\newblock Phys. Rev. Lett. \textbf{122}, 250603 (2019),
\newblock \doi{10.1103/PhysRevLett.122.250603}.

\bibitem{alba2021diffusion}
V.~Alba,
\newblock \emph{Diffusion and operator entanglement spreading},
\newblock Phys. Rev. B \textbf{104}, 094410 (2021),
\newblock \doi{10.1103/PhysRevB.104.094410}.

\bibitem{buca2021rule}
B.~Bu\v{c}a, K.~Klobas and T.~Prosen,
\newblock \emph{Rule 54: Exactly solvable model of nonequilibrium statistical
  mechanics},
\newblock J. Stat. Mech. \textbf{2021}(7), 074001 (2021),
\newblock \doi{10.1088/1742-5468/ac096b}.

\bibitem{klobas2020exactPhD}
K.~Klobas,
\newblock \emph{Exact time-dependent solutions of interacting systems},
\newblock Ph.D. thesis, Univerza v Ljubljani, Fakulteta za matematiko in fiziko
  (2020),
  \eprint{https://repozitorij.uni-lj.si/IzpisGradiva.php?id=120396&lang=eng}.

\bibitem{rubel1956necessary}
L.~A. Rubel,
\newblock \emph{Necessary and sufficient conditions for {C}arlson's theorem on
  entire functions},
\newblock Trans. Am. Math. Soc. \textbf{83}(2), 417 (1956),
\newblock \doi{10.2307/1992882}.

\bibitem{bertini2018entanglementand}
B.~Bertini, E.~Tartaglia and P.~Calabrese,
\newblock \emph{Entanglement and diagonal entropies after a quench with no pair
  structure},
\newblock J. Stat. Mech. \textbf{2018}(6), 063104 (2018),
\newblock \doi{10.1088/1742-5468/aac73f}.

\bibitem{bastianello2018spreading}
A.~Bastianello and P.~Calabrese,
\newblock \emph{Spreading of entanglement and correlations after a quench with
  intertwined quasiparticles},
\newblock SciPost Phys. \textbf{5}, 33 (2018),
\newblock \doi{10.21468/SciPostPhys.5.4.033}.

\bibitem{cao2019entanglement}
X.~Cao, A.~Tilloy and A.~D. Luca,
\newblock \emph{Entanglement in a fermion chain under continuous monitoring},
\newblock SciPost Phys. \textbf{7}, 24 (2019),
\newblock \doi{10.21468/SciPostPhys.7.2.024}.

\bibitem{alba2021spreading}
V.~Alba and F.~Carollo,
\newblock \emph{Spreading of correlations in {Markovian} open quantum systems},
\newblock Phys. Rev. B \textbf{103}, L020302 (2021),
\newblock \doi{10.1103/PhysRevB.103.L020302}.

\bibitem{carollo2021emergent}
F.~Carollo and V.~Alba,
\newblock \emph{Emergent dissipative quasi-particle picture in noninteracting
  {Markovian} open quantum systems},
\newblock {a}rXiv:2106.11997 (2021), \eprint{https://arxiv.org/abs/2106.11997}.

\bibitem{calabrese2017lecture}
P.~Calabrese,
\newblock \emph{Entanglement and thermodynamics in non-equilibrium isolated
  quantum systems},
\newblock Physica A \textbf{504}, 31 (2018),
\newblock \doi{10.1016/j.physa.2017.10.011}.

\bibitem{bonnes2014light}
L.~Bonnes, F.~H.~L. Essler and A.~M. L\"auchli,
\newblock \emph{``{L}ight-cone'' dynamics after quantum quenches in spin
  chains},
\newblock Phys. Rev. Lett. \textbf{113}, 187203 (2014),
\newblock \doi{10.1103/PhysRevLett.113.187203}.

\bibitem{alba2017renyi}
V.~Alba and P.~Calabrese,
\newblock \emph{R{\'e}nyi entropies after releasing the {N}{\'e}el state in the
  {XXZ} spin-chain},
\newblock J. Stat. Mech. \textbf{2017}(11), 113105 (2017),
\newblock \doi{10.1088/1742-5468/aa934c}.

\bibitem{prosen2007operator}
T.~Prosen and I.~Pi{\v z}orn,
\newblock \emph{Operator space entanglement entropy in a transverse {Ising}
  chain},
\newblock Phys. Rev. A \textbf{76}(3) (2007),
\newblock \doi{10.1103/physreva.76.032316}.

\bibitem{zanardi2001entanglement}
P.~Zanardi,
\newblock \emph{Entanglement of quantum evolutions},
\newblock Phys. Rev. A \textbf{63}(4), 040304 (2001),
\newblock \doi{10.1103/PhysRevA.63.040304}.

\bibitem{zhou2020entanglement}
T.~Zhou and A.~Nahum,
\newblock \emph{Entanglement membrane in chaotic many-body systems},
\newblock Phys. Rev. X \textbf{10}, 031066 (2020),
\newblock \doi{10.1103/PhysRevX.10.031066}.

\bibitem{descartes1954geometry}
R.~Descartes, D.~E. Smith and M.~L. Latham,
\newblock \emph{The geometry of Rene Descartes},
\newblock Dover Publications, Inc.,
\newblock ISBN 9780486600680 (1954).

\bibitem{rudin1987real}
W.~Rudin,
\newblock \emph{Real and complex analysis},
\newblock Higher Mathematics series. McGraw-Hill,
\newblock ISBN 9780070542341 (1987).

\end{thebibliography}

\end{document}